\documentclass[11pt]{article}
\usepackage[a4paper, margin=1in]{geometry}
\usepackage{alltt}
\usepackage[utf8]{inputenc}
\usepackage[T1]{fontenc}
\usepackage{./packages/isabelle,./packages/isabellesym}
\usepackage{mathtools}
\usepackage{amsmath} % official of American Mathematical Society
\usepackage{amssymb} % adds useful symbols like \Cap and \Cup
\usepackage{amsthm}
\usepackage{longtable}
\usepackage{manyfoot}
\usepackage{multicol}
\usepackage{stmaryrd}
\usepackage{enumerate} % to change indices in enumerates
\usepackage{lipsum}  % dummy text
\usepackage{color}
\usepackage[colorinlistoftodos]{todonotes} % comments on the paper with 
\usepackage{xspace}
\usepackage{bussproofs}
\usepackage{alltt}
\usepackage{subcaption}
\usepackage[colorlinks=true]{hyperref} % for hyperlinks in PDF, ensure it is last package

\hypersetup{
    colorlinks,
    linkcolor={red!50!black},
    citecolor={blue!50!black},
    urlcolor={blue!80!black}
  }
\graphicspath{ {./images/} }

\theoremstyle{thmstylethree}

\newtheorem{example}{Example}

\theoremstyle{thmstyleone}

\DeclareMathOperator{\Pow}{\mathcal{P}}

\newcommand{\nats}{\mathbb{N}}
\newcommand{\bools}{\mathbb{B}}
\newcommand{\reals}{\mathbb{R}}
\newcommand{\id}{\mathit{id}}

\newcommand{\downcl}[2]{{#2}{{\downarrow}_{#1}}}
\renewcommand{\implies}{\Rightarrow}
\renewcommand{\iff}{\Leftrightarrow}

\newcommand{\prog}{\alpha}
\newcommand{\ivpsols}{\mathit{ivp}\texttt{-}\mathit{sols}}
\newcommand{\diffinv}{\mathit{diff}\texttt{-}\mathit{inv}}

\newcommand{\seqcomp}{\mathbin{;}}
\newcommand{\IF}[3]{\texttt{if } #1\texttt{ then } #2\texttt{ else } #3}
\newcommand{\WHILE}[2]{\texttt{while } #1\texttt{ do } #2}
\newcommand{\LOOP}[1]{\texttt{loop } #1}
\newcommand{\INV}[2]{#1\texttt{ inv }#2}
\newcommand{\SKIP}{\texttt{skip} }
\newcommand{\ABORT}{\texttt{abort}}
\newcommand{\mquestiondown}{\mbox{\textquestiondown}}
\newcommand{\mtest}[1]{\mquestiondown #1 ?}
\newcommand{\fddia}[2]{\mathop{|#1\rangle}#2}

\newcommand{\fdbox}[2]{\mathop{|#1]}#2}
\newcommand{\wlp}{\textit{\textsf{wlp}}}

\renewcommand{\wp}{\textit{\textsf{wp}}}
\newcommand{\hoare}[3]{\left\{#1\right\}\,#2\,\left\{#3\right\}}

\newcommand{\dL}{$\mathsf{d}\mathcal{L}$\xspace}

\newcommand{\lens}{x}
\newcommand{\lto}{\Rightarrow}
\newcommand{\lput}{\textit{\textsf{put}}}
\newcommand{\lget}{\textit{\textsf{get}}}
\newcommand{\lindep}{\mathop{\,\bowtie\,}}
\newcommand{\view}{\mathcal{V}}
\newcommand{\src}{\mathcal{S}}
\newcommand{\csrc}{\mathcal{C}}
\newcommand{\lsubst}[3]{#1[#2/#3]}
\newcommand{\lsugarget}[2]{#2_{#1}}
\newcommand{\ldowngr}[3]{#1{\downarrow}^{#2}_{#3}}
\newcommand{\lupgr}[3]{#1{\uparrow}^{#2}_{#3}}
\newcommand{\assigns}[1]{\langle #1 \rangle}
\newcommand{\fderiv}[3]{\mathcal{D}^{#1}_{#3}#2}
\newcommand{\lcomp}{\fatsemi}
\newcommand{\lquot}{\mathop{\!\sslash\!}}
\newcommand{\lplus}{\oplus}
\newcommand{\lequiv}{\cong}
\newcommand{\lone}{\mathbf{1}}

\newcommand{\unrest}{\mathop{\sharp\,}}

\newcommand{\substlk}[1]{\langle #1 \rangle_s}

\newcommand{\smapsto}{\leadsto}

\newcommand{\nmods}{\mathop{\textit{\textsf{nmods}}\,}}
\newcommand{\hoaretriple}[3]{\left\{#1\right\}\,#2\,\left\{#3\right\}}

\newcommand{\relsemi}{\mathop{\fatsemi\,}}
\newcommand{\ebrack}[1]{(#1)_{\skey{e}}}
\newcommand{\ebracku}[1]{(#1)^{\skey{e}}}

\definecolor{gscolor}{cmyk}{1,0,0,0}
\definecolor{jhcolor}{cmyk}{0.36,0.44,0,0.07}
\definecolor{sfcolor}{cmyk}{0,1,0,0}
\definecolor{rccolor}{cmyk}{1,0,0,0}
\definecolor{mgcolor}{cmyk}{1,0,0,0}
\newcounter{law}
\newcommand{\nextlaw}{\refstepcounter{law}}
\makeatletter

\def\orcidID#1{\relax{\href{https://orcid.org/#1}{\protect\raisebox{-1.25pt}{\protect\includegraphics{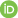}}}}}

\newcommand{\leqnomode}{\tagsleft@true\let\veqno\@@leqno}
\newcommand{\reqnomode}{\tagsleft@false\let\veqno\@@eqno}
\newcommand\customtag[2]{\nextlaw#1\def\@currentlabel{#1}\label{#2}}
\makeatother
\definecolor{IsaBlue}{cmyk}{1.0,0.33,0,.4}
\definecolor{IsaGreen}{cmyk}{1.0,0,1.0,.2}
\definecolor{IsaRed}{cmyk}{0,0.69,0.69,0}
\definecolor{IsaDarkRed}{cmyk}{0,1.0,1.0,0.15}

\newcommand{\isalogo}{\includegraphics[width=9pt]{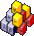}}
\newcommand{\isalink}[1]{\href{#1}{\isalogo}}

\newcommand{\isakwmaj}[1]{\textcolor{IsaBlue}{\textbf{\texttt{#1}}}}
\newcommand{\isakwmin}[1]{\textcolor{IsaGreen}{\textbf{\texttt{#1}}}}

\newcommand{\isacmt}[1]{\textcolor{IsaDarkRed}{\texttt{#1}}}

\isabellestylett

\renewcommand{\isastyle}{\isastyleminor}

\newcommand{\isakw}[1]{{\bfseries\ttfamily\def\isachardot{.}\def\isacharunderscore{\isacharunderscorekeyword}%
\def\isacharbraceleft{\{}\def\isacharbraceright{\}}#1}}

\renewcommand{\isacommand}[1]{\textcolor{IsaBlue}{\isakw{#1}}}
\renewcommand{\isakeyword}[1]{\textcolor{IsaGreen}{\isakw{#1}}}

\renewcommand{\isastyletext}{\normalsize\normalfont}

\newcommand{\labelthis}[2]{%
  \def\@currentlabel{#2}\label{#1}{\scriptsize \textnormal{(#2)}}%
}
\makeatother

\newcommand{\real}{\mathbb{R}}
\newcommand{\skey}[1]{\textit{\textsf{#1}}}

\newcommand{\syneq}{\mathop{\,\rightleftharpoons\,}}

\newtheorem{nonformat-algorithm}{Algorithm}%
\usepackage{algorithm2e}

\begin{document}

\title{Scalable Automated Verification for Cyber-Physical Systems in Isabelle/HOL}
\author{Jonathan Juli\'an Huerta y Munive\orcidID{0000-0003-3279-3685}\\
Czech Institute of Informatics, Robotics and Cybernetics\\
Prague, Czechia
\and
Simon Foster\orcidID{0000-0002-9889-9514}\\
University of York\\
United Kingdom
\and
Mario Gleirscher\orcidID{0000-0002-9445-6863}\\
University of Bremen\\
Germany
\and
Georg Struth\orcidID{0000-0001-9466-7815}\\
University of Sheffield\\
United Kingdom
\and
Christian Pardillo Laursen\orcidID{0000-0001-7838-2764}\\
University of York\\
United Kingdom
\and
Thomas Hickman\orcidID{0000-0001-7216-1696}\\
Genomics PLC\\
United Kingdom
}

\date{22/January/2024}

\maketitle

\begin{abstract}
We formally introduce IsaVODEs (Isabelle verification with Ordinary Differential Equations), a framework for the verification of cyber-physical systems. We describe the semantic foundations of the framework's formalisation in the Isabelle/HOL proof assistant. A user-friendly language specification based on a robust state model makes our framework flexible and adaptable to various engineering workflows. New additions to the framework increase both its expressivity and proof automation. Specifically, formalisations related to forward diamond correctness specifications, certification of unique solutions to ordinary differential equations (ODEs) as flows, and invariant reasoning for systems of ODEs contribute to the framework's scalability and usability. Various examples and an evaluation validate the effectiveness of our framework.
\vspace{\baselineskip}

\end{abstract}

% Additional contributions beyond FM2021 paper: 
% (1) new dG laws & Darboux
% (2) new procedure to check uniqueness
% (3) Diamond laws, with forward and backward variants
% (4) CAS Integration for providing solutions to ODEs
% (5) Evaluation using the ARCH2022 benchmarks

% Additional contributions beyond FM2021 paper: 
% (1) new dG laws & Darboux
% (2) new procedure to check uniqueness
% (3) Diamond laws, with forward and backward variants
% (4) CAS Integration for providing solutions to ODEs
% (5) Evaluation using the ARCH2022 benchmarks

%%%% INTRODUCTION %%%%
\section{Introduction}

%\looseness=-1
Cyber-physical systems (CPSs) are computerised systems whose software (the ``cyber'' part) interacts with their physical environment. The software is modelled as a variable-updating, potentially non-deterministic program, while the environment is modelled by a system of ordinary differential equations (ODEs). When CPSs interact with humans, for example through robotic manipulators, they are invariably safety-critical, which makes their design verification imperative.

% Robotic systems are an important example and subclass of CPSs: physical robots are often designed to accomplish non-trivial tasks through software-controlled motion of various manipulators.

\looseness=-1
However, CPS verification is challenging because of the complex interactions between the software, hardware, and the physical environment. These produce uncountably infinite state spaces, which makes verification generally intractable, and so requires the use of abstractions. A deductive verification approach uses symbolic logics, which support the encoding of solutions and invariants of ODEs. Such an approach has been implemented with the KeYmaera X tool (KYX)~\cite{KeYmaera}, which models CPSs via hybrid programs, and verifies their behaviour with differential dynamic logic (\dL). Numerous case studies and competitions support the applicability of this approach~\cite{JeanninGKSGMP17, LoosPN11, ARCH22, XiangFC21}.

\looseness=-1
General-purpose interactive theorem provers (ITPs), like Coq, Lean, and Isabelle, also support CPS verification~\cite{MuniveS22, FosterMS20}. Their track record for large-scale deductive verification is well-documented~\cite{AlmeidaBBBGLOPS17, BodinCFGMNSS14, Klein2009, Lammich19LLVM, LeroyBKSPF2016}. A significant reason for these successes is the combination of expressive logical languages, and plugins to enhance reasoning capacity and usability. Different from bespoke axiomatic provers like KYX, these tools are inherently extensible since the mathematics is constructed from first principles, rather than postulated, with accompanying soundness guarantees. For example, the recent addition of transcendental functions to KYX necessitated an extension of the tool, whereas these have been available in HOL, Coq, and Isabelle for many years. Moreover, the foundational mathematical libraries are under constant development by the community\footnote{See Isabelle's Archive of Formal Proofs (\url{https://www.isa-afp.org/statistics/}) and Lean's Mathematical Library (\url{https://leanprover-community.github.io/mathlib_stats.html}).}, and so ITP-based verification tools benefit from an ever-growing library of definitions and theorems.

Our ITP of choice, Isabelle/HOL, provides a framework supporting assured software engineering~\cite{Klein2009, BasinDHHMKKMRST22}, through an open extensible architecture; a flexible syntax pipeline~\cite{MakariusIsarManual, Paulson96}; integration of external analysis tools in \textsf{Sledgehammer}, \textsf{Nitpick}, and cousins; and output of software artefacts in a variety of languages~\cite{Haftman2010-CodeGen}. Isabelle's rich library for multi-variate analysis (\textsf{HOL-Analysis}) can be applied to CPS verification~\cite{HolzlIH13, ImmlerT16}. Thanks to its higher-order logic, Isabelle supports several modelling facilities essential to control engineering, such as vectors, matrices, and transcendental functions (e.g. $\sin$, $\cos$, and $e^x$)~\cite{AransayD17, Munive20}, and it has extended reasoning capability through its integration with SMT solvers (CVC4, Z3, Vampire)~\cite{PaulsonB10}. Isabelle is therefore ideally suited to reasoning about CPSs, but to harness its facilities we need an accessible and powerful CPS modelling language and verification framework.

\looseness=-1
In this paper we contribute an Isabelle-based verification framework called IsaVODEs (Isabelle Verification with Ordinary Differential Equations)~\cite{MuniveS22, FosterMS20}. IsaVODEs provides a textual language for modelling CPSs, which is constructed as a shallow embedding in Isabelle. The language is equipped with nondeterministic state-transformer-based program semantics and a flexible hybrid store model based on lenses~\cite{Foster2020-IsabelleUTP, FosterMGS21}. The program model provides several verification calculi, including Hoare logic and dynamic logic, with modalities for specifying both safety and reachability properties. The store model allows software models to benefit from the full generality of the Isabelle type system, including continuous structures like vectors and matrices, and discrete structures like algebraic data types. Our language can therefore scale to systems with both realistic dynamics, and complex control structures. We harness Isabelle's syntax translation mechanisms to provide a user-friendly frontend for the language, including declarative context, program notation, and operators for arithmetic, vectors, and matrices. Our technical solution is extensible 
and endeavours to resemble languages like Modelica, Mathematica, and MATLAB.
% Given all these benefits, in this paper, we contribute to scaling the automation of an Isabelle-based CPS verification
% framework~\cite{MuniveS22, FosterMS20} through various orthogonal formalisations, packages and integrations.

\looseness=-1
Verification of models in IsaVODEs is supported by a library of deduction rules. For reasoning about ODEs, we support both the use of solutions with flow functions, and differential induction, which avoids the need for solutions. We integrate two Computer Algebra Systems (CASs), namely Mathematica and SageMath, into Isabelle for the supply of symbolic solutions to Lipschitz-continuous ODEs in an analogous way to \textsf{sledgehammer}. We also support both a \dL-style differential ghost rule and Darboux rules that 
enhance IsaVODEs' invariant reasoning capabilities for complex continuous dynamics.

\looseness=-1
Orthogonal to this, we also contribute novel laws to support local reasoning in the style of separation logic. We can infer the frame of a program (i.e. the mutable variables), and use a frame rule to prove invariants over variables outside the frame. This is accompanied by a novel form of differentiation called ``framed Fr\'echet derivatives'', which allow us to differentiate with respect to a strict subset of the store variables, particularly ignoring any discrete variables lacking a topological structure. When variables are outside of the frame of a system of ODEs, those variables are unchanged during continuous evolution. Local reasoning further facilitates scalability of our tool, by allowing a verification task to be decomposed into modules supported by individual assumptions and guarantees.

\looseness=-1
We supply several proof methods to support automated proof and verification condition generation (VCG), including the application of differential induction, certification of ODE solutions, and the various Hoare-logic laws. Care is taken to ensure that the resulting proof obligations are presented in a way that preserves abstraction, and elides irrelevant details of the Isabelle mechanisation, to maintain the user experience. Harnessing Isabelle's Eisbach tool~\cite{MatichukMW16}, we also employ various high-level search-based proof methods, which exhaustively apply the proof rules to compute a complete set of VCs. The VCs can finally be tackled in the usual way using tools like \textsf{auto} and \textsf{sledgehammer}.

\looseness=-1
The framework has been tested successfully on a large set of hybrid verification benchmarks~\cite{MitschMJZWZ20}, and many larger examples. Our enhancements yield at least the same performance in essential verification tasks as that of domain-specific provers. Initial case studies suggest a simplification of user interaction. Our new components can be found online\footnote{\href{https://github.com/isabelle-utp/Hybrid-Verification}{\texttt{github.com/isabelle-utp/Hybrid-Verification}}, also by clicking our icons\ \isalogo.}. Our contributions are highlighted through examples, while additional contributions are noted throughout the text. 

\looseness=-1 This article is a substantial extension of previously published work at the Formal Methods symposium
(FM2021)~\cite{FosterMGS21}. Our additional contributions include generalisations of the laws for
frames~(\S\ref{subsec:frames}), differential ghosts, and Darboux rules~(\S\ref{subsec:darboux}); proof methods for
certification of continuity, Lipschitz continuity, and uniqueness of solutions~(\S\ref{subsec:lipschitz}); extensions to
our previous proof methods~(\S\ref{sec:tactics}); the CAS integrations~(\S\ref{sec:cas}); a more substantial
evaluation~(\S\ref{sec:eval}); and several additional examples~(\S\ref{sec:planar-flight}, \S\ref{sec:ex}). As a result of these enhancements, the
usability of the tool is more mature and polished.

\looseness=-1
We begin our paper by illustrating the framework with a small 
case study (\S\ref{sec:planar-flight}). We then expound our formalisation of 
dynamical systems (\S\ref{subsec:dynsys}), the
state-transformer hybrid program model (\S\ref{subsec:hp}), the foundations of the store model (\S\ref{subsec:stores}), the specification of continuous evolution (\S\ref{subsec:evolcoms}), and finally VCG though our framework's algebraic foundations (\S\ref{subsec:vcg}). This includes both safety, and also reachability and termination (\S\ref{subsec:fdia}).

\looseness=-1
Next, in \S\ref{sec:hs}, we introduce our framework's hybrid modelling language. We supply commands for generating hybrid stores, with variables, constants, and foundational properties (\S\ref{subsec:dataspaces}), and user-friendly notation for expressions (\S\ref{subsec:subst}), matrices, and vectors (\S\ref{subsec:vecmat}). The notation is automatically processed via rewriting rules supplied to 
Isabelle's simplifier, hiding implementation details. 
% Our technical solution is meant to be extensible 
% and resembles modelling languages such as Modelica or computer algebra systems (CAS) such as 
% Mathematica. Our store model supports local reasoning, and 
% therefore, it allows us to refine the framework's deductive and calculational rules, including 
% inference rules \`a la \dL~\cite{Platzer18}, a new frame rule \`a la separation logic,

In \S\ref{sec:frames} we present laws for local reasoning about frames (\S\ref{subsec:frames}), the
% With these additions we can individually verify components of a CPS model that satisfy disjoint parts of the 
% state space, which promotes scalable reasoning.
 seamless integration of framed ODEs into our hybrid programs (\S\ref{subsec:dynevol}), and automatic certification of ODE invariants through framed Fr\'echet derivatives (\S\ref{subsec:dinv}). These also serve to supply new \dL-style ghost and Darboux rules that 
enhance IsaVODEs' invariant reasoning (\S\ref{subsec:darboux}).

% It uses lenses, a refined hybrid program store model, to reason about 
% mutable and immutable, logical and program variables locally~\cite{Foster2020-IsabelleUTP,FosterMGS21}. This allows IsaVODEs to use Isabelle's syntax translation mechanisms modularly, creating a more 
% user-friendly specification language for the framework (Section~\ref{subsec:stores}). Lenses also provide support for 
% local reasoning, enabling a distinction between continuous and discrete variables, and integration of both types
% of variables into hybrid programs (Section~\ref{subsec:evolcoms}). Finally, the framework's algebraic foundations 
% naturally give rise to deductive rules for verification condition generation (VCG, Section~\ref{subsec:vcg}). 

\looseness=-1 We contribute increased IsaVODEs proof-capabilities for the verification process. We formalise theorems
necessary for increased automation, like the fact that differentiable functions are Lipschitz-continuous
(\S\ref{subsec:lipschitz}) or the first and second derivative test laws (\S\ref{subsec:deriv-tests}).  We provide
methods to automate certification of differentiation (\S\ref{subsec:auto-deriv}), uniqueness of solutions
(\S\ref{subsec:auto-flow}), invariants for ODEs (\S\ref{subsec:auto-dinduct}), and VCG
(\S\ref{subsec:auto-flow-vcg}). Additional automation is provided through the integration of Mathematica and SageMath
(\S\ref{sec:cas}). We then bring all of the results together for the evaluation (\S\ref{sec:eval}) using benchmarks and
examples (\S\ref{sec:ex}). In the remaining sections we review related work (\S\ref{sec:related-work}), provide
concluding statements, and discuss future research avenues (\S\ref{sec:conclusions}).

% \sfin{Add more details about what was achieved in the evaluation. Our tool covers most the features, and goes beyond them.}
% \sfin{Think about motivation in terms of reachability analysis (cf. SpaceEx etc.)}
% \sfin{Adapt existing FM introduction, and make argument about how our tool can support software engineering with good theoretical foundations}

%\jhin{Formally present IsaVODEs?}
%List of contributions:
%\begin{itemize}
%\item Forward diamond laws
%\item Differentiability implies Lipschitz (more automation for flow-certification)
%\item First and second derivatives tests (more reasoning techniques for real-arithmetic)
%\item Generalised ghost rule (more problems can be tackled via differential induction)
%\item Mathematica integration
%\item Improved tactics (more automation overall)
%\end{itemize}

\section{Motivating Case Study: Flight Dynamics}\label{sec:planar-flight}

\begin{figure}
    \centering
    \begin{subfigure}{0.45\textwidth}
    \includegraphics[width=\textwidth]{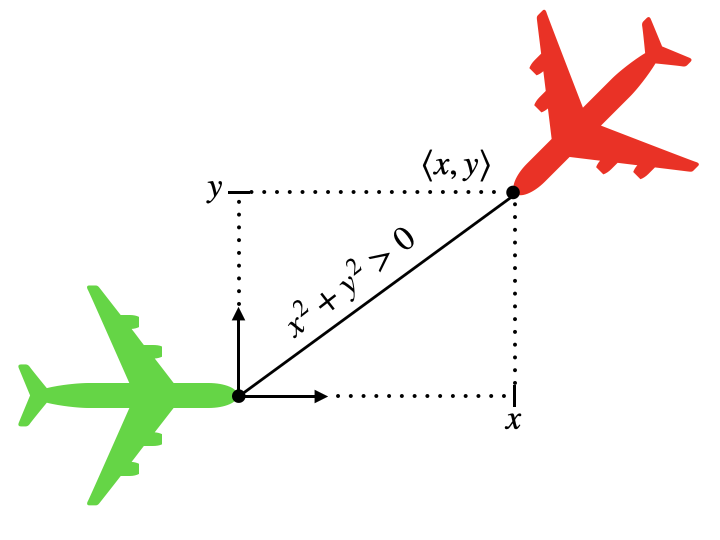}
    \caption{Plane x and y position}
    \label{fig:plane-dist}
    \end{subfigure}
    \hfill
    \begin{subfigure}{0.45\textwidth}
        \includegraphics[width=\textwidth]{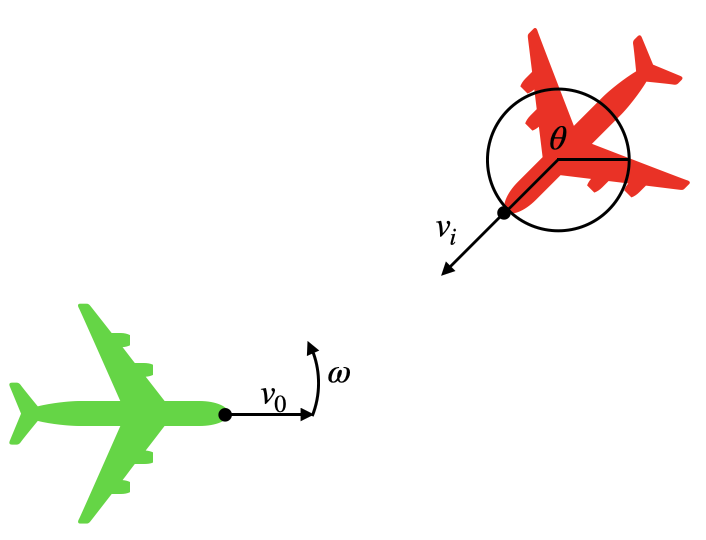}
        \caption{Velocities and angle}
        \label{fig:velocities}
    \end{subfigure}
    \hfill
    \begin{subfigure}{0.45\textwidth}
        \includegraphics[width=\textwidth]{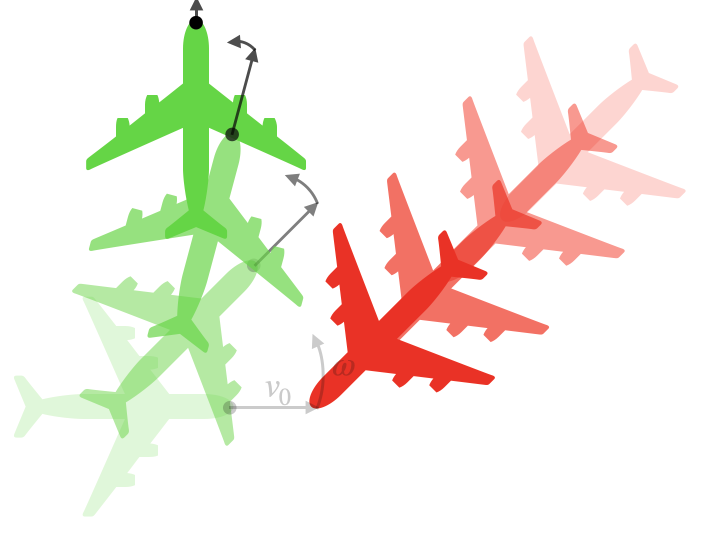}
        \caption{Evasive maneuvers}
        \label{fig:evasion}
    \end{subfigure}
    \hfill
    \begin{subfigure}{0.45\textwidth}
        \includegraphics[width=\textwidth]{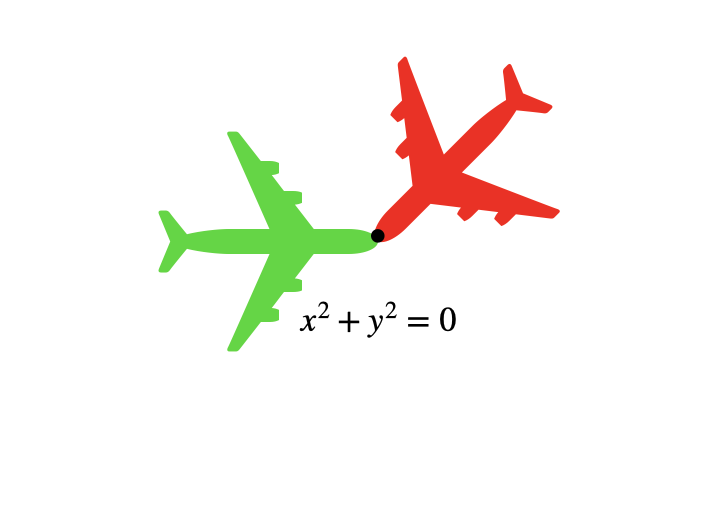}
        \caption{Collision}
        \label{fig:collision}
    \end{subfigure}
    \caption{Diagrams illustrating the flight dynamics example}
    \label{fig:plane-diagrams}
\end{figure}

In this section, we motivate and demonstrate our framework's usage with a worked example: an aircraft collision avoidance scenario that was first presented by Mitsch et. al.~\cite{Mitsch2018}. It describes an aircraft trying to avoid a collision with a nearby intruding aircraft travelling at the same altitude. We can model this intruding aircraft by considering its coordinates \(x\) and \(y\), and angle \(\vartheta\), in the reference frame of our own ship. The intruding plane has fixed velocity \(v_i > 0\), and our plane has fixed velocity \(v_o > 0\) and variable angular velocity \(\omega\). This is illustrated in Figure~\ref{fig:plane-dist} and~\ref{fig:velocities}.
%\cpin{Add flight diagram}

With our tool, we can model this using a \isakwmaj{dataspace} command, as shown below:\hfill\isalink{https://github.com/isabelle-utp/Hybrid-Verification/blob/7981dd8beae46d5220b30d5e59a9d5a62ce353f0/Hybrid_Programs/Verification_Examples/HS_Lens_Examples.thy\#L754}
\vspace{1ex}
\begin{center}
\begin{alltt}
  \isakwmaj{dataspace} planar_flight =
    \isakwmin{constants}
      v\(\sb{o}\) :: real \isacmt{(* own_velocity *)}
      v\(\sb{i}\) :: real \isacmt{(* intruder velocity *)}
    \isakwmin{assumes}
      v\(\sb{o}\)_pos: "v\(\sb{o}\) > 0" and
      v\(\sb{i}\)_pos: "v\(\sb{i}\) > 0"
    \isakwmin{variables} \isacmt{(* Coordinates in reference frame of own ship *)}
      x :: real \isacmt{(* Intruder x *)}
      y :: real \isacmt{(* Intruder y *)}
      \(\theta\) :: real \isacmt{(* Intruder angle *)}
      \(\omega\) :: real \isacmt{(* Angular velocity *)}
\end{alltt}
\end{center}
\vspace{1ex}
\noindent This command allows us to define our constants, any assumptions about the verification problem, and the system's variables. In this case, we postulate two constants $v_o$ and $v_i$, both of type \texttt{real}, for the velocity of the aircraft and intruder, respectively. Here, \texttt{real} is the type of precise mathematical real numbers, as opposed to floating points or rationals.%, which are modelled using Cauchy sequences in Isabelle/HOL.

As per the problem statement, we assume that both of these constants are strictly positive. This is specified by the assumptions called \texttt{v${}_o$\_pos} and \texttt{v${}_i$\_pos}, respectively, which can be used as hypotheses in proofs. We supply the variables of this system, which give the relative position of the intruder (\texttt{x} and \texttt{y}), its orientation $\theta$, and its angular velocity $\omega$. 

We next specify the ODEs that model the system as a constant called \texttt{plant}:
\vspace{1ex}
\begin{center}
\begin{alltt}
  \isakwmaj{definition} "plant \(\equiv\) \{x` = v\(\sb{i}\) * cos \(\theta\) - v\(\sb{o}\) + \(\omega\) * y,
                       y` = v\(\sb{i}\) * sin \(\theta\) - \(\omega\) * x,
                       \(\theta\)` = -\(\omega\)\}"
\end{alltt}
\vspace{1ex}                     
\end{center}

\noindent The ODEs can be specified in a user-friendly manner, as physicists and engineers would informally state them, in terms of the constants and variables of the system \texttt{x}, \texttt{y}, and $\varphi$.

Next, we define a simple controller to avoid collisions, as explained below: 
\vspace{1ex}
\begin{center}
\begin{alltt}
  \isakwmaj{abbreviation} "I \(\equiv\) (v\(\sb{i}\) * sin \(\theta\) * x - (v\(\sb{i}\) * cos \(\theta\) - v\(\sb{o}\)) * y
                     > v\(\sb{o}\) + v\(\sb{i}\))\(\sp{e}\)"

  \isakwmaj{abbreviation} "J \(\equiv\) (v\(\sb{i}\) * \(\omega\) * sin \(\theta\) * x - v\(\sb{i}\) * \(\omega\) * cos \(\theta\) * y 
                    + v\(\sb{o}\) * v\(\sb{i}\) * cos \(\theta\) 
                    > v\(\sb{o}\) * v\(\sb{i}\) + v\(\sb{i}\) * \(\omega\))\(\sp{e}\)"

  \isakwmaj{definition} "ctrl \(\equiv\) (\(\omega\) ::= 0; ¿I?) \(\sqcap\) (\(\omega\) ::= 1; ¿J?)"
\end{alltt}
\end{center}
\vspace{1ex}
This is based on two invariants, $I$ and $J$, for two different scenarios. These are properties that remain true after the unfolding of each scenario. The invariant $I$ holds when going straight is safe, while $J$ holds when an evasion manoeuvre is allowed. Our controller selects whether to set our aircraft's angular velocity to 0 or 1 depending on which invariant holds ($I$ and $J$ respectively). The evasive manoeuvres that occur when the angular velocity is set to 1 are illustrated in figure \ref{fig:evasion}.

Finally, our model follows the usual structure of an iteration ($^*$) of control choices (\texttt{ctrl}) followed by a nondeterministic evolution of the system dynamics (\texttt{plant}):

\vspace{1ex}
\begin{center}
\begin{alltt}
  \isakwmaj{definition} "flight \(\equiv\) (ctrl; plant)\(\sp{*}\)"
\end{alltt}
\end{center}
\vspace{1ex}

\noindent The behaviour of the system is characterised by all states reachable by executing the controller followed by the plant a finite number of times.

% \cpin{Comment on solving the differential equation with find\_local\_flow}

Next, we show how we can formally verify the collision avoidance of this system. We do this by specifying a Hoare triple within an Isabelle lemma:\hfill\isalink{https://github.com/isabelle-utp/Hybrid-Verification/blob/7981dd8beae46d5220b30d5e59a9d5a62ce353f0/Hybrid_Programs/Verification_Examples/HS_Lens_Examples.thy\#L786}

\begin{isabellebody}
\isanewline
\isacommand{lemma}\isamarkupfalse%
\ flight{\isacharunderscore}{\kern0pt}safe{\isacharcolon}{\kern0pt}\ {\isachardoublequoteopen}\isactrlbold {\isacharbraceleft}{\kern0pt}x\isactrlsup {\isadigit{2}}\ {\isacharplus}{\kern0pt}\ y\isactrlsup {\isadigit{2}}\ {\isachargreater}{\kern0pt}\ {\isadigit{0}}\isactrlbold {\isacharbraceright}{\kern0pt}\ flight\ \isactrlbold {\isacharbraceleft}{\kern0pt}x\isactrlsup {\isadigit{2}}\ {\isacharplus}{\kern0pt}\ y\isactrlsup {\isadigit{2}}\ {\isachargreater}{\kern0pt}\ {\isadigit{0}}\isactrlbold {\isacharbraceright}{\kern0pt}{\isachardoublequoteclose}\isanewline
\isacommand{proof}\isamarkupfalse%
\ {\isacharminus}{\kern0pt}\isanewline
\ \ \isacommand{have}\isamarkupfalse%
\ ctrl{\isacharunderscore}{\kern0pt}post{\isacharcolon}{\kern0pt}\ {\isachardoublequoteopen}\isactrlbold {\isacharbraceleft}{\kern0pt}x\isactrlsup {\isadigit{2}}\ {\isacharplus}{\kern0pt}\ y\isactrlsup {\isadigit{2}}\ {\isachargreater}{\kern0pt}\ {\isadigit{0}}\isactrlbold {\isacharbraceright}{\kern0pt}\ ctrl\ \isactrlbold {\isacharbraceleft}{\kern0pt}{\isacharparenleft}{\kern0pt}{\isasymomega}\ {\isacharequal}{\kern0pt}\ {\isadigit{0}}\ {\isasymand}\ {\isacharat}{\kern0pt}I{\isacharparenright}{\kern0pt}\ {\isasymor}\ {\isacharparenleft}{\kern0pt}{\isasymomega}\ {\isacharequal}{\kern0pt}\ {\isadigit{1}}\ {\isasymand}\ {\isacharat}{\kern0pt}J{\isacharparenright}{\kern0pt}\isactrlbold {\isacharbraceright}{\kern0pt}{\isachardoublequoteclose}\isanewline
\ \ \ \ \isacommand{unfolding}\isamarkupfalse%
\ ctrl{\isacharunderscore}{\kern0pt}def\ \isacommand{by}\isamarkupfalse%
\ wlp{\isacharunderscore}{\kern0pt}full\isanewline
\isanewline
\ \ \isacommand{have}\isamarkupfalse%
\ plant{\isacharunderscore}{\kern0pt}safe{\isacharunderscore}{\kern0pt}I{\isacharcolon}{\kern0pt}\ {\isachardoublequoteopen}\isactrlbold {\isacharbraceleft}{\kern0pt}{\isasymomega}\ {\isacharequal}{\kern0pt}\ {\isadigit{0}}\ {\isasymand}\ {\isacharat}{\kern0pt}I\isactrlbold {\isacharbraceright}{\kern0pt}\ plant\ \isactrlbold {\isacharbraceleft}{\kern0pt}x\isactrlsup {\isadigit{2}}\ {\isacharplus}{\kern0pt}\ y\isactrlsup {\isadigit{2}}\ {\isachargreater}{\kern0pt}\ {\isadigit{0}}\isactrlbold {\isacharbraceright}{\kern0pt}{\isachardoublequoteclose}\isanewline
\ \ \ \ \isacommand{unfolding}\isamarkupfalse%
\ plant{\isacharunderscore}{\kern0pt}def\ \isacommand{apply}\isamarkupfalse%
\ {\isacharparenleft}{\kern0pt}dInv\ {\isachardoublequoteopen}{\isacharparenleft}{\kern0pt}{\isachardollar}{\kern0pt}{\isasymomega}\ {\isacharequal}{\kern0pt}\ {\isadigit{0}}\ {\isasymand}\ {\isacharat}{\kern0pt}I{\isacharparenright}{\kern0pt}\isactrlsup e{\isachardoublequoteclose}{\isacharcomma}{\kern0pt}\ dWeaken{\isacharparenright}{\kern0pt}\isanewline
\ \ \ \ \isacommand{using}\isamarkupfalse%
\ v\isactrlsub o{\isacharunderscore}{\kern0pt}pos\ v\isactrlsub i{\isacharunderscore}{\kern0pt}pos\ sum{\isacharunderscore}{\kern0pt}squares{\isacharunderscore}{\kern0pt}gt{\isacharunderscore}{\kern0pt}zero{\isacharunderscore}{\kern0pt}iff\ \isacommand{by}\isamarkupfalse%
\ fastforce\isanewline
\isanewline
\ \ \isacommand{have}\isamarkupfalse%
\ plant{\isacharunderscore}{\kern0pt}safe{\isacharunderscore}{\kern0pt}J{\isacharcolon}{\kern0pt}\ {\isachardoublequoteopen}\isactrlbold {\isacharbraceleft}{\kern0pt}{\isasymomega}\ {\isacharequal}{\kern0pt}\ {\isadigit{1}}\ {\isasymand}\ {\isacharat}{\kern0pt}J\isactrlbold {\isacharbraceright}{\kern0pt}\ plant\ \isactrlbold {\isacharbraceleft}{\kern0pt}x\isactrlsup {\isadigit{2}}\ {\isacharplus}{\kern0pt}\ y\isactrlsup {\isadigit{2}}\ {\isachargreater}{\kern0pt}\ {\isadigit{0}}\isactrlbold {\isacharbraceright}{\kern0pt}{\isachardoublequoteclose}\isanewline
\ \ \ \ \isacommand{unfolding}\isamarkupfalse%
\ plant{\isacharunderscore}{\kern0pt}def\ \isacommand{apply}\isamarkupfalse%
\ {\isacharparenleft}{\kern0pt}dInv\ {\isachardoublequoteopen}{\isacharparenleft}{\kern0pt}{\isasymomega}{\isacharequal}{\kern0pt}{\isadigit{1}}\ {\isasymand}\ {\isacharat}{\kern0pt}J{\isacharparenright}{\kern0pt}\isactrlsup e{\isachardoublequoteclose}{\isacharcomma}{\kern0pt}\ dWeaken{\isacharparenright}{\kern0pt}\isanewline
\ \ \ \ \isacommand{by}\isamarkupfalse%
\ {\isacharparenleft}{\kern0pt}smt\ {\isacharparenleft}{\kern0pt}z{\isadigit{3}}{\isacharparenright}{\kern0pt}\ cos{\isacharunderscore}{\kern0pt}le{\isacharunderscore}{\kern0pt}one\ mult{\isacharunderscore}{\kern0pt}if{\isacharunderscore}{\kern0pt}delta\ mult{\isacharunderscore}{\kern0pt}le{\isacharunderscore}{\kern0pt}cancel{\isacharunderscore}{\kern0pt}iff{\isadigit{2}}\ \isanewline
\ \ \ \ \ \ \ \ \ \ mult{\isacharunderscore}{\kern0pt}left{\isacharunderscore}{\kern0pt}le\ sum{\isacharunderscore}{\kern0pt}squares{\isacharunderscore}{\kern0pt}gt{\isacharunderscore}{\kern0pt}zero{\isacharunderscore}{\kern0pt}iff\ v\isactrlsub i{\isacharunderscore}{\kern0pt}pos\ v\isactrlsub o{\isacharunderscore}{\kern0pt}pos{\isacharparenright}{\kern0pt}\isanewline
\isanewline
\ \ \isacommand{show}\isamarkupfalse%
\ {\isacharquery}{\kern0pt}thesis\isanewline
\ \ \ \ \isacommand{unfolding}\isamarkupfalse%
\ flight{\isacharunderscore}{\kern0pt}def\isanewline
\ \ \ \ \isacommand{apply}\isamarkupfalse%
\ {\isacharparenleft}{\kern0pt}intro\ hoare{\isacharunderscore}{\kern0pt}kstar{\isacharunderscore}{\kern0pt}inv\ hoare{\isacharunderscore}{\kern0pt}kcomp{\isacharbrackleft}{\kern0pt}OF\ ctrl{\isacharunderscore}{\kern0pt}post{\isacharbrackright}{\kern0pt}{\isacharparenright}{\kern0pt}\isanewline
\ \ \ \ \isacommand{by}\isamarkupfalse%
\ {\isacharparenleft}{\kern0pt}rule\ hoare{\isacharunderscore}{\kern0pt}disj{\isacharunderscore}{\kern0pt}split{\isacharbrackleft}{\kern0pt}OF\ plant{\isacharunderscore}{\kern0pt}safe{\isacharunderscore}{\kern0pt}I\ plant{\isacharunderscore}{\kern0pt}safe{\isacharunderscore}{\kern0pt}J{\isacharbrackright}{\kern0pt}{\isacharparenright}{\kern0pt}\isanewline
\isacommand{qed}\isamarkupfalse%
\isanewline
\end{isabellebody}
%\vspace{1ex}
%\begin{center}
%    \includegraphics[width=\textwidth]{flight-safe}
%\end{center}
%\vspace{1ex}

\noindent We formulate collision avoidance using \(x^2 + y^2 > 0\) as the invariant in the Hoare triple. A collision occurs when \(x^2 + y^2 = 0\), as illustrated in figure \ref{fig:collision}. We use Isabelle's Isar scripting language to break down the verification into several intermediate properties via the Isar command ``\isakwmaj{have}'', which takes a label followed by a property specification.

We first calculate the postconditions that arise from running the controller (\texttt{ctrl}) by using our tactic \texttt{wlp\_full} (see Section~\ref{sec:tactics}). This gives us two possible execution branches -- one where \(I \land \omega = 0\) holds, and one where \(J \land \omega=1\) holds. We give this first property the name \texttt{ctrl\_post}.

The postconditions provide two possible initial states for the plant, which we consider using the properties \texttt{plant\_safe\_I} and \texttt{plant\_safe\_J}. We show that both preconditions guarantee the problem's postcondition $x^2+y^2>0$ by applying differential induction, a technique that proves an invariant of a system of ODEs without computing a solution (see Section~\ref{subsec:dinv}). In each case, we use our Eisbach-designed method \texttt{dInv}, which takes as a parameter the invariant we wish to prove. The invariant is simply the precondition of each Hoare triple. However, we then need to prove that this invariant implies the postcondition, which is done using a further method called \texttt{dWeaken}. The remaining proof obligations can be solved using the \isakwmaj{Sledgehammer} tool, which calls external automated theorem provers to find a solution and reconstructs it using the names of previously proven theorems in Isabelle's libraries. In particular, the property \texttt{plant\_safe\_J} is discharged using Z3, which uses trigonometric identities formalised in \textsf{HOL-Analysis}. We display the proof state resulting from applying differential induction below:

\vspace{1ex}
\begin{center}
    \includegraphics[width=0.8\textwidth]{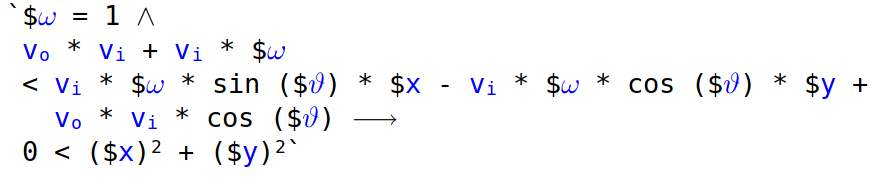}
\end{center}
\vspace{1ex} 

In the case that \isakwmaj{Sledgehammer} is unable to find a proof, this proof obligation is quite readable, and so a manual proof or refutation could be given (See Section~\ref{sec:ex}). 

Finally, we put everything together to show that the whole system is safe, using the Isar command \isakwmaj{show}, which is used to conclude proofs and restate the overall goal of the lemma (\texttt{?thesis}). The proof uses a couple of high-level Hoare logic laws, and the properties that were proven to complete the proof. This final step can be completely automated using Isabelle's classical reasoner, but we leave the details for the purpose of demonstration. 

This completes our overview of our tool and its capabilities. In the remainder of the paper, we will expound the technical foundations of the tool, and our key results.

%%%%%%%%%% SEMANTIC PRELIMINARIES %%%%%%%%%%%%%
\section{Semantics for hybrid systems verification}\label{sec:prelim}
We start the section by recapitulating basic concepts
from the theory of dynamical systems. We use these 
notions to describe our approach~\cite{MuniveS22,FosterMGS21}
to hybrid systems verification in general-purpose proof 
assistants. Specifically, we present hybrid programs and 
their state transformer semantics. We then introduce our 
store model and provide intuitions for deriving state 
transformers for program assignments and ODEs relative to 
this model. We extend our approach to a 
predicate transformer semantics to derive laws 
for verification condition generation (VCG)~\cite{MuniveS22,ArmstrongGS16}. 
That is, we introduce the forward box predicate 
transformer and use its properties to derive the rules of 
Hoare logic. Finally, we introduce the forward diamond
predicate transformer which serves to prove reachability
and progress properties of hybrid systems. Our 
formalisation of these concepts as a shallow embedding in
Isabelle/HOL maximises the availability of the prover's 
proof automation in VCG.

\subsection{Dynamical Systems}\label{subsec:dynsys}

In this section, we consider two ways to specify
continuous dynamical systems~\cite{Teschl12}: explicitly 
via flows and implicitly via systems of ordinary differential
equations (ODEs). \emph{Flows} are functions
$\varphi:T\to\csrc\to\csrc$, where $T$ is a non-discrete
submonoid of the real numbers $\reals$ representing time.
Similarly, $\csrc$ is a set with some topological 
structure like a vector space or a metric space. We 
emphasise this intuition using an overhead arrow for 
elements $\vec{c}\in \csrc$ and refer to $\csrc$ as a \emph{continuous} state space. By definition, flows are 
continuously differentiable ($C^1$) functions and monoid 
actions: they satisfy the laws 
$\varphi (t_1 + t_2)
= \varphi\, t_1 \circ \varphi\, t_2$ and $\varphi\, 0 = \id_\csrc$. 
Given a state $\vec{c}\in\csrc$, the \emph{trajectory} 
$\varphi_{\vec{c}}:T\to\csrc$, such that 
$\varphi_{\vec{c}}\, t = \varphi\, t\, \vec{c}$, is a 
curve modelling the continuous dynamical system's 
evolution in time and passing through $\vec{c}$. 
The \emph{orbit map} $\gamma^\varphi: \csrc\to \Pow\, \csrc$,
such that $\gamma^\varphi\, \vec{c} = \Pow\, \varphi_{\vec{c}}\, T$, 
gives the graph of this curve for $\vec{c}$, where $\Pow$
denotes both the powerset operation and the direct image operator. 

Systems of ODEs are related to flows through their solutions. 
Formally, systems of ODEs are specified by \emph{vector fields} 
$f:T\rightarrow \csrc\rightarrow \csrc$, functions assigning 
vectors to points in space-time. A \emph{solution} $X:T\to\csrc$ 
to the system of ODEs specified by $f$ is then a $C^1$-function 
such that $X'\, t= f\, t\, (X\, t)$ for all $t$ in some interval
$U\subseteq T$. This solution also solves the associated 
\emph{initial value problem} (IVP) given by $(t_0,\vec{c})$ 
(denoted by $X\in\ivpsols\, U\, f\, t_0\, \,\vec{c}$) if it 
satisfies $X\, t_0 = \vec{c}$, with $t_0\in U$. The existence of 
solutions to IVPs is guaranteed for continuous vector fields by 
the Peano theorem, albeit on an interval $U\, \vec{c}$ depending 
on the initial condition $(t_0,\vec{c})$. Similarly, the 
Picard-Lindel\"of theorem states that all solutions to the IVP 
$X'\, t= f\, t\, (X\, t)$ with $X\, t_0 = \vec{c}$ coincide 
in some interval $U\, \vec{c} \subseteq T$ (around $t_0$) 
if $f$ is Lipschitz continuous on $T$. In other words, it states
that there is a unique solution to the IVP on $U\, \vec{c}$. 
Thanks to this, when $f$ is Lipschitz continuous on $T$, $t_0 =0$, 
and $U\, \vec{c} = T$ for all $\vec{c}\in\csrc$, the solutions to 
the associated IVPs are exactly a flow's trajectories 
$\varphi_{\vec{c}}$, that is, 
$\varphi_{\vec{c}}'\, t= f\, t\, (\varphi_{\vec{c}}\, t)$ 
and $\varphi_{\vec{c}}\, 0 = \vec{c}$. Therefore, the flow 
$\varphi$ is the function mapping to each $\vec{c}\in\csrc$
the unique $\varphi_{\vec{c}}$ such that 
$\varphi_{\vec{c}}\in\ivpsols\, T\, f\, 0\, \,\vec{c}$.\vspace{1ex}

\begin{example}\label{ex:odes} 
We illustrate the above properties about flows and ODEs with
the differential equation $y'\, t = a \cdot y\,t + b$,
where $a\neq0$ and $t\in\reals$. This is an important ODE 
modelling for instance idealised bacterial growth~\cite{Stanescu09},
radioactive decay~\cite{Weinert09} or concentration of glucose
in the blood (without the intervention of insulin)~\cite{AckermanGRM65}. 
First, given that differentiability implies Lipschitz 
continuity~\cite{Teschl12}, we can verify that this equation 
has unique solutions (by the Picard-Lindel\"of theorem) simply
by noticing that $f\, t=a \cdot y\,t + b$ is differentiable 
with derivative $f'\, t = a\cdot y'\, t$. The solution to the 
associated IVP with initial condition $y\, 0=c$ is 
$\varphi_{c}\, t = \frac{b}{a}\cdot e^{a\cdot t}+c\cdot e^{a\cdot t}-\frac{b}{a}$. 
Indeed, 
\begin{align*}
\varphi_{c}'\, t 
	&= \left(\frac{b}{a}\cdot e^{a\cdot t}+c\cdot e^{a\cdot t}-\frac{b}{a}\right)'\\
	&=\frac{b}{a}\cdot a\cdot e^{a\cdot t}+c\cdot a\cdot e^{a\cdot t}\\
	&= a\cdot \left(\frac{b}{a}\cdot e^{a\cdot t}+c\cdot e^{a\cdot t}-\frac{b}{a}\right)+b\\
	&=a\cdot \varphi_{c}\, t + b.
\end{align*}
It is also easy to check that $\varphi_{c}\, 0= c$. Hence, the
mapping $\varphi:\reals\to\reals\to\reals$, such that 
$\varphi\, t\, c = \varphi_c\, t$, is the unique flow associated
to the ODE $y'\, t = a \cdot y\,t + b$. Indeed, for a fixed 
$\tau\in\reals$, the function $g:\reals\to\reals$ such that 
$g\, t =\varphi\, (t+\tau)\, c = \varphi_c\, (t+\tau)$ satisfies
$g\, 0 = \varphi_c\, \tau$ and also $g'\, t 
= \varphi_c'(t+\tau)
=a\cdot \varphi_c\, (t+\tau)+b
=a\cdot (g\, t)+b$. However, by uniqueness, the only function 
satisfying these two equations is $\varphi_{\varphi_c\, \tau}$. 
Hence $g\, t = \varphi_{\varphi_c\, \tau}\, t$, thus, the monoid 
law $\varphi\, (t+\tau)\, c =\varphi\, t\, (\varphi\, \tau\, c)$ holds. Thus, the function $\varphi$ mapping points $c$ to 
IVP-solutions $\varphi_{c}$ of the ODE $y'\, t = a \cdot y\,t + b$ 
satisfies the monoid action laws, and is therefore, a flow. \qed
\end{example}

\subsection{State transformer semantics for hybrid programs}\label{subsec:hp}

Having introduced some basic concepts from the theory of dynamical 
systems, we present our hybrid systems model. We represent these 
via hybrid programs which are traditionally defined 
syntactically~\cite{HarelKT00,Platzer18}. Yet, our approach is
purely semantic and we merely provide the recursive definition 
below as a guide to what our semantics should model:
\begin{equation*}
\alpha\, ::=\, x:=e \mid x' = f \, \&\, G \mid  \mquestiondown P?
\mid \alpha\seqcomp \alpha\mid \alpha\sqcap\alpha\mid \alpha^*.
\end{equation*}
Typically, $x$ denotes variables; $e$ and $f$ are terms, and $G$ 
and $P$ are assertions. In dynamic logic~\cite{HarelKT00}, the 
statement $x:=e$ represents an \emph{assignment} of variable $x$ 
to expression $e$, $\mquestiondown P?$ models testing whether $P$
holds, while $\alpha\seqcomp \beta$, $\alpha\sqcap \beta$ and 
$\alpha^*$ are the \emph{sequential composition}, 
\emph{nondeterministic choice}, and \emph{finite iteration} of
programs $\alpha$ and $\beta$. Well-known while-programs emerge 
via the equations 
$\IF{P}{\alpha}{\beta} \equiv (\mquestiondown P?\seqcomp \alpha) 
\sqcap (\mquestiondown\neg P?\seqcomp \beta)$ 
and $\WHILE{P}{\alpha}\equiv (\mquestiondown P?\seqcomp \alpha)^*\seqcomp \mquestiondown\neg P?$. 
Beyond these, differential dynamic logic (\dL)~\cite{Platzer18} 
introduces \emph{evolution commands} $x' = f \, \&\, G$ 
that represent systems of ODEs with \emph{boundary conditions}
or \emph{guards} $G$ that delimit the solutions' range to the region described by $G$.

We use \emph{nondeterministic state transformers} 
$\alpha:\src\rightarrow \Pow \src$ as our semantic 
representation for hybrid programs. 
Thus, our ``hybrid programs'' are really arrows 
in the Kleisli category of the powerset monad. Observe that 
a subset of these arrows also model ``assertions'', namely 
the \emph{subidentities} of the monadic unit $\eta_\src$, 
such that $\eta_\src\, s = \{s\}$ for all $s\in\src$. That is, the 
functions mapping each state $s$ either to $\{s\}$ or
to $\emptyset$ model assertions where $P\, s = \{s\}$ 
represents that $P$ holds for $s$, and $P\, s = \emptyset$ 
that $P$ does not hold for $s$. Henceforth, we abuse notation 
and identify predicates $P:\src\rightarrow \bools$ (or $P\in\bools^\src$), sets $\Pow S$, and subidentities of $\eta_\src$, where $\bools$ denotes the Booleans. 
We also treat predicates as logic
formulae by writing, for instance, $P\land Q$ and $P\lor Q$ 
instead of $\lambda s.\ P\, s\land Q\, s$ and 
$\lambda s.\ P\, s\lor Q\, s$. Thus, we denote the constantly
true and constantly false predicates by $\top$ and $\bot$ 
respectively. They coincide with the $\SKIP$ and $\ABORT$ 
programs such that $\SKIP = \eta_\src$ and $\ABORT\, s = \emptyset$
for all $s\in\src$. In this state transformer semantics, 
sequential compositions correspond to (forward) 
\emph{Kleisli compositions} $(\alpha\seqcomp\beta)\, s 
= \bigcup\{\beta\, s'\mid s' \in \alpha\, s\}$, 
nondeterministic choices are pointwise unions 
$(\alpha\sqcap\beta)\, s = \alpha\, s \cup \beta\, s$, and 
finite iterations are the functions
$\alpha^{\ast}\, s = \bigcup_{i\in\mathbb{N}} \alpha^{i}\, s$, 
with $\alpha^{i+1}=\alpha;\alpha^i$ and $\alpha^{0}=\SKIP$. 

\subsection{Store and expressions model}\label{subsec:stores}

To introduce our state transformer semantics of assignments
and ODEs, we first describe our store model. We use 
lenses~\cite{Oles82,BackW98,Foster09} to algebraically 
characterise the hybrid store. Through an axiomatic approach 
to accessing and mutating functions, lenses allow us to locally 
manipulate program stores~\cite{Foster2020-LocalVars} and 
algebraically reason about 
program variables~\cite{Foster2020-IsabelleUTP,Optics-AFP}. 
Formally, a \emph{lens} $\lens$ with \emph{source} $\src$ 
and \emph{view} $\view$, denoted $\lens :: \view \lto \src$, 
is a pair of functions $(\lget_\lens, \lput_\lens)$ with 
$\lget_\lens : \src \rightarrow \view$ and 
$\lput_\lens : \view\rightarrow \src \rightarrow \src$ 
such that
\hfill\isalink{https://github.com/isabelle-utp/Optics/blob/main/Lens_Laws.thy}
\begin{equation*}
  \lget_\lens~(\lput_\lens~v~s) = v, \qquad \lput_\lens~v \circ
\lput_\lens~v' = \lput_\lens~v,\quad\text{ and }\quad
\lput_\lens~(\lget_\lens s)~s = s,
\end{equation*}
for all $v,v'\in \view$ and $s\in \src$. Usually, a lens 
$\lens$ represents a variable, $\src$ is the system's state 
space and $\view$ is the value domain for $\lens$. Under 
this interpretation, $\lget_\lens$ returns the value of 
variable $\lens$ while $\lput_\lens$ updates it. Yet, we 
sometimes interpret $\view$ as a subregion of $\src$, 
making $\lget_\lens$ a projection or restriction and 
$\lput_\lens$ an overwriting or substituting function. 
For other state models using diverse variable lenses, 
such as arrays, maps and local variables, 
see~\cite{Foster2020-IsabelleUTP,Foster2020-LocalVars}. 
Lenses $\lens,\lens'::\view \lto \src$ are \emph{independent}, 
$\lens \lindep \lens'$, if 
$\lput_\lens~u \circ \lput_{\lens'}~v 
= \lput_{\lens'}~v \circ \lput_\lens~u$, for all $u,v\in \view$, 
that is, these operations commute on all states.

We model expressions and assertions used in program syntax as
functions $e:\src\rightarrow \view$, which semantically are queries 
over the store $\src$ returning a value of type $\view$. We can 
use the $\lget$ function to perform such queries by ``looking up 
the value of variables''. For instance, if 
$x, y :: \reals \lto \src$ model independent program variables, 
and $c\in\reals$ is a constant, then the function 
$\lambda s.\ (\lget_x~s)^2 + (\lget_y~s)^2 \le c^2$ represents
the ``expression'' $x^2 + y^2 \le c^2$. Then, function evaluation 
corresponds to computing the value of the expression at
state $s\in\src$.
% When applied to a concrete state, this obtains the values for $x$ and $y$, using $\lget$, and then evaluates the expression.

With this representation, the state transformer 
$\lambda s.\ \{\lput_x\, (e\, s)\, s\}$ models a program 
assignment, denoted by $x := e$. More generally, we turn 
deterministic functions $\sigma:\src\to\src$ into state 
transformers via function composition with the Kleisli unit,
which we denote by $\assigns{\sigma} = \eta_\src\circ \sigma$. 
Then, representing expressions $e$ as functions $e:\src\to\view$,
our model for variable assignments becomes 
$(x:=e) = \assigns{\lambda s.\, \lput_x\, (e\, s)\, s}$. 

\subsection{Model for evolution commands}\label{subsec:evolcoms}
To derive in our framework a state transformer semantics 
$\src\to\Pow \src$ for evolution commands $x' = f \, \&\, G$,
observe that a flow's orbit map 
$\gamma^\varphi:\csrc\to\Pow\csrc$ with
$\gamma^\varphi\, \vec{c}=\{\varphi\, t\, \vec{c}\mid t\in T\}$
is already a state transformer on $\csrc$. It sends each 
$\vec{c}$ in the continuous state space $\csrc$ to the reachable 
states of the trajectory $\varphi_{\vec{c}}$. Based on the 
relationship between flows and the solutions to IVPs from 
Subsection~\ref{subsec:dynsys}, we can generalise this state
transformer to a set of all points $X\, t$ of the solutions
$X$ of the system of ODEs represented by $f$, i.e. $X'\, t = f\, t\, (X\, t)$, with initial condition $X\, t_0 = \vec{c}$
over an interval $U\, \vec{c}$ around $t_0$ ($t_0\in U\,\vec{c}$). Moreover, in line 
with \dL, the solutions should remain within the guard or 
boundary condition $G$: 
$\forall\tau\in U\, \vec{c}.\ \tau\leq t\Rightarrow G\, (X\, \tau)$. 
Thus, to specify dynamical systems via ODEs $f$ instead of flows $\varphi$, we define the \emph{generalised guarded orbits} 
map~\cite{MuniveS22}
\begin{equation*}
\gamma^f\, G\, U\, t_0\, \vec{c} = \{X\, t\mid t \in U\, \vec{c}
	\land X\in\ivpsols\, U\, f\, t_0\, \,\vec{c}
	\land \Pow\, X\, (\downcl{U\, \vec{c}}{t})\subseteq G\},
\end{equation*}
% $\ivl{t_0}{t} = [t_0, t]=\{\tau\mid t_0\leq\tau\leq t\}$, if $t_0 \leq t$, and $\ivl{t_0}{t} = [t, t_0]$
that also introduces guards $G$, initial conditions $t_0$, 
and intervals $U\, \vec{c}$. The set $\downcl{T}{t}$ is a 
downward closure $\downcl{T}{t} =\{\tau\in T\mid \tau\leq t\}$
which in applications becomes the interval $[0,t]=\{\tau\mid 0\leq\tau\leq t\}$ because we usually fix 
$U\, \vec{c}=\reals_{\geq 0}=\{\tau\mid \tau\geq 0\}$ for all 
$\vec{c}\in\csrc$. This is why we also abuse notation and write 
constant interval functions $\lambda\vec{c}.\ T$ simply as $T$. 
Notice that when the flow $\varphi$ for $f$ exists, $\gamma^f\, \top\, T\, 0\, \vec{c} = \gamma^\varphi\, \vec{c}$. 

Lenses support algebraic reasoning about variable frames: 
the set of variables that a hybrid program can modify. In 
particular, they allow us to split the state space into 
continuous and discrete parts. We explain a lens-based 
lifting of guarded orbits $\gamma^f\, G\, U\, t_0$ from 
the continuous space $\csrc\to\Pow\csrc$ to the full state
space $\src$ in Subsection~\ref{subsec:dynevol}. This 
produces our state transformer for evolution 
commands $(x' = f\, \&\, G)_U^{t_0}:\src\to\Pow\src$. 
Intuitively, it maps each state $s\in\src$ to all 
$X$-reachable states within the region $G$, where $X$ solves 
the system of ODEs specified by $f$, and leaves intact the 
non-continuous part of $\src$. Having the same type as the 
above-described state transformers enables us to seamlessly 
use the same operations on $(x' = f\, \&\, G)_U^{t_0}$. 
This also enables us to do modular verification condition 
generation (VCG) of hybrid systems as described below.\vspace{1ex}

% example hybrid program
\begin{example}\label{ex:hp}
In this example, we use a hybrid program \texttt{blood\_sugar} to model an idealised 
machine controlling the concentration of glucose in a patient's body. Hybrid 
programs are often split into discrete control \texttt{ctrl} and continuous 
dynamics \texttt{dyn}. Their composition is then wrapped in a finite iteration: 
$\texttt{blood\_sugar}=(\texttt{ctrl}\seqcomp\texttt{dyn})^{*}$.
For the control, we use a conditional statement reacting to the patient's
blood-glucose. Concretely, the program
\begin{equation*}
\texttt{ctrl} = \IF{\texttt{g}\leq g_m}{g:=g_M}{\SKIP}
\end{equation*}
states that if the value of the patient's blood-glucose concentration $g$ 
is below a certain warning threshold $g_m\geq 0$, the maximum healthy dose 
of insulin, represented as an immediate spike to the patient's glucose $g:=g_M$,
is injected into the patient's body. Otherwise, the patient 
is fine and the machine does nothing. The continuous variable $g$ follows the 
dynamics in Example~\ref{ex:odes}: $y'\, t=a\cdot y\, t+b$. We assume $a=-1$ 
and $b=0$ so that the concentration of glucose decreases over time. This results 
in the evolution command 
$\texttt{dyn}=(g' = -g\, \&\, \top)^0_{\reals_{\geq 0}}$ which we abbreviate 
as $\texttt{dyn} = (g' = -g)$. Formally, the assignment $g:=g_M$ is the state transformer 
$\lambda s.\ \{\lput_g\, g_M\, s\}$, the test $g\leq g_m$ is the 
predicate $\lambda s.\ \lget_g\, s\leq g_m$, and the evolution command $g' = -g$
is the orbit map $\gamma^\varphi$ lifted to the whole space $\src$, 
where $\varphi\, t\, c=c\cdot e^{- t}$ for all $t\in\reals$. \qed
\end{example}

\subsection{Predicate transformer semantics}\label{subsec:vcg}
Finally, we extend our state transformer semantics to a 
\emph{predicate transformer} ($\bools^\src\to\bools^\src$) 
semantics for verification condition generation (VCG). 
Concretely, we define two predicate transformers and use 
the definition of the first one for deriving partial 
correctness laws, including the rules of Hoare logic, and 
the definition of the second one for deriving reachability 
laws. We also exemplify the application of these to VCG.

\subsubsection{Forward boxes}\label{subsec:fdbox}
We define dynamic logic's \emph{forward box} 
or \emph{weakest liberal precondition} (wlp)
$\fdbox{-}{-}$ operator
$\fdbox{\alpha}{}:\bools^\src\to\bools^\src$
%predicate transformer, $\wlp\, \alpha\, Q$, defined by 
as $\fdbox{\alpha}{Q}\, s \Leftrightarrow 
\left(\forall s'.\ s'\in \alpha\, s\Rightarrow Q\, s'\right)$,
for $\alpha:\src\to\Pow\, \src$ and $Q:\src\to\bools$. 
It is true for those initial system's states that lead 
to a state satisfying $Q$ after executing $\alpha$, if $\alpha$
terminates. Well-known $\wlp$-laws are derivable 
and simple consequences from this and our previous definitions~\cite{MuniveS22}. These laws allow us to automate
and do VCG much more efficiently than by doing Hoare 
Logic since all of them, except for the loop rule, are 
equational simplifications from left to right:\vspace{1ex}

\begin{center}
\begin{tabular}{l r c l r}
(\customtag{wlp-skip}{eq:wlp-skip})
& $\fdbox{\SKIP}{Q}$                            & $=$                       & $Q$                         
&\isalink{https://github.com/isabelle-utp/Hybrid-Verification/blob/e766a6b4744f37e176cd289e9e80120b6238ad81/Hybrid_Programs/Regular_Programs.thy\#L19}\\
(\customtag{wlp-abort}{eq:wlp-abort})
& $\fdbox{\ABORT}{Q}$                        & $=$                       & $\top$
&\isalink{https://github.com/isabelle-utp/Hybrid-Verification/blob/e766a6b4744f37e176cd289e9e80120b6238ad81/Hybrid_Programs/Regular_Programs.thy\#L33}\\
(\customtag{wlp-test}{eq:wlp-test})
& $\fdbox{\mquestiondown P?}{Q}$      & $=$                      & $P\Rightarrow Q$
&\isalink{https://github.com/isabelle-utp/Hybrid-Verification/blob/e766a6b4744f37e176cd289e9e80120b6238ad81/Hybrid_Programs/Regular_Programs.thy\#L54}\\
(\customtag{wlp-assign}{eq:wlp-assign})
& $\fdbox{x := e}{Q}$                             & $=$                       & $\lsubst{Q}{e}{\lens}$
&\isalink{https://github.com/isabelle-utp/Hybrid-Verification/blob/e766a6b4744f37e176cd289e9e80120b6238ad81/Hybrid_Programs/Regular_Programs.thy\#L79}\\
(\customtag{wlp-seq}{eq:wlp-seq})
& $\fdbox{\alpha\seqcomp\beta}{Q}$    & $=$                       & $\fdbox{\alpha}{\fdbox{\beta}{Q}}$
&\isalink{https://github.com/isabelle-utp/Hybrid-Verification/blob/e766a6b4744f37e176cd289e9e80120b6238ad81/Hybrid_Programs/Regular_Programs.thy\#L168}\\
(\customtag{wlp-choice}{eq:wlp-choice})
& $\fdbox{\alpha\sqcap\beta}{Q}$                 & $=$                        & $\fdbox{\alpha}{Q}\land\fdbox{\beta}{Q}$
&\isalink{https://github.com/isabelle-utp/Hybrid-Verification/blob/e766a6b4744f37e176cd289e9e80120b6238ad81/Hybrid_Programs/Regular_Programs.thy\#L120}\\
(\customtag{wlp-loop}{eq:wlp-loop})
& $\fdbox{\LOOP{\alpha}}{Q}$         & $=$   & $\forall n.\ \fdbox{\alpha^n}{Q}$
&\isalink{https://github.com/isabelle-utp/Hybrid-Verification/blob/e766a6b4744f37e176cd289e9e80120b6238ad81/Hybrid_Programs/Regular_Programs.thy\#L362}\\
(\customtag{wlp-cond}{eq:wlp-cond}) 
& $\fdbox{\IF{T}{\alpha}{\beta}}{Q}$      & $=$ 
& $(T\Rightarrow\fdbox{\alpha}{Q})\land(\neg T\Rightarrow\fdbox{\beta}{Q})$.
&\isalink{https://github.com/isabelle-utp/Hybrid-Verification/blob/e766a6b4744f37e176cd289e9e80120b6238ad81/Hybrid_Programs/Regular_Programs.thy\#L221}\\
\end{tabular}
\end{center}\vspace{1ex}
%\begin{align*}
%\fdbox{\SKIP}{Q}				& = Q,\label{eq:wlp-skip}\tag{wlp-skip}\\%\isalink{https://github.com/isabelle-utp/Hybrid-Verification/blob/8c2760f4ad5006edfbbcae95762f9ae816467302/HS_Lens_Spartan.thy\#L136}\\
%\fdbox{\ABORT}{Q}				& = \top,\label{eq:wlp-abort}\tag{wlp-abort}\\
%\fdbox{\mquestiondown P?}{Q}		& = P\Rightarrow Q,\label{eq:wlp-test}\tag{wlp-test}\\
%\fdbox{x := e}{Q}\, s 				& \Leftrightarrow Q\, \lsubst{s}{e\, s}{\lens},\label{eq:wlp-assign}\tag{wlp-assign}\\
%\fdbox{x := e}{Q}\, s 				& \Leftrightarrow \lsubst{Q}{e}{\lens},\label{eq:wlp-assign}\tag{wlp-assign}\\
%\fdbox{\alpha\seqcomp\beta}{Q}	&=\fdbox{\alpha}{\fdbox{\beta}{Q}},\label{eq:wlp-seq}\tag{wlp-seq}\\
%\fdbox{\alpha+\beta}{Q}			&=\fdbox{\alpha}{Q}\land\fdbox{\beta}{Q},\label{eq:wlp-choice}\tag{wlp-choice}\\
%\fdbox{\LOOP{\alpha}}{Q}\, s		&\Leftrightarrow\forall n.\ \fdbox{\alpha^n}{Q}\, s,\label{eq:wlp-loop}\tag{wlp-loop}\\
%\fdbox{\IF{T}{\alpha}{\beta}}{Q} 		&= (T\preceq\fdbox{\alpha}{Q})\land(\neg T\preceq\fdbox{\beta}{Q})\label{eq:wlp-cond}\tag{wlp-cond}
%\end{align*}

\noindent Here, %$s\in\src$, %$n\in\nats$, $\LOOP{\alpha}=\alpha^*$,
$n$ is a natural number (i.e. $n\in\nats$), 
%$P\preceq Q\Leftrightarrow(\forall s.\ P\, s\Rightarrow Q\, s)$,
$\LOOP{\alpha}$ is simply $\alpha^*$, 
and $\lsubst{Q}{e}{\lens}$ is our abbreviation for the function
$\lambda s.\ Q\, (\lput_\lens\, (e\, s)\, s)$ that represents the 
value of $Q$ after variable $x$ has been updated by the value of 
evaluating $e$ on $s\in\src$. We write this semantic operation 
as a substitution to resemble
Hoare logic (see Section~\ref{subsec:subst}). Similarly, the 
$\wlp$-law for evolution commands informally corresponds 
to\vspace{1ex}

\begin{center}
\begin{tabular}{l r c l r}
(\customtag{wlp-evol}{eq:wlp-evol})
& $\fdbox{(x' = f\, \&\, G)_U^{t_0}}{Q}\, s$ & $\Leftrightarrow$ & 
    \begin{tabular}{@{}c@{}}
    $\forall X{\in}\ivpsols\, U\, f\, t_0\, s.\ \forall t\in U\, s.$ \\ 
    $(\forall\tau\in\downcl{Us}{t}.\ G\, (X\, \tau))\Rightarrow Q\, (X\, t)$.
    \end{tabular}
&\isalink{https://github.com/isabelle-utp/Hybrid-Verification/blob/e766a6b4744f37e176cd289e9e80120b6238ad81/Hybrid_Programs/Evolution_Commands.thy\#L87}\\
\end{tabular}
\end{center}\vspace{1ex}

\noindent That is, a postcondition $Q$ holds after an evolution
command starting at $(t_0,s)$, if and only if, the postcondition
holds $Q\, (X\, t)$ for all solutions to the IVP 
$X'\, t=f\, t\, (X\, t)$, $X\, t_0=s$, for all times in the interval
$t\in U\, s$ whose previous times respect $G$. Notice that, if there 
is a flow $\varphi:T\to\csrc\to\csrc$ for %$\ldowngr{f}{s}{\lens}$ 
$f$ and $U=T=\reals_{\geq 0}$, this simplifies to\vspace{1ex}

\begin{center}
\begin{tabular}{@{}l r c l r@{}}
(\customtag{wlp-flow}{eq:wlp-flow})
& $\fdbox{x' = f\, \&\, G}{Q}\, s$ & $\Leftrightarrow$ 
& $(\forall t\geq 0.\ (\forall\tau{\in}[0,t].\ G\, (\varphi_{s}\, \tau))\Rightarrow Q\, (\varphi_{s}\, t)$.                         
&\isalink{https://github.com/isabelle-utp/Hybrid-Verification/blob/e766a6b4744f37e176cd289e9e80120b6238ad81/Hybrid_Programs/Evolution_Commands.thy\#L113}\\
\end{tabular}
\end{center}\vspace{1ex}

\noindent See Section~\ref{sec:frames} for the formal version of these laws.

%\begin{equation*}
%\fdbox{x' = f\, \&\, G}{Q}\, s \Leftrightarrow (\forall t\geq 0.\ (\forall\tau\in[0,t].\ \lsubst{G}{\varphi_{s}\, \tau}{\lens})\Rightarrow \lsubst{Q}{\varphi_{s}\, t}{\lens}). \label{eq:wlp-flow}\tag{wlp-flow}
%\fdbox{x' = f\, \&\, G}{Q}\, s \Leftrightarrow (\forall t\geq 0.\ (\forall\tau\in[0,t].\ G\, \lsubst{s}{\varphi_{\lsugarget{s}{\lens}}\, \tau}{\lens})\Rightarrow Q\, \lsubst{s}{\varphi_{\lsugarget{s}{\lens}}\, t}{\lens}). \label{eq:wlp-flow}\tag{wlp-flow}
%\end{equation*}

\subsubsection{Hoare triples}\label{subsec:hoare}

It is well-known that Hoare logic can be derived from the forward
box operator of dynamic logic~\cite{HarelKT00}. Thus, we can also
write partial correctness specifications as Hoare-triples with our 
forward box operators via 
$\hoare{P}{\alpha}{Q}\Leftrightarrow (P\Rightarrow\wlp\, \alpha\, Q)$.
From our $\wlp$-laws and definitions, the Hoare logic rules below hold:\vspace{1ex}

\begin{center}
% >{$}c prepend centered cell with $
% c<{$} append centered cell with $
% @{}c space before column should be empty
% c@{} space after column should be empty
\begin{longtable}{l r c l r}
(\customtag{h-skip}{eq:h-skip})
&
&                       & $\hoare{P}{\SKIP}{P}\label{eq:h-skip}$          
&\isalink{https://github.com/isabelle-utp/Hybrid-Verification/blob/e766a6b4744f37e176cd289e9e80120b6238ad81/Hybrid_Programs/Regular_Programs.thy\#L25}\\
(\customtag{h-abort}{eq:h-abort})
&
&                       & $\hoare{P}{\ABORT}{Q}$
&\isalink{https://github.com/isabelle-utp/Hybrid-Verification/blob/e766a6b4744f37e176cd289e9e80120b6238ad81/Hybrid_Programs/Regular_Programs.thy\#L39}\\
(\customtag{h-test}{eq:h-test})
&
&                       & $\hoare{P}{\mquestiondown Q?}{P\land Q}$
&\isalink{https://github.com/isabelle-utp/Hybrid-Verification/blob/e766a6b4744f37e176cd289e9e80120b6238ad81/Hybrid_Programs/Regular_Programs.thy\#L60}\\
(\customtag{h-assign}{eq:h-assign})
&
&                       & $\hoare{\lsubst{Q}{e}{\lens}}{x:=e}{Q}$
&\isalink{https://github.com/isabelle-utp/Hybrid-Verification/blob/e766a6b4744f37e176cd289e9e80120b6238ad81/Hybrid_Programs/Regular_Programs.thy\#L82}\\
(\customtag{h-seq}{eq:h-seq})
& $\hoare{P}{\alpha}{R}\land\hoare{R}{\beta}{Q}$ 
& $\Rightarrow$ & $\hoare{P}{\alpha\seqcomp\beta}{Q}$
&\isalink{https://github.com/isabelle-utp/Hybrid-Verification/blob/e766a6b4744f37e176cd289e9e80120b6238ad81/Hybrid_Programs/Regular_Programs.thy\#L171}\\
(\customtag{h-choice}{eq:h-choice})
& $\hoare{P}{\alpha}{Q}\land\hoare{P}{\beta}{Q}$
& $\Rightarrow$ & $\hoare{P}{\alpha\sqcap\beta}{Q}$
&\isalink{https://github.com/isabelle-utp/Hybrid-Verification/blob/e766a6b4744f37e176cd289e9e80120b6238ad81/Hybrid_Programs/Regular_Programs.thy\#L129}\\
(\customtag{h-loop}{eq:h-loop})
& $\hoare{I}{\alpha}{I}$
& $\Rightarrow$ & $\hoare{I}{\LOOP{\alpha}}{I}$
&\isalink{https://github.com/isabelle-utp/Hybrid-Verification/blob/e766a6b4744f37e176cd289e9e80120b6238ad81/Hybrid_Programs/Regular_Programs.thy\#L380}\\
(\customtag{h-cons}{eq:h-cons})
& \begin{tabular}{@{}c@{}}
    $(P_1\Rightarrow P_2)\land (Q_2\Rightarrow Q_1)$ \\ 
    $\land \hoare{P_2}{\alpha}{Q_2}$ 
    \end{tabular}
& $\Rightarrow$ & $\hoare{P_1}{\alpha}{Q_1}$
&\isalink{https://github.com/isabelle-utp/Hybrid-Verification/blob/e766a6b4744f37e176cd289e9e80120b6238ad81/Hybrid_Programs/Correctness_Specs.thy\#L161}\\
(\customtag{h-cond}{eq:h-cond})
& \begin{tabular}{@{}c@{}}
    $\hoare{T\land P}{\alpha}{Q}$ \\ 
    $\land \hoare{\neg T\land P}{\beta}{Q}$ 
    \end{tabular}
& $\Rightarrow$ & $\hoare{P}{\IF{T}{\alpha}{\beta}}{Q}$
&\isalink{https://github.com/isabelle-utp/Hybrid-Verification/blob/e766a6b4744f37e176cd289e9e80120b6238ad81/Hybrid_Programs/Regular_Programs.thy\#L225}\\
(\customtag{h-while}{eq:h-while})
& $\hoare{T\land I}{\alpha}{I}$
& $\Rightarrow$ & $\hoare{I}{\WHILE{T}{\alpha}}{\neg T\land I}$
&\isalink{https://github.com/isabelle-utp/Hybrid-Verification/blob/e766a6b4744f37e176cd289e9e80120b6238ad81/Hybrid_Programs/Regular_Programs.thy\#L738}\\
(\customtag{h-whilei}{eq:h-whilei})
& \begin{tabular}{@{}c@{}}
    $(P\Rightarrow I)\land (I\land\neg T\Rightarrow Q)$ \\ 
    $\land \hoare{I\land T}{\alpha}{I}$ 
    \end{tabular}
& $\Rightarrow$ & $\hoare{P}{\INV{\WHILE{T}{\alpha}}{I}}{Q}$
&\isalink{https://github.com/isabelle-utp/Hybrid-Verification/blob/e766a6b4744f37e176cd289e9e80120b6238ad81/Hybrid_Programs/Regular_Programs.thy\#L751}\\
(\customtag{h-loopi}{eq:h-loopi})
& \begin{tabular}{@{}c@{}}
    $(P\Rightarrow I)\land (I\Rightarrow Q)$ \\ 
    $\land \hoare{I}{\alpha}{I}$ 
    \end{tabular}
& $\Rightarrow$ & $\hoare{P}{\INV{\LOOP{\alpha}}{I}}{Q}$
&\isalink{https://github.com/isabelle-utp/Hybrid-Verification/blob/e766a6b4744f37e176cd289e9e80120b6238ad81/Hybrid_Programs/Regular_Programs.thy\#L709}\\
(\customtag{h-evoli}{eq:h-evoli})
& \begin{tabular}{@{}c@{}}
    $(P\Rightarrow I)\land (I\land G\Rightarrow Q)$ \\ 
    $\land \hoare{I}{(x' = f\, \&\, G)_U^{t_0}}{I}$ 
    \end{tabular}
& $\Rightarrow$ & $\hoare{P}{\INV{(x' = f\, \&\, G)_U^{t_0}}{I}}{Q}$
&\isalink{https://github.com/isabelle-utp/Hybrid-Verification/blob/e766a6b4744f37e176cd289e9e80120b6238ad81/Hybrid_Programs/Evolution_Commands.thy\#L448}\\
(\customtag{h-conji}{eq:h-conji})
& $\hoare{I}{\alpha}{I}\land \hoare{J}{\alpha}{J}$ 
& $\Rightarrow$ & $\hoare{I\land J}{\alpha}{I\land J}$
&\isalink{https://github.com/isabelle-utp/Hybrid-Verification/blob/e766a6b4744f37e176cd289e9e80120b6238ad81/Hybrid_Programs/Correctness_Specs.thy\#L197}\\
(\customtag{h-disji}{eq:h-disji})
& $\hoare{I}{\alpha}{I}\land \hoare{J}{\alpha}{J}$ 
& $\Rightarrow$ & $\hoare{I\lor J}{\alpha}{I\lor J}$
&\isalink{https://github.com/isabelle-utp/Hybrid-Verification/blob/e766a6b4744f37e176cd289e9e80120b6238ad81/Hybrid_Programs/Correctness_Specs.thy\#L197}\\
\end{longtable}
\end{center}\vspace{-1ex}

\noindent where $\INV{\alpha}{I}$ is simply $\alpha$ with the
annotated invariant $I$ and it binds less than any other 
program operator, e.g. 
$\hoare{P}{\INV{\LOOP{\alpha}}{I}}{Q}=\hoare{P}{\INV{(\LOOP{\alpha})}{I}}{Q}$.

For automating VCG, the $\wlp$-laws are preferable over the 
Hoare-style rules since the laws can be added to the proof 
assistant's simplifier which rewrites them automatically. 
However, when loops and ODEs are involved, we use the rules 
(\ref{eq:h-whilei}), (\ref{eq:h-loopi}) and (\ref{eq:h-evoli}). 
In particular, two workflows emerge for discharging ODEs. 
If Picard-Lindel\"of holds, that is, if there is a unique 
solution to the system of ODEs and it is known, the law 
(\ref{eq:wlp-flow}) is the best choice. Otherwise, we employ 
the rule (\ref{eq:h-evoli}) if an invariant is known. 
See Section~\ref{subsec:lipschitz} for a procedure guaranteeing 
the existence of flows or Section~\ref{subsec:dinv} for a
procedure determining invariance for evolution commands.\vspace{1ex}

% example vcg 
\begin{example}\label{ex:vcg}
We prove that $I\, s\Leftrightarrow \lget_g\, s\geq 0$, or simply $g\geq 0$, is 
an invariant for the program 
$\texttt{blood\_sugar}= \LOOP{(\texttt{ctrl}\seqcomp\texttt{dyn})}$
from Example~\ref{ex:hp}. That is, we show that $\hoare{I}{\texttt{blood\_sugar}}{I}$. 
We start applying~(\ref{eq:h-loopi}) and 
proceed with $\wlp$-laws:
\begin{align*}
&\hoare{I}{\INV{\LOOP{(\texttt{ctrl}\seqcomp\texttt{dyn})}}{I}}{I} \\
& \Leftarrow (I\Rightarrow I)\land (I\Rightarrow\fdbox{\texttt{ctrl}\seqcomp\texttt{dyn}}{I})\land (I\Rightarrow I)\\
& = \left(\forall s.\ I\, s\Rightarrow\fdbox{\IF{g\leq g_m}{g:=g_M}{\SKIP}}{\fdbox{g'=-g}{I}}\, s\right)\\
& = \left(\forall s.\ I\, s\Rightarrow
\left(g\leq g_m\Rightarrow \fdbox{g:=g_M}{\fdbox{g'=-g}}{I}\, s\right)
\land \left(g> g_m\Rightarrow\fdbox{g'=-g}{I}\, s\right)\right),
% \left(\lget_g\, s\leq g_m\Rightarrow \fdbox{g:=g_M}{\fdbox{g'=-g}}{I}\, s\right)
% \land \left(\lget_g\, s> g_m\Rightarrow\fdbox{g'=-g}{I}\, s\right)\right),
\end{align*}
where the first equality applies~(\ref{eq:wlp-seq}) and unfolds
the definition of %$\preceq$, 
\texttt{ctrl} and \texttt{dyn}.
The second follows by~(\ref{eq:wlp-cond}). Next, given that
$\varphi\, t\, c=c\cdot e^{- t}$ is the flow for $g'=-g$:
\begin{equation*}
\fdbox{g'=-g}{I}\, s
%\Leftrightarrow (\forall t\geq 0.\ I\, (\lsubst{s}{\lsugarget{s}{g}\cdot e^{- t}}{g}))
\Leftrightarrow (\forall t\geq 0.\ \lsubst{I}{\varphi\, t\, c}{g})
%\Leftrightarrow (\forall t\geq 0.\  \lsugarget{s}{g}\cdot e^{- t}\geq 0)
\Leftrightarrow (\forall t\geq 0.\  g\cdot e^{- t}\geq 0)
%\Leftrightarrow \lsugarget{s}{g}\geq 0,
\Leftrightarrow g\geq 0,
\end{equation*}
for all $s\in\src$ by~(\ref{eq:wlp-flow}), the lens laws, 
and because $G=\top$ and 
$k\cdot e^{- t}\geq 0\Leftrightarrow k\geq 0$. Thus, the 
conjuncts above simplify to
\begin{equation*}
\begin{aligned}[c]
%& \lget_\texttt{g}\, s\leq g_m\Rightarrow \fdbox{g'=-g}{I}\, (\lsubst{s}{g_M}{g})\\
& g\leq g_m\Rightarrow \lsubst{(\fdbox{g'=-g}{I})}{g_M}{g}\\
%& = \lget_\texttt{g}\, s\leq g_m\Rightarrow \lsugarget{(\lsubst{s}{g_M}{g})}{g}\geq 0\\
& = g\leq g_m\Rightarrow \lsubst{(g\geq 0)}{g_M}{g}\\
%& = (\lget_\texttt{g}\, s\leq g_m\Rightarrow g_M\geq 0) = \top,
& = (g\leq g_m\Rightarrow g_M\geq 0) = \top,
\end{aligned}
\qquad\qquad
\begin{aligned}[c]
%& \lget_g\, s> g_m\Rightarrow\fdbox{g'=-g}{I}\, s\\
& g> g_m\Rightarrow\fdbox{g'=-g}{I}\, s\\
%& = \lsugarget{s}{g}> g_m\Rightarrow\lsugarget{s}{g}\geq 0\\
& = g> g_m\Rightarrow g\geq 0\\
& = \top,
\end{aligned}
\end{equation*}
by~(\ref{eq:wlp-assign}), and because $g_m,g_M\geq0$. Thus, $\left(I\Rightarrow\fdbox{\texttt{ctrl}\seqcomp\texttt{dyn}}{I}\right)=\top$. \qed
\end{example}

%%%%%%%%%%%% FWD DIAMONDS %%%%%%%%%%%%%
\subsubsection{Forward diamonds}\label{subsec:fdia}
% Talk about the four operators?
% Categorically, we know that Kleisli extensions $\alpha^\dagger 
% = \bigcup\circ\Pow$ model backward diamonds $\bddia{\alpha}{}$~\cite{MuniveS22}. 
% Their right adjoints on the boolean algebra of tests are the forward boxes $\fdbox{\alpha}{}$.
Here we add to our VCG approach by including forward diamonds 
in our verification framework. Our VCG laws from 
Sections~\ref{subsec:fdbox} and~\ref{subsec:hoare} 
help users prove partial correctness specifications. Yet, 
our approach is generic and extensible and can cover other types of 
specifications~\cite{GomesS16,MuniveS22,afp:transem}. For instance, 
we have integrated the \emph{forward diamond} $\fddia{-}{-}$ 
predicate transformer, defined as
$\fddia{\alpha}{Q}\, s\Leftrightarrow(\exists s'.\ s'\in\alpha\, s\land Q\, s')$.
It holds if there is a $Q$-satisfying state of $\alpha$ 
reachable from $s$.
%For deterministic programs, Dijkstra's total correctness \emph{weakest precondition} $\wp$ corresponds to $\wp\, \alpha\, Q\, s =\fddia{\alpha}{\top}\, s \land \fdbox{\alpha}{Q}\, s$ and similarly, total correctness Hoare triples satisfy $\totalHoare{P}{\alpha}{Q}\Leftrightarrow (P\preceq\fddia{\alpha}{\top}\land P\preceq\fdbox{\alpha}{Q})$.
Due to their semantics, forward diamonds enable us to reason 
about progress and reachability properties. In applications, 
this implies that our tool supports proofs describing worst 
and best case scenarios stating that the modelled system can 
evolve into an undesired/desired state. See Section~\ref{sec:ex} 
for an example showing progress for a dynamical system. We 
formalise and prove the forward diamonds laws below which are 
also direct consequences of the duality law 
$\fddia{\alpha}{Q}=\neg\fdbox{\alpha}{\neg Q}$. 
The example immediately after them
merely illustrates their seamless application.\vspace{1ex}

\begin{center}
\begin{tabular}{l r c l r}
(\customtag{fdia-skip}{eq:fdia-skip})
& $\fddia{\SKIP}{Q}$                            & $=$                       & $Q$                         
&\isalink{https://github.com/isabelle-utp/Hybrid-Verification/blob/e766a6b4744f37e176cd289e9e80120b6238ad81/Hybrid_Programs/Regular_Programs.thy\#L22}\\
(\customtag{fdia-abort}{eq:fdia-abort})
& $\fddia{\ABORT}{Q}$                        & $=$                       & $\bot$
&\isalink{https://github.com/isabelle-utp/Hybrid-Verification/blob/e766a6b4744f37e176cd289e9e80120b6238ad81/Hybrid_Programs/Regular_Programs.thy\#L36}\\
(\customtag{fdia-test}{eq:fdia-test})
& $\fddia{\mquestiondown P?}{Q}$      & $=$                       & $P\land Q$
&\isalink{https://github.com/isabelle-utp/Hybrid-Verification/blob/e766a6b4744f37e176cd289e9e80120b6238ad81/Hybrid_Programs/Regular_Programs.thy\#L57}\\
(\customtag{fdia-assign}{eq:fdia-assign})
& $\fddia{x := e}{Q}$                             & $=$                       & $\lsubst{Q}{e}{\lens}$
&\isalink{https://github.com/isabelle-utp/Hybrid-Verification/blob/e766a6b4744f37e176cd289e9e80120b6238ad81/Hybrid_Programs/Regular_Programs.thy\#L94}\\
(\customtag{fdia-seq}{eq:fdia-seq})
& $\fddia{\alpha\seqcomp\beta}{Q}$    & $=$                       & $\fddia{\alpha}{\fddia{\beta}{Q}}$
&\isalink{https://github.com/isabelle-utp/Hybrid-Verification/blob/e766a6b4744f37e176cd289e9e80120b6238ad81/Hybrid_Programs/Regular_Programs.thy\#L183}\\
(\customtag{fdia-choice}{eq:fdia-choice})
& $\fddia{\alpha\sqcap\beta}{Q}$                 & $=$                        & $\fddia{\alpha}{Q}\lor\fddia{\beta}{Q}$
&\isalink{https://github.com/isabelle-utp/Hybrid-Verification/blob/e766a6b4744f37e176cd289e9e80120b6238ad81/Hybrid_Programs/Regular_Programs.thy\#L133}\\
(\customtag{fdia-loop}{eq:fdia-loop})
& $\fddia{\LOOP{\alpha}}{Q}$         & $=$   & $\exists n.\ \fddia{\alpha^n}{Q}$
&\isalink{https://github.com/isabelle-utp/Hybrid-Verification/blob/e766a6b4744f37e176cd289e9e80120b6238ad81/Hybrid_Programs/Regular_Programs.thy\#L366}\\
(\customtag{fdia-cond}{eq:fdia-cond}) 
& $\fddia{\IF{T}{\alpha}{\beta}}{Q}$      & $=$ 
& $(T\land\fddia{\alpha}{Q})\lor(\neg T\land\fddia{\beta}{Q})$.
&\isalink{https://github.com/isabelle-utp/Hybrid-Verification/blob/e766a6b4744f37e176cd289e9e80120b6238ad81/Hybrid_Programs/Regular_Programs.thy\#L237}\\
\end{tabular}
\end{center}\vspace{1ex}

Additionally, the (informal) diamond law for evolution commands is\vspace{1ex}

\begin{center}
\begin{tabular}{l r c l r}
(\customtag{fdia-evol}{eq:fdia-evol})
& $\fddia{(x' = f\, \&\, G)_U^{t_0}}{Q}\, s$ & $\Leftrightarrow$ & 
    \begin{tabular}{@{}c@{}}
    $\exists X\in\ivpsols\, U\, f\, t_0\, s.\ \exists t\in U\, s.$ \\ 
    $(\forall\tau\in\downcl{U\, s}{t}.\ G\, (X\, \tau))\Rightarrow Q\, (X\, t)$.
    \end{tabular}
    %\begin{tabular}{@{}c@{}}
    % $ \exists X\in\ivpsols\, U\, (\ldowngr{f}{s}{\lens})\, t_0\, \lsugarget{s}{\lens}.\ \exists t\in U\, \lsugarget{s}{\lens}.$ \\ 
    % $(\forall\tau\in\downcl{U\, \lsugarget{s}{\lens}}{t}.\ \lsubst{G}{X\, \tau}{\lens})\Rightarrow Q\, \lsubst{s}{X\, t}{\lens}$.
    %\end{tabular}
&\isalink{https://github.com/isabelle-utp/Hybrid-Verification/blob/e766a6b4744f37e176cd289e9e80120b6238ad81/Hybrid_Programs/Evolution_Commands.thy\#L92}\\
\end{tabular}
\end{center}\vspace{1ex}
%\begin{align*}
%\isalink{https://github.com/isabelle-utp/Hybrid-Verification/blob/8c2760f4ad5006edfbbcae95762f9ae816467302/HS_Lens_Spartan.thy\#L972}
%\qquad\fddia{(x' = f\, \&\, G)_U^{t_0}}{Q}\, s \Leftrightarrow\ & \exists X\in\ivpsols\, U\, (\ldowngr{f}{s}{\lens})\, t_0\, \lsugarget{s}{\lens}.\ \exists t\in U\, \lsugarget{s}{\lens}. \label{eq:fdia-evol}\tag{fdia-evol}\\ & (\forall\tau\in\downcl{U\, \lsugarget{s}{\lens}}{t}.\ G\, \lsubst{s}{X\, \tau}{\lens})\Rightarrow Q\, \lsubst{s}{X\, t}{\lens},
%\end{align*}
and the corresponding law for flows $\varphi:T\to\csrc\to\csrc$ of $f$ 
%$\ldowngr{f}{s}{\lens}$ 
and $U=T=\reals_{\geq 0}$ is

\begin{center}
\begin{tabular}{@{}l r c l r@{}}
(\customtag{fdia-flow}{eq:fdia-flow})
& $\fddia{x' = f\, \&\, G}{Q}\, s$ & $\Leftrightarrow$ 
& $(\exists t\geq 0.\ (\forall\tau{\in}[0,t].\ G\, (\varphi_{s}\, \tau))\Rightarrow Q\, (\varphi_{s}\, t))$.          
&\isalink{https://github.com/isabelle-utp/Hybrid-Verification/blob/e766a6b4744f37e176cd289e9e80120b6238ad81/Hybrid_Programs/Evolution_Commands.thy\#L139}\\
\end{tabular}
\end{center}\vspace{1ex}

%\begin{equation*}
%\isalink{https://github.com/isabelle-utp/Hybrid-Verification/blob/8c2760f4ad5006edfbbcae95762f9ae816467302/HS_Lens_Spartan.thy\#L1019}
%\quad\fddia{x' = f\, \&\, G}{Q}\, s \Leftrightarrow (\exists t\geq 0.\ (\forall\tau\in[0,t].\ G\, \lsubst{s}{\varphi_{\lsugarget{s}{\lens}}\, \tau}{\lens})\Rightarrow Q\, \lsubst{s}{\varphi_{\lsugarget{s}{\lens}}\, t}{\lens}). \label{eq:fdia-flow}\tag{fdia-flow}
%\end{equation*}

% example forward diamond
\begin{example}\label{ex:fddia} A similar argument as that in 
Example~\ref{ex:vcg} allows us to prove the inequality
$I\Rightarrow \fddia{\texttt{blood\_sugar}}{I}$ where 
$I\, s\Leftrightarrow (g\geq 0)$ and 
$\texttt{blood\_sugar}= \LOOP{(\texttt{ctrl}\seqcomp\texttt{dyn})}$. 
Namely, we observe that by~(\ref{eq:fdia-evol}), the forward diamond
of $g'=-g$ and $I$ becomes
\begin{equation*}
\fddia{g'=-g}{I}\, s
\Leftrightarrow (\exists t\geq 0.\ (\lsubst{I}{g\cdot e^{- t}}{g}))
\Leftrightarrow (\exists t\geq 0.\  g\cdot e^{- t}\geq 0)
\Leftrightarrow g\geq 0,
\end{equation*}
for all $s\in\src$. Hence, the conjuncts below simplify as shown:
\begin{equation*}
\begin{aligned}[c]
& g\leq g_m\land \lsubst{(\fddia{g'=-g}{I})}{g_M}{g}\\
& = g\leq g_m\land\lsubst{(g\geq 0)}{g_M}{g}\\
& = g\leq g_m\land g_M\geq 0 = g\leq g_m,
\end{aligned}
\qquad\qquad
\begin{aligned}[c]
& \lsugarget{s}{g}> g_m\land\fddia{g'=-g}{I}\, s\\
& = \lsugarget{s}{g}> g_m\land \lsugarget{s}{g}\geq 0\\
& = \lsugarget{s}{g}> g_m.
\end{aligned}
\end{equation*}
Therefore, by backward reasoning with the diamond laws, we have
\begin{align*}
&I\Rightarrow\fddia{\INV{\LOOP{(\texttt{ctrl}\seqcomp\texttt{dyn})}}{I}}{I} \\
& \Leftarrow I\Rightarrow\fddia{\texttt{ctrl}\seqcomp\texttt{dyn}}{I}\\
& = \left(\forall s.\ I\, s\Rightarrow\fddia{\IF{g\leq g_m}{g:=g_M}{\SKIP}}{\fddia{g'=-g}{I}}\, s\right)\\
& = \left(\forall s.\ I\, s
\Rightarrow \left(g\leq g_m\land \fddia{g:=g_M}{\fddia{g'=-g}}{I}\, s\right)
	\lor \left(g> g_m\land\fddia{g'=-g}{I}\, s\right)\right)\\
& = (\forall s.\ I\, s\Rightarrow g\leq g_m\lor g> g_m) = \top,
\end{align*}
where the first implication follows by a rule 
analogous to~(\ref{eq:h-loopi}) for
diamonds.\hfill\isalink{https://github.com/isabelle-utp/Hybrid-Verification/blob/c00d1b16490ce79c481a7be298ba43ba44837531/Hybrid_Programs/Regular_Programs.thy\#L691}

\noindent Thus, we have shown that $I\Rightarrow\fddia{\texttt{blood\_sugar}}{I}$. \qed
\end{example}\vspace{1ex}

We have summarised our approach to hybrid systems verification
in general purpose proof assistants~\cite{FosterMGS21,MuniveS22}. 
This is the basis for describing our contributions for the
rest of this article, included among them, the formalisation
of forward diamonds into IsaVODEs. Although a similar 
formalisation has 
been done before~\cite{afp:transem}, our implementation is 
more automated due to its use of standard types, e.g. Isabelle 
predicates ($\src\to\bools$), that have had more support over 
time. Thus, our formalisation increases the proof capabilities 
of our Isabelle-based framework and its expressivity, since 
the forward diamonds enable us to assert the progress of hybrid
programs. Other extensions to our framework not described here 
are the addition of \dL's nondeterministic assignments and their 
corresponding partial correctness and progress laws, as well as 
the formalisation of variant-based rules on the reachability of 
finite iterations and while-loops. See previous 
work~\cite{ArmstrongGS16, GomesS16} %section~\ref{sec:ex} for an example using 
% nondeterministic assignments and 
for examples with while-loops that our verification framework could
tackle.\hfill\isalink{https://github.com/isabelle-utp/Hybrid-Verification/blob/e766a6b4744f37e176cd289e9e80120b6238ad81/Hybrid_Programs/Regular_Programs.thy\#L518}

\section{Hybrid Modelling Language}\label{sec:hs}

Here, we describe our implementation of a hybrid modelling language, which takes advantage of lenses and Isabelle's
flexible syntax processing features. Beyond the advantages already mentioned, lenses enhance our hybrid store models in
several ways. They allow us to model frames---sets of mutable variables---and thus support local reasoning. They also
allow us to project parts of the global store onto vector spaces to describe continuous dynamics. These projections can
be constructed using three lens combinators: composition, sum and quotient.

The projections particularly allow us to use hybrid state spaces, consisting of both continuous components with a
topological structure (e.g. $\reals^n$), and discrete components using Isabelle's flexible type system. This in turn
allows our tool to support more general software engineering notations, which typically make use of
object-oriented data structures~\cite{Miyazawa2019-RoboChart}. Moreover, the projections allow us to reason locally
about the continuous variables, since discrete variables are outside of the frame during continuous evolution.

%%%%%%%%%% STATES AND VARIABLES %%%%%%%%%%%
\subsection{Dataspaces}\label{subsec:dataspaces}

Most modelling and programming languages support modules with local variables, constant declarations, and
assumptions. We have implemented an Isabelle command that automates the creation of hybrid stores, which provide a local
theory context for hybrid programs. We call these \emph{dataspaces}, since they are state spaces that can make use of rich data
structures in the program variables. \hfill\isalink{https://github.com/isabelle-utp/Optics/blob/main/Dataspace_Example.thy}

\vspace{1ex}
{
\begin{alltt}
\isakwmaj{dataspace} store = [parent_store +]
  \isakwmin{constants} c\(\sb{1}\)::C\(\sb{1}\) ... c\(\sb{n}\)::C\(\sb{n}\)
  \isakwmin{assumes} a\(\sb{1}\):P\(\sb{1}\) ... a\(\sb{n}\):P\(\sb{n}\)
  \isakwmin{variables} x\(\sb{1}\)::T\(\sb{1}\) ... x\(\sb{n}\)::T\(\sb{n}\)
\end{alltt}}
\vspace{1ex}

\noindent A dataspace has constants $c_i : C_i$, named constraints $a_i : P_i$ and state variables $x_i : T_i$. In its
context, we can create local definitions, theorems and proofs, which are hidden, but accessible using its
namespace. Internally, the dataspace command creates an Isabelle \isakwmaj{locale} with fixed constants and assumptions,
inspired by previous work by Schirmer and Wenzel~\cite{Schirmer2009}. Like locales, dataspaces support a form of
inheritance, whereby constants, assumptions, and variables can be imported from an existing dataspace
(e.g. \texttt{parent\_store}) and extended with further constants, assumptions, and variables.

Each declared state variable is assigned a lens $x_i :: T_i \lto \src$, using the abstract store type $\src$ with the lens axioms from Secion~\ref{subsec:stores}
as locale assumptions. We also generate independence assumptions, e.g. $x_i\lindep x_j$ for $x_i\neq x_j$, that distinguish different variables semantically~\cite{Foster2020-IsabelleUTP}. Formally, $x\lindep y$ if
$\lput_x\,  u\circ \lput_y\, v = \lput_y\, v\circ\lput_x\, u$ for all $u,v\in\view$. That is, two lenses are independent if their $\lput$ operations commute on all states.

\subsection{Lifted Expressions}

As discussed in Section~\ref{subsec:stores}, expressions in our hybrid modelling language are modelled by functions of type $\src\rightarrow
\view$. Assertions are therefore state predicates, or ``expressions'' where $\view = \bools$. Discharging VCs requires showing that
assertions hold for all states. For example, the law~(\ref{eq:wlp-test}) requires us to prove a VC of the form $P \implies Q$, that is, $\forall s.\ P s\implies Q\, s$. 
Also, if we have state variables $x$ and $y$, then proving the assertion
$x + y \ge x$ corresponds to proving the HOL predicate $\lget_x~s + \lget_y~s \ge \lget_x~s$ for some
arbitrary-but-fixed state $s$, which can readily by discharged using one of Isabelle's proof methods (\skey{simp},
\skey{auto} etc.). This process is automated by methods \skey{expr-simp} and \skey{expr-auto}.

Nevertheless, there remains a gap between the syntax used in typical programming languages and its semantic
representation. Namely, users would prefer writing $x^2 + y^2 \le c^2$ over
$\lambda s.\ (\lget_x~s)^2 + (\lget_y~s)^2 \le c^2$, and so, the main technical challenge is to seamlessly transform
between the two. This can be achieved using Isabelle's syntax pipeline, which significantly improves the usability of
our tool.
 
Isabelle's multi-stage syntax pipeline parses Unicode strings
and transforms them into ``pre-terms''~\cite{Kuncar2019-IsabelleFoundation}:
elements of the ML \texttt{term} type containing syntactic constants. These must
be mapped to previously defined semantic constants by syntax translations,
before they can be checked and certified in the Isabelle kernel. Printing
reverses this pipeline, mapping terms to strings.

We automate the translation between the expression syntax (pre-terms) and semantics using parse and print translations implemented in Isabelle/ML, as part of our \textsf{Shallow-Expressions} component. It lifts pre-terms by replacement of free variables
and constants, and insertion of store variables ($\skey{s}$) and $\lambda$-binders. Its implementation uses the syntactic annotation $\ebrack{t}$ to lift the syntactic term $t$ to a semantic expression in the syntax translation rules. The syntax translation is described by the following equations:
$$\begin{array}{ccc}
  \ebrack{t} \syneq [\ebracku{t}]_{\skey{e}},  &\quad \ebracku{n} \syneq
          \begin{cases}
            \lambda \skey{s}.\ \lget_n~\skey{s} & \text{if } n \text{ is a lens},\\
            \lambda \skey{s}.\ n~\skey{s} & \text{if } n \text { is an expression},\\
            \lambda \skey{s}.\ n & \text{otherwise},
          \end{cases} &\quad \ebracku{f~t} \syneq \lambda \skey{s}.\ f~(\ebracku{t}~\skey{s}),
  \end{array}$$
where $p \syneq q$ means that pre-term $p$ is translated to term
$q$, and $q$ printed as $p$. Moreover, $[-]_{\skey{e}}$ is a
constant that marks lifted expressions that are embedded in
terms. When the translation encounters a name $n$ (i.e. a free variable or constant), it checks whether $n$ has a definition in the context. If it 
is a lens (i.e. $n :: \view \lto \src)$, then it inserts a $\lget$. If it is an
expression (i.e. $n : \src \to \view$), then it is applied to the state. Otherwise, it leaves the name unchanged, assuming it to be a constant. Function applications are also left unchanged by $\rightleftharpoons$. For instance,
%, except for expression constructs like $e[v/x]$.  
% substitution not yet introduced
$\ebrack{(x + y)^2 / z} \rightleftharpoons [\lambda \skey{s}.\, (\lget_x~\skey{s} + \lget_y~\skey{s})^2 / z]_{\skey{e}}$ for variables (lenses) $x$ and $y$ and constant $z$. Once an expression has been processed, the resulting $\lambda$-term is enclosed in $[-]_{\skey{e}}$. The pretty printer can then recognise a lifted term and print it. This process is fully automated, so that users see only the sugared expression, without the $\lambda$-binders, in both the parser and terms' output during the proving process. \hfill\isalink{https://github.com/isabelle-utp/Shallow-Expressions/blob/bddbf31a67859011c81e16ad3d6723d66ed9591e/Expressions.thy\#L58}

%  Intuitively, $\ebrack{t}$ is processed as follows. The syntax
%  processor first parses a pre-term from string $t$. Then our parse
%  translation traverses its syntax tree. Whenever it encounters a free
%  variable $x$, the type system determines from the context whether it
%  is a lens, in which case a $\lget_x$ is inserted. Otherwise it is
%  left unchanged as a logical variable. Function applications are left
%  unchanged by $\rightleftharpoons$, except for expression constructs
%  like $e[v/x]$.  For program variables $x$ and $y$ and logical
%  variable $z$, e.g.,
%  $\ebrack{(x + y)^2 / z} \rightleftharpoons [\lambda \skey{s}.\,
%  (\lget_x~\skey{s} + \lget_y~\skey{s})^2 / z]_{\skey{e}}$. Once an
%  expression has been processed, the resulting $\lambda$-term is
%  enclosed in $[-]_{\skey{e}}$.

%%%%%%%%%%%% SUBSTITUTIONS %%%%%%%%%%%%%
\subsection{Substitutions}\label{subsec:subst}

In our semantic approach, substitutions correspond to functions $\sigma : \src \to \src$ on the store $\src$. This interpretation allows us to denote updates as a sequence of variable assignments. 
That is, instead of directly manipulating the store $s : \src$ with the lens functions,
we provide more user-friendly program specifications with the notation
$\sigma(x\!\smapsto\!e) = \lambda s.\,\lput_x\, (e\, s)\, (\sigma\,s)$. It allows us to describe assignments as
sequences of updates: $[x_1 \leadsto e_1, x_2 \leadsto e_2, \cdots] = id(x_1 \leadsto e_1)(x_2 \leadsto e_2)\cdots$, for
variable lenses $x_i :: \view_i \lto \src$ and ``expressions'' $e_i : \src \rightarrow \view_i$.%, which as usual in a shallow embedding are modelled as functions.

Implicitly, any variable $y$ not mentioned in such a substitution is left unchanged: $y \smapsto y$. We further write
$e[v/x] = e \circ [x \smapsto v]$, for $x :: \view \lto \src$, $e : \src \rightarrow \view'$, and
$v : \src \rightarrow \view$, for the application of substitutions to expressions. This yields standard notations for
program specifications, e.g.  $(x := e) = \assigns{[x \leadsto e]}$ and $\wlp~\assigns{[x\leadsto e]}~Q = Q[e/x]$. Using
an Isabelle simplification procedure (a ``simproc''), the simplifier can reorder assignments alphabetically according to
their variable name, and otherwise reduce and evaluate substitutions during VCG~\cite{Foster2020-IsabelleUTP}. We can
extract assignments for $x$ writing $\substlk{\sigma}\,x = \lget_x \circ \sigma$ so that, e.g.
$\substlk{[x \smapsto e_1, y \smapsto e_2]}\, x$ reduces to $e_1$ when $x \lindep
y$.\hfill\isalink{https://github.com/isabelle-utp/Shallow-Expressions/blob/main/Substitutions.thy}

\vspace{1ex}
\begin{example} We continue our blood glucose running example and formalise Example~\ref{ex:hp}. First, we declare our problem variables and assumptions via our \isa{\isacommand{dataspace}} command. We name this dataspace \isa{glucose}, and assume that there is a minimal warning threshold $g_m>0$ and a maximum dosage $g_M>g_m$. The patient's glucose is represented via the ``continuous'' variable $g$.

\begin{isabellebody}\isanewline
\isacommand{dataspace}\isamarkupfalse%
\ glucose\ {\isacharequal}{\kern0pt}\ \isanewline
\ \ \isakeyword{constants}\ g\isactrlsub m\ {\isacharcolon}{\kern0pt}{\isacharcolon}{\kern0pt}\ real\ g\isactrlsub M\ {\isacharcolon}{\kern0pt}{\isacharcolon}{\kern0pt}\ real\isanewline
\ \ \isakeyword{assumes}\ ge{\isacharunderscore}{\kern0pt}{\isadigit{0}}{\isacharcolon}{\kern0pt}\ {\isachardoublequoteopen}g\isactrlsub m\ {\isachargreater}{\kern0pt}\ {\isadigit{0}}{\isachardoublequoteclose}\ \isakeyword{and}\ ge{\isacharunderscore}{\kern0pt}gm{\isacharcolon}{\kern0pt}\ {\isachardoublequoteopen}g\isactrlsub M\ {\isachargreater}{\kern0pt}\ g\isactrlsub m{\isachardoublequoteclose}\isanewline
\ \ \isakeyword{variables}\ g\ {\isacharcolon}{\kern0pt}{\isacharcolon}{\kern0pt}\ real\isanewline
\end{isabellebody}

Next, inside the glucose context we declare, via Isabelle's \isa{\isacommand{abbreviation}} command, the definition of the controller and the dynamics. Our shallow expressions hide the lens infrastructure and, from the user's perspective, the definitions are Isabelle abbreviations. Notice also, that our recently introduced ``substitution'' notation allows us to explicitly specify the flow's behaviour on the continuous variable $g$. It also occurs implicitly in our declaration of the differential equation $g'=g$ (see Section~\ref{subsec:dynevol}).

\begin{isabellebody}\isanewline
\isacommand{context}\isamarkupfalse%
\ glucose\isanewline
\isakeyword{begin}\isanewline
\isanewline
\isacommand{abbreviation}\isamarkupfalse%
\ {\isachardoublequoteopen}ctrl\ {\isasymequiv}\ IF\ g\ {\isasymle}\ g\isactrlsub m\ THEN\ g\ {\isacharcolon}{\kern0pt}{\isacharcolon}{\kern0pt}{\isacharequal}{\kern0pt}\ g\isactrlsub M\ ELSE\ skip{\isachardoublequoteclose}\isanewline
\isanewline
\isacommand{abbreviation}\isamarkupfalse%
\ {\isachardoublequoteopen}dyn\ {\isasymequiv}\ {\isacharbraceleft}{\kern0pt}g{\isacharbackquote}{\kern0pt}\ {\isacharequal}{\kern0pt}\ {\isacharminus}{\kern0pt}g{\isacharbraceright}{\kern0pt}{\isachardoublequoteclose}\isanewline
\isanewline
\isacommand{abbreviation}\isamarkupfalse%
\ {\isachardoublequoteopen}flow\ {\isasymtau}\ {\isasymequiv}\ {\isacharbrackleft}{\kern0pt}g\ {\isasymleadsto}\ g\ {\isacharasterisk}{\kern0pt}\ exp\ {\isacharparenleft}{\kern0pt}{\isacharminus}{\kern0pt}\ {\isasymtau}{\isacharparenright}{\kern0pt}{\isacharbrackright}{\kern0pt}{\isachardoublequoteclose}\isanewline
\isanewline
\isacommand{abbreviation}\isamarkupfalse%
\ {\isachardoublequoteopen}blood{\isacharunderscore}{\kern0pt}sugar\ {\isasymequiv}\ LOOP\ {\isacharparenleft}{\kern0pt}ctrl{\isacharsemicolon}{\kern0pt}\ dyn{\isacharparenright}{\kern0pt}\ INV\ {\isacharparenleft}{\kern0pt}g\ {\isasymge}\ {\isadigit{0}}{\isacharparenright}{\kern0pt}{\isachardoublequoteclose}\isanewline

\isakeyword{end}\isanewline
\end{isabellebody}

\noindent Thus, our lens integrations provide a seamless way to formalise hybrid system verification problems in Isabelle. We explore their verification condition generation in Section~\ref{subsec:lipschitz}.\qed
\end{example}

%%%%%%%%%% VECTORS AND MATRICES %%%%%%%%%%
\subsection{Vectors and matrices}\label{subsec:vecmat}
Vectors and matrices are ubiquitous in engineering applications and users of 
our framework would appreciate using familiar concepts and notations to them.
This is possible due to our modelling language. In particular, vectors are 
supported by \textsf{HOL-Analysis} using finite Cartesian product types,
\texttt{(A,n) vec} with the notation \texttt{A\textasciicircum n}. Here, \texttt{A} is the element type, and \texttt{n} is a numeral type denoting the dimension. The type of vectors is isomorphic to $[n] \rightarrow A$ where  $[n]=\{1,\dots,n\}$.
A matrix is simply a vector of vectors, \texttt{A\textasciicircum m\textasciicircum n}, hence a map $[m] \rightarrow [n] \rightarrow A$. 
Building on this, we supply notation
\texttt{[[x11,...,x1n],...,[xm1,...,xmn]]} for matrices and means for accessing coordinates of vectors via hybrid
program variables~\cite{FosterGC20}. This notation supports the inference of vector and matrices' dimensions conveyed by the type variables.

Vectors and matrices are often represented as composite objects consisting of several values, e.g. $p = (p_x, p_y)\in\reals^2$. When writing specifications, it is often convenient to refer to and
manipulate these components individually. We can denote such variables using component lenses and the lens composition
operator. We write $\lens_1 \lcomp \lens_2::\src_1\lto \src_3$, for $\lens_1::\src_1 \lto \src_2$,
$\lens_2::\src_2\lto\src_3$, for the forward composition and $\lone_\src :: \src \lto \src$ for the units in the lens
category, but do not show formal definitions~\cite{Foster2020-IsabelleUTP}. Intuitively, the composition $\lcomp$ selects part of a
larger store as illustrated below.

We model vectors in $\reals^n$ as part of larger hybrid stores, lenses $v::\reals^n\lto \src$, and project onto
coordinate $v_k::\reals\lto\src$ using lens composition and a \emph{vector lens} $\Pi(i)::\reals\lto\reals^n$:\hfill\isalink{https://github.com/isabelle-utp/Hybrid-Library/blob/6b37e352181fb5613cc9a960df6aed12d68cf370/Cont_Lens.thy\#L117}
%\begin{equation*}
%\Pi(k : [n]) = ((\lambda s.\, \textit{vec-nth}~s~k) : A^n \rightarrow A, (\lambda v~s.\,
%\textit{vec-upd}~s~k~v) : A \rightarrow A^n \rightarrow A^n),
%\end{equation*} 
\begin{align*}
\Pi(i) &= (\lget_{\Pi(i)}, \lput_{\Pi(i)}),\text{ where }\\
\lget_{\Pi(i)} &= (\lambda s.\, \textit{vec-nth}~s~i), \\
\lput_{\Pi(i)} &= (\lambda v~s.\, \textit{vec-upd}~s~i~v),
\end{align*}
and $i\in [n]=\{1,\dots,n\}$.%, and $A^n$ is the vector type \texttt{A\textasciicircum n}. 
The lookup function
$\textit{vec-nth} : A^n \rightarrow [n] \rightarrow A$ and update function
$\textit{vec-upd} : A^n \rightarrow [n] \rightarrow A \rightarrow A^n$ come from \textsf{HOL-Analysis} and satisfy the
lens axioms (Section~\ref{subsec:stores}). Then, as an example, $p_x = \Pi(1) \lcomp p$ and $p_y = \Pi(2) \lcomp p$ for $p :: \reals^2 \lto \src$, using
$\lcomp$ to first select the variable $p$ and then the vector-part of the hybrid store.  Intuitively, two vector
elements are independent, $\Pi(i) \lindep \Pi(j)$ iff they have different indices, $i \neq j$.

\vspace{1ex}

\begin{example} \label{ex:boat}

To illustrate the use of vector variables, we model the dynamics and 
a controller for an autonomous boat. We refer readers to previous 
publications for the verification of an invariant for this system~\cite{FosterGC20,FosterMGS21}. 
The boat is manoeuvrable in $\real^2$ and has a rotatable thruster generating a positive propulsive force $f$ with maximum $f_{max}$. The boat's state is determined by its position $p=(p_x,p_y)$, velocity $v=(v_x,v_y)$, and acceleration $a=(a_x,a_y)$. We describe this state with the following dataspace:\hfill\isalink{https://github.com/isabelle-utp/Hybrid-Verification/blob/e766a6b4744f37e176cd289e9e80120b6238ad81/Hybrid_Programs/Verification_Examples/AMV.thy\#L220}

\vspace{1ex}
\begin{alltt}
\isakwmaj{dataspace} AMV =
  \isakwmin{constants} S::\(\real\) f\(\sb{max}\)::\(\real\) \isakwmin{assumes} fmax:"f\(\sb{max}\) \(\ge\) 0"
  \isakwmin{variables} p::"\(\real\) vec[2]" v::"\(\real\) vec[2]" a::"\(\real\) vec[2]" \(\phi\)::\(\real\) s::\(\real\)
  wps::"(\(\real\) vec[2]) list" org::"(\(\real\) vec[2]) set" rs::\(\real\) rh::\(\real\)
\end{alltt}
\vspace{1ex}

\noindent This store model combines discrete and continuous variables and uses the alternative notation $\real~\texttt{vec[n]}$ for a real-valued vector of dimension $n$. The \isakwmaj{dataspace} specifies a variable for linear speed $s$, and a constant $S$ for the boat's maximum speed. We also provide discrete variable \texttt{wps} for a list of points to pass through in the vehicle's path (way-point path), \texttt{org} for a set of points where obstacles are located (obstacle register), and the requested speed and heading (\texttt{rs} and \texttt{rh}). 
Our \isakwmaj{dataspace} allows us to declare variables $p, v, a:\real\ \texttt{vec[2]}$ and manipulate them using operations for vectors (see Section~\ref{subsec:dynevol}).\qed %and transcendental functions.\qed
\end{example}

%%%%%%%%%%%%%% FRAMES %%%%%%%%%%%%%%%
\section{Local Reasoning}\label{sec:frames}

In this section, we describe our framework's support for local reasoning, which allows us to consider only parts of the state that are changed by a component in the verification. This improves the scalability of our approach, since we can decompose verification tasks into smaller manageable tasks, in an analogous way to separation logic~\cite{Reynolds2002}. We show how lenses can be used to characterise a program's frame: the set of variables which may be modified. We then explain how frames extend to evolution commands, such that variables with no derivative (or derivative 0) are outside of the frame. Next, we develop a framed version of differentiation, called \emph{framed Fr\'echet derivatives}, which allows us to perform local differentiation with respect to a strict subset of the store variables. This, in turn, supports a method, framed differential induction, for proving invariants in the continuous part of the state space. Finally, we introduce a corresponding implementation of \dL's differential ghost rule~\cite{Platzer18} that augments systems of ODEs with fresh equations to aid invariant reasoning. This rule likewise supports frames.

\subsection{Frames}\label{subsec:frames}

Lenses support algebraic manipulations of variable frames. A \emph{frame} is the set of variables that a program is permitted
to change. Variables outside of the frame are immutable. We first show how variable sets can be modelled via lens
sums. Then we recall a predicate characterising immutable program variables~\cite{Foster2021-IsaSACM}. Most importantly,
we derive a frame rule \`a la separation logic for local reasoning with framed variables.

Variable lenses $\lens_1 :: \view_1\lto \src$ and $\lens_2 ::\view_2 \lto \src$ can be combined into lenses for variable sets with \emph{lens sum}~\cite{Foster2020-IsabelleUTP},
$\lens_1 \lplus \lens_2:: \view_1\times \view_2 \lto \src$ if $\lens_1 \lindep \lens_2$ via $\lget_{\lens_1 \lplus \lens_2}\, (s_1,s_2)= (\lget_{\lens_1}\,s_1, \lget_{\lens_2}\,s_2)$ 
and $\lput_{\lens_1 \lplus \lens_2}\, (v_1,v_2)=\lput_{\lens_1}\,v_1 \circ\lput_{\lens_2}\,v_2$
%$(\lambda (s_1, s_2).\, (\lget_{\lens_1}\,s_1, \lget_{\lens_2}\,s_2), (\lambda (v_1, v_2).\, \lput_{\lens_1}\,v_1 \circ \lput_{\lens_2}\,v_2))$. 
This combines two independent lenses into a single lens with a product view. It can be used to
model composite variables, for example, $(x \oplus y) := (e, f)$ is a simultaneous assignment to $x$ and $y$. We can
decompose such a composite update into two atomic updates, with
$[(x, y)\!\smapsto\!(e_1, e_2)] = [x\!\smapsto\!e_1, y\!\smapsto\!e_2]$. We can also use lens sums to model finite sets, for example $\{x, y, z\}$ is modelled as $x \oplus (y \oplus z)$. Each variable in such a sum may have a different type, e.g. $\{v_x, \vec{p}\}$ is a valid and well-typed
construction. \hfill\isalink{https://github.com/isabelle-utp/Optics/blob/6e24cde61989a79f7601acc537dd2ee9fdf3f4f6/Lens_Algebra.thy\#L38}

Lens sums are only associative and commutative up-to isomorphism of cartesian products. We need heterogeneous orderings
and equivalences between lenses to capture this. We define a \emph{lens preorder}~\cite{Foster2020-IsabelleUTP},
$\lens_1 \preceq \lens_2 \Leftrightarrow \exists \lens_3.\, \lens_1 = \lens_3 \lcomp \lens_2$ that captures the part-of
relation between $\lens_1 :: \view_1 \lto \src$ and $\lens_2 :: \view_2 \lto \src$, e.g.  $v_x \preceq \vec{v}$ and
$\vec{p} \preceq \vec{p} \lplus \vec{v}$.  \emph{Lens equivalence} ${\lequiv} = {\preceq}\cap {\succeq }$ then
identifies lenses with the same shape in the store. Then, for variable set lenses up-to $\lequiv$, $\lplus$ models
$\cup$, $\lindep$ models $\notin$, %\mgsb{$\lindep$
 % models $\notin$}{just a detail: in general or only if the left
 % parameter corresponds to a single variable?}\jhsb{}{For var lens $x$
 %and set lens $A$, $x\lindep A$ models $x\notin A$ and $A\lindep x$ models
 % $x\notin A$, but I am not sure its worth giving this lengthy explanation}, 
and $\preceq$ models $\subseteq$ or $\in$. Since $\lens_1 \preceq \lens_1 \lplus \lens_2$ and
$\lens_1 \lplus \lens_2 \lequiv \lens_2 \lplus \lens_1$, with our variable set interpretation, we can show, e.g., that
$x \in \{x, y, z\}$, $\{x, y\} \subseteq \{x, y, z\}$, and $\{x, y\} = \{y, x\}$. Hence we can use these lens
combinators to construct and reason about variable
frames. \hfill\isalink{https://github.com/isabelle-utp/Optics/blob/main/Lens_Order.thy}

We can use variable set lenses to capture the frame of a program. Let $A::\view\lto\src$ be a lens modelling a variable
set. For $s_1,s_2\in \src$ let $s_1\approx_A s_2$ hold if $s_1=s_2$ up-to the values of variables in 
$A$, that is $\lget_A~s_1 = \lget_A~s_2$. Local reasoning within $A$ uses the 
\emph{lens quotient}~\cite{Foster2020-LocalVars} $\lens \lquot A$, which
localises a lens $\lens :: \view \lto \src$ to a lens $\view \lto \csrc$. Assuming $\lens \preceq A$, 
it yields $\lens_1 :: \view \lto \csrc$ such that $\lens = \lens_1 \lcomp A$. For example, $p_x \lquot p 
= \Pi(1)$ with $\csrc =\real^n$. \hfill\isalink{https://github.com/isabelle-utp/Optics/blob/6e24cde61989a79f7601acc537dd2ee9fdf3f4f6/Lens_Algebra.thy\#L83}.

We can also use lenses to describe when a variable does not occur freely in an expression or predicate with the
unrestriction property:
$A \unrest e \Leftrightarrow \forall v.\, e \circ (\lput_A~v) = e$~\cite{Foster2020-IsabelleUTP}. A variable $x$ is
unrestricted in $e$, written $x \unrest e$, provides that $e$ does not semantically depend on $x$ for its
evaluation. For example, $x \unrest (y + 1)$, when $x \lindep y$, since $y + 1$ does not mention $x$. We also define
$(- A) \unrest e \Leftrightarrow \forall s_1\, s_2\, v.\, e\,(\lput_A\,v\,s_1) = e\, (\lput_A\,v\,s_2)$ as the converse,
which requires that $e$ does not depend on variables outside of $A$.

Next, we capture the non-modification of variables by a program. For $\alpha:\src\rightarrow \Pow \src$ and an
expression (or predicate) $e$ we define
$\alpha\nmods e \Leftrightarrow \left(\forall s_1\in\src.\ e(s_1) = e(s_2)\right)$, which describes when $e$ does not
depend on the mutable variable of $\alpha$.  The expression $e$ can characterise a set of variables giving the set of
immutable variables. For example, we have it that $(x := x + 1) \nmods (y, z)$, when $x \lindep y$ and $x \lindep z$,
since this assignment changes only $x$ and no other variables.

Intuitively, non-modification $\alpha \nmods x$, where $x$ is a variable lens, is equivalent to the specification for $\hoare{x = v}{\alpha}{x = v}$ for
fresh logical variable $v$. This means that $x$ retains its initial value in any final state of $\alpha$. We prove the following laws for
non-modification:

\vspace{2ex}
\begin{center}
\begin{tabular}{cccc}
  \AxiomC{$A \unrest x$}
  \UnaryInfC{$(x := e) \nmods A$}
  \DisplayProof
  &
  \AxiomC{--}
  \UnaryInfC{$\mtest{P} \nmods A$}
  \DisplayProof
  &
  \AxiomC{$\alpha\nmods A$}
  \AxiomC{$\beta \nmods A$}
  \BinaryInfC{$(\alpha \relsemi \beta) \nmods A$}
  \DisplayProof
\end{tabular}

\vspace{2ex}

\begin{tabular}{ccc}
  \AxiomC{$\alpha\nmods A$}
  \AxiomC{$\beta \nmods A$}
  \BinaryInfC{$(\alpha\sqcap \beta) \nmods A$}
  \DisplayProof
  &
  \AxiomC{$\alpha\nmods A$}
  \UnaryInfC{$\alpha^* \nmods A$}
  \DisplayProof
  &
  \AxiomC{$\alpha \nmods B$}
  \AxiomC{$A \preceq B$}
  \BinaryInfC{$\alpha \nmods A$}
  \DisplayProof  
\end{tabular}
\end{center}

\vspace{2ex}

\noindent The variables in $A$ are immutable for assignment $x := e$ provided $x$ is not in $A$. A test $\mtest{P}$
modifies no variables, and therefore any set $A$ is immutable. For the programming operators, non-modification is
inherited from the parts. The final law shows that we can always shrink the specified set of immutable variables.

With these concepts in place, we derive two frame rules for local reasoning:
\hfill\isalink{https://github.com/isabelle-utp/Hybrid-Verification/blob/7baedd092dff0182336ef0bb6251fd8beff6a1cc/HS_Lens_Spartan.thy\#L456}
% \begin{equation} \label{eq:frame-rule}

\vspace{2ex}
\begin{center}
\begin{tabular}{cc}
\AxiomC{$\prog \nmods I$}
% \AxiomC{$(-A) \unrest I$}
\AxiomC{$\hoaretriple{P}{\prog}{Q}$}
\BinaryInfC{$\hoaretriple{P \land I}{\prog}{Q \land I}$}
\DisplayProof &
\AxiomC{$\prog \nmods A$}
\AxiomC{$(-A) \unrest I$}
\AxiomC{$\hoaretriple{P}{\prog}{Q}$}
\TrinaryInfC{$\hoaretriple{P \land I}{\prog}{Q \land I}$}
\DisplayProof
\end{tabular}
\end{center}

\vspace{2ex}

\noindent If program $\prog$ does not modify any variables mentioned in $I$, then $I$ can be added as an invariant of
$\prog$. In the first law, non-modification is checked directly of the variables used by $I$. In the second, which is an
instance of the first, we instead infer the immutable variables of $A$ and check that $I$ does not depend on variables
outside of $A$. With these laws, we can import invariants for a program fragment that refer to only those variables that
are left unchanged. This allows us to perform modular verification, whereby we need only consider invariants of
variables used in a component. In the following section, we show how this can be applied to systems of ODEs.

% \end{equation}

%%%%%%%%%% EVOLUTION COMMANDS %%%%%%%%%%%
\subsection{Framed evolution commands}\label{subsec:dynevol}
We extend previous components~\cite{MuniveS22} for continuous dynamics with function framing 
techniques that project onto parts of the store. That is, we formally describe the implementation of the
evolution command state transformer using the lens infrastructure described so
far~\cite{FosterMGS21}. Specifically, we use framing to derive continuous vector fields ($\csrc\to\csrc$)
and flows from state-wide ``substitutions'' ($\src\to\src$). We also add a non-modification rule for evolution
commands. This supports local reasoning where evolution commands modify only 
continuous variables and leave discrete ones---outside a frame---unchanged.

Framing uses the second interpretation of lenses where the frame $\csrc$ is 
a subregion of $\src$ that we can access through $\lens::\csrc\lto\src$. We
view the store as divided into its continuous $\csrc$ and discrete parts and localise 
continuous variables to the former. The continuous part must have sufficient topological structure 
to support derivatives and is thus restricted to certain type constructions like normed vector spaces or the real numbers. However, the 
discrete part may use any type defined in HOL. With this view, we can use 
$\lget_\lens$ and $\lput_\lens$ to lift entities defined on $\csrc$ or project those in $\src$. 
For instance, given any $s\in\src$ and a predicate $G:\src\to\bools$ (like the guards in evolution commands), there is a corresponding 
restriction $\ldowngr{G}{s}{\lens}:\csrc\to\bools$ such that $\ldowngr{G}{s}{\lens}\, \vec{c}
\Leftrightarrow \lsubst{G}{\vec{c}}{\lens}\Leftrightarrow G\, (\lput_\lens\, \vec{c}\, s)$. 
Conversely, for $s\in\src$ and $X$, a set of vectors in $\csrc$, the set 
$\lupgr{X}{s}{\lens} = \Pow\, (\lambda\vec{c}.\ \lput_\lens\, \vec{c}\, s)\, X$ has values in $\src$.

More importantly, we can specify ODEs and flows via time-dependent deterministic 
functions (Section~\ref{subsec:subst}'s substitutions). Given a lens $x :: \csrc \lto \src$ from 
global store $\src$ onto local continuous store $\csrc$ and $s\in\src$, we can 
turn any state-wide function $f: T\to\src\to\src$ into a vector field 
$\ldowngr{f}{s}{\lens}:T\rightarrow \csrc\rightarrow \csrc$ by framing it via 
$\ldowngr{f}{s}{\lens}\, t\, \vec{c} = \lget_\lens \left((f\, t)\, (\lput_\lens \vec{c}\, s)\right)$.

\vspace{1ex}
\begin{example}\looseness=-1
Suppose $\src = \real^2 \times \real^2\times \real^2\times \src'$ and
$p, v, a :: \reals^2 \lto\src$. The variable set lens 
$A=(p \lplus v\lplus a) :: \real^2\times \real^2\times \real^2 \lto \src$ frames the continuous
part of the state space $\src$. The substitution $f:T\to\src\to\src$ such that 
$f\, t=[p\!\leadsto\!v, v\!\leadsto\!a,a\!\leadsto\!0]$ then behaves as the identity function 
on $\src'$ and becomes the vector field 
$\ldowngr{f}{s}{A}:T\rightarrow \real^2\times \real^2\times \real^2\rightarrow 
\real^2\times \real^2\times \real^2$. Hence, $f$ naturally describes 
the ODEs $p'\, t=v\, t, v'\, t=a\, t, a'\, t=0$ after 
framing. %Though ostensibly syntactic objects, these substitutions are semantic functions and can be used with Isabelle's ODE components~\cite{Immler12,HolzlIH13}. 
\qed
\end{example}
\vspace{1ex}

Using the previously described liftings and projections, we formally define the semantics
of evolution commands. For this, we only need to lift the definition of generalised guarded
orbits maps (Section~\ref{sec:prelim}) on the continuous $\csrc$ to the larger space $\src$. 
Thus, for substitution $f:T\to\src\to\src$, predicate $G:\src\to\bools$, 
interval function $U:\csrc\to\Pow\, \reals$, and $t_0\in\reals$ the state transformer 
$\src\to\Pow\src$ modelling evolution commands is 
$(x' = f\, \&\, G)_U^{t_0}\, s = \lupgr{(\gamma^{\ldowngr{f}{s}{\lens}}\, (\ldowngr{G}{s}{\lens})\, U\, t_0\, \lsugarget{s}{\lens})}{s}{\lens}$, or equivalently\hfill\isalink{https://github.com/isabelle-utp/Hybrid-Verification/blob/7baedd092dff0182336ef0bb6251fd8beff6a1cc/HS_Lens_ODEs.thy\#L30}
\begin{equation*}
(x' = f\, \&\, G)_U^{t_0}\, s = \left\{\lput_\lens\, (X\, t)\, s \,\middle|\,
        \begin{array}{l} t \in U\, \lsugarget{s}{\lens}
	\land X\in\ivpsols\, U\, (\ldowngr{f}{s}{\lens})\, t_0\, \,\lsugarget{s}{\lens} \\
	\land \Pow\, X\, (\downcl{U\, \lsugarget{s}{\lens}}{t})\subseteq \ldowngr{G}{s}{\lens}
        \end{array}\right\},
%	\land (\forall\tau\in\ivl{t_0}{t}.\ G\, \lsubst{s}{X\, t}{\lens})\},
%	\land (\forall\tau\in\downcl{U\, \lsugarget{s}{\lens}}{t}.\ G\, \lsubst{s}{X\, t}{\lens})\},
\end{equation*}
where we abbreviate $\lget_\lens\, s$ with $\lsugarget{s}{\lens}$. That is, evolution commands are state transformers that output those states whose discrete part remains unchanged from $s$ but whose continuous part changes according to the ODEs' solutions within $G$. With this,
the law~(\ref{eq:wlp-evol}) formally becomes\hfill\isalink{https://github.com/isabelle-utp/Hybrid-Verification/blob/e766a6b4744f37e176cd289e9e80120b6238ad81/Hybrid_Programs/Evolution_Commands.thy\#L87}
\begin{center}
\begin{tabular}{r c l}
$\fdbox{(x' = f\, \&\, G)_U^{t_0}}{Q}\, s$ & $\Leftrightarrow$ & 
    \begin{tabular}{@{}c@{}}
    $\forall X{\in}\ivpsols\, U\, (\ldowngr{f}{s}{\lens})\, t_0\, \lsugarget{s}{\lens}.\ \forall t{\in} U\, \lsugarget{s}{\lens}.$ \\ 
    $(\forall\tau{\in}\downcl{U\, \lsugarget{s}{\lens}}{t}.\ \lsubst{G}{X\, \tau}{\lens})\Rightarrow \lsubst{Q}{X\, t}{\lens}$.
    \end{tabular}\\
\end{tabular}
\end{center}
%\begin{align*}
%& \fdbox{(x' = f\, \&\, G)_U^{t_0}}{Q}\, s \\ & \quad\Leftrightarrow\  \forall X{\in}\ivpsols\, U\, (\ldowngr{f}{s}{\lens})\, t_0\, \lsugarget{s}{\lens}.\ \forall t{\in} U\, \lsugarget{s}{\lens}.\ (\forall\tau{\in}\downcl{U\, \lsugarget{s}{\lens}}{t}.\ \lsubst{G}{X\, \tau}{\lens})\Rightarrow \lsubst{Q}{X\, t}{\lens}.
%\end{align*}
This says that the postcondition $Q$ holds after an evolution command $(x' = f\, \&\, G)_U^{t_0}$ for $s\in\src$ if every solution $X$ to the IVP corresponding to $(t_0, \lens_s)$ satisfies $Q$ on every time $t$, provided $G$ holds from the beginning of the interval $t_0\in U\, x_s$ until $t$. Thus, VCG follows our description in Section~\ref{sec:prelim}: users must supply flows
and evidence for Lipschitz continuity in order to obtain $\wlp$s. We provide tactics that 
automate these processes in Section~\ref{sec:tactics}. 

We use Isabelle's syntax translations to provide a natural syntax for specifying evolution
commands. Users can write
$\{x_1' = e_1, \cdots, x_n' = e_n ~|~ G \mathop{\,\skey{on}\,} U\,V\, \mathop{\,@\,} t_0\}$
directly into the prover where each $x_i::\view_i\lto\src$ is a summand of the frame 
lens $x=\{x_1, \cdots, x_n\}::\csrc\lto\src$. Users can thus declare the ODEs in evolution 
commands coordinate-wise with lifted 
expressions $e_i:\src\to\view_i$. They can also omit the parameters 
$G$, $U$, $V$ and $t_0$ which defaults them to $\top$, $\reals_{\geq 0}$, $\csrc$ 
and $0$, respectively. If desired, they can also use
product syntax $(x_1', \cdots, x_n') = (e_1, \cdots, e_n)$ or vector syntax 
$x' = e$, and specify evolution commands using flows instead of 
ODEs with the notation 
$\{\texttt{EVOL}\, x = e\, \tau \,|\, G\}$.\hfill\isalink{https://github.com/isabelle-utp/Hybrid-Verification/blob/7baedd092dff0182336ef0bb6251fd8beff6a1cc/HS_Lens_Spartan.thy\#L533}

With these, non-modification of variables naturally extends to ODEs with the law

\vspace{1ex}
\begin{center}
  \AxiomC{$x \unrest A$}
  \UnaryInfC{$\{ x' = e \, |\, G \} \nmods A$}
  \DisplayProof
\end{center}
\vspace{1ex}
\noindent Specifically, any set of variables ($A$) without assigned derivatives in a system of ODEs is immutable. Then,
by application of the frame rule, we can demonstrate that any assertion $I$ that uses only variables outside of $x$ is
an invariant of the system of ODEs.

\vspace{1ex}
\begin{example}
  We use the autonomous boat from Example~\ref{ex:boat} to illustrate the use of non-modification. A system of ODEs for the boat's state $p,v,a$ may be specified as follows:

\vspace{1ex}
\begin{alltt}
\isakwmaj{abbreviation} "ODE \(\equiv\) \{ p` = v, v` = a, a` = 0, \(\phi\)` = \(\omega\), 
                       s` = if s \(\neq\) 0 then (v \(\cdot\) a) / s else \(\lVert\)a\(\rVert\) 
                     | s *\(\sb{R}\) [[sin(\(\phi\)), cos(\(\phi\))]] = v \(\land\) 0 \(\le\) s \(\land\) s \(\le\) S \}"
\end{alltt}
\vspace{1ex}
\noindent We also write derivatives for $\phi$ and $s$. The derivative of the former is the angular velocity $\omega$, which has the value
$\mathit{arcos}((v + a)\cdot v / (\lVert v + a \rVert \cdot \lVert v \rVert))$ when
$\lVert v \rVert \neq 0$ and $0$ otherwise~\cite{FosterGC20}. The linear acceleration ($s'$) is calculated using the inner product of $v$
and $a$. If the current speed is $0$, then $s'$ is $\lVert a \rVert$.
Immediately after the derivatives, we also specify the guard or boundary condition that serves to constrain the relationship between the velocity vector and
the heading $\phi$. The guard states that the velocity vector $v$ is equal to $s$ multiplied with the heading unit-vector using scalar multiplication (\texttt{*\(\sb{R}\)})
and our vector syntax. We also require that $0 \le s \le S$, i.e. that the linear speed is between $0$ and the maximum speed.

All other variables in the store remain outside the evolution frame and do not need to be specified. In particular, notice that the $\textit{ODE}$ above does not mention the requested speed variable $\textit{rs}$. This is a discrete variable
  that is unchanged during evolution. Therefore, we can show: $\textit{ODE} \nmods \textit{rs}$. Moreover, using the frame rule we can also demonstrate that $\textit{rs} > 0$ is an
  invariant, i.e. $\hoaretriple{\textit{rs} > 0}{\textit{ODE}}{\textit{rs} > 0}$~\cite{FosterMGS21}. \qed
\end{example}

%%%%%%%%% FRAMED FRECHET DERIVS %%%%%%%%%%
\subsection{Frames and invariants for ODEs}\label{subsec:dinv}
As discussed in Section~\ref{sec:prelim}, an alternative to using flows 
for verification of evolution commands is finding and certifying 
invariants for them. Mathematically, evolution commands' 
invariants coincide with \emph{invariant sets} for dynamical systems or \dL's
\emph{differential invariants}~\cite{MuniveS22}. We abbreviate the statement ``$I$ is an invariant for the evolution command $(x' = f\, \&\, G)_U^{t_0}$'' with the notation $\diffinv\, x\, f\, G\, U\, t_0\, I$. In terms of Hoare logic, invariants for evolution commands satisfy\hfill\isalink{https://github.com/isabelle-utp/Hybrid-Verification/blob/fa0d8409cffdc04b0b9769762124d360c0bdd07c/Hybrid_Programs/Evolution_Commands.thy\#L389}
\begin{equation*}
\diffinv\, x\, f\, G\, U\, t_0\, I
\Leftrightarrow \hoare{I}{(x' = f\, \&\, G)_U^{t_0}}{I}.
\end{equation*}
Informally, $\diffinv\, x\, f\, G\, U\, t_0\, I$ asserts that all states in the generalised guarded orbit $(x' = f\, \&\, G)_U^{t_0}\, s$ of $s\in\src$ such that $I\, s$, will also satisfy $I$. In dynamical systems parlance, the orbits of the system of ODEs within the region characterised by $I$ remain within $I$. %Formally, if $I\, s$, then $(\gamma^{\ldowngr{f}{s}{\lens}}\, (\ldowngr{G}{s}{\lens})\, U\, t_0\, \lsugarget{s}{\lens})\subseteq I$.

A common approach in hybrid program verification for certifying invariants for evolution commands is \emph{differential induction}~\cite{Platzer10}. It establishes sufficient conditions for guaranteeing that simple predicates, such as (in)equalities, are invariants. From these, more complex predicates like conjunctions or disjunctions of these (in)equalities can be shown to be invariants using the rules~(\ref{eq:h-conji}) and~(\ref{eq:h-disji}).\vspace{1ex}

\begin{example}\label{ex:diffinduct} To prove that the conjunction $x>c\land y\geq x$ is an invariant of the pair of ODEs
$x'=1, y'=2$ with $c\in\reals$ (a constant) we need to show that
\begin{equation*}
\hoaretriple{x>c\land y\geq x}{(x'=1, y'=2)}{x>c\land y\geq x}.
\end{equation*}

An application of the rule~(\ref{eq:h-conji}) yields the two proof obligations
\begin{align*}
\hoaretriple{x>c}{(x'=1, y'=2)}{x>c},\text{ and }
\hoaretriple{y\geq x}{(x'=1, y'=2)}{y\geq x}.
\end{align*}

\noindent We conclude the proof informally to provide an intuition for how to proceed: 

\noindent Since the derivative of $x$ is greater than $0$, its magnitude is increasing. Hence, for all time $t\geq0$, the value of $x\, t$ is greater or equal to its original value $x\, 0>c$. This means that the ``values'' of $x$ remain above $c$. Similarly, since the derivative of $y$ is greater than that of $x$ and positive, $y$ ``grows'' faster than $x$. Hence, the value of $y$ remains greater or equal to that of $x$. Thus, the predicate $x>c\land y\geq d$ is an invariant of $x'=1, y'=2$.\qed
\end{example}\vspace{1ex}

Formally, if a predicate $I$ is in negation normal form (NNF) in a first-order language for the real numbers 
$\mathcal{L}_\reals\langle 0,1,+,-,\cdot,<,\leq\rangle$, to show that it is an invariant, we can apply the rules~(\ref{eq:h-conji}) 
and~(\ref{eq:h-disji}) until the only remaining proof obligations are 
Hoare triples of literals. The negated literals can also be converted 
into positive ones via the equivalences $\neg (x<y)\iff y\leq x$, 
$\neg (x\leq y)\iff y<x$, and $\neg (x=y)\iff(y<x\lor x<y)$. The remaining proof obligations 
can be discharged by analysing the derivatives of the magnitudes represented in them as done in Example~\ref{ex:diffinduct}. In the sequel, we present the theory to do this analysis formally in our setting.

The discussion in Example~\ref{ex:diffinduct} compares derivatives of expressions depending on the ODEs' variables. In our semantic approach, the system of ODEs is modelled by a function $f\, t:\src\to\src$ that becomes a vector field ($\csrc\to\csrc$) after framing via some $x::\csrc\lto\src$, where $\csrc$ is the continuous part of $\src$. Similarly, our ``expressions'' are really functions $e:\src\to\mathcal{U}$ (see Section~\ref{subsec:stores}), and we can assume that $\mathcal{U}$ is a continuous state space to get ``continuous expressions''. We frame these functions to the continuous part $\csrc$ of $\src$ to obtain our \emph{framed expressions}
$\ldowngr{e}{\lens}{s}:\csrc\to\mathcal{U}$ such that $\ldowngr{e}{\lens}{s}
= e\circ(\lambda\vec{c}.\ \lput_\lens\, \vec{c}\, s)$. We wish to ``differentiate''
these expressions as informally done in Example~\ref{ex:diffinduct}.
Hence, for a general and formal treatment of our semantic entities, we use the Fr\'echet derivative of these framed expressions. More specifically, recall that, if a function $F:\csrc\to\mathcal{U}$ between normed spaces $\csrc,\mathcal{U}$ is Fr\'echet differentiable at $\vec{c}$, the Fr\'echet derivative of $F$ at $\vec{c}$ is the 
bounded linear operator $D\, F\, \vec{c}: \csrc\to\mathcal{U}$ that attests this. In the finite-dimensional case, e.g. 
$F:\reals^n\to\reals^m$ with $m, n\in\nats$, the Fr\'echet derivative 
$D\, F\, \vec{c}$ is the Jacobian. It is well-known that if $\vec{e}_i$ 
is the $i$-th unit 
vector of the canonical ordered base, the function 
$\lambda \vec{x}.\ (D\, F\, \vec{x})\, \vec{e}_i$ provides the $i$th partial 
derivatives of $F$ while the directional derivative of $F$ in the direction 
of $\vec{c}$ is $\lambda \vec{x}.\ (D\, F\, \vec{x})\, \vec{c}$. With these ideas in mind, we define our \emph{framed Fr\'echet derivatives} $\fderiv{f}{e}{x}:\src\to\mathcal{U}$ of expression $e:\src\to\mathcal{U}$ in the direction of 
$f:\src\to\src$ with respect to $x::\csrc\lto\src$ as~\cite{FosterMGS21} \hfill\isalink{https://github.com/isabelle-utp/Hybrid-Verification/blob/7aa6a58e057586085f9eb6d646d523c6b7c85cb5/Framed_Derivatives.thy\#L139}
\begin{equation*}
(\fderiv{f}{e}{\lens})\, s =  \left(D~\ldowngr{e}{\lens}{s}\, (\lget_\lens\, s)\right)\, (\lget_\lens\, (f\, s)).
\end{equation*}
\noindent That is, they are the directional 
derivatives of framed expressions
$\ldowngr{e}{\lens}{s}$ in the direction of the
projection of $f$ onto the continuous space $\csrc$. These framed Fr\'echet derivatives capture the intuitive analysis performed in Example~\ref{ex:diffinduct}. In fact, the following rules are sound:\vspace{1ex}
\begin{center}
\begin{tabular}{l r c l r}
(\customtag{dinv-eq}{eq:dinv-eq})
& $(G\Rightarrow\fderiv{f}{e_1}{x}=\fderiv{f}{e_2}{x})$
& $\Rightarrow$         & $\diffinv\, x\, f\, G\, \reals_{\geq 0}\, 0\, (e_1 = e_2)$                         
&\isalink{https://github.com/isabelle-utp/Hybrid-Verification/blob/fa0d8409cffdc04b0b9769762124d360c0bdd07c/Framed_Derivatives.thy\#L410}\\
(\customtag{dinv-leq}{eq:dinv-leq})
& $(G\Rightarrow\fderiv{f}{e_1}{x}\leq\fderiv{f}{e_2}{x})$
& $\Rightarrow$         & $\diffinv\, x\, f\, G\, \reals_{\geq 0}\, 0\, (e_1 \leq e_2)$
&\isalink{https://github.com/isabelle-utp/Hybrid-Verification/blob/fa0d8409cffdc04b0b9769762124d360c0bdd07c/Framed_Derivatives.thy\#L424}\\
(\customtag{dinv-less}{eq:dinv-less})
& $(G\Rightarrow\fderiv{f}{e_1}{x}\leq\fderiv{f}{e_2}{x})$
& $\Rightarrow$         & $\diffinv\, x\, f\, G\, \reals_{\geq 0}\, 0\, (e_1 <e_2)$                         
&\isalink{https://github.com/isabelle-utp/Hybrid-Verification/blob/fa0d8409cffdc04b0b9769762124d360c0bdd07c/Framed_Derivatives.thy\#L438}\\
\end{tabular}
\end{center}\vspace{1ex}

%\begin{align*}
%\isalink{https://github.com/isabelle-utp/Hybrid-Verification/blob/fa0d8409cffdc04b0b9769762124d360c0bdd07c/Framed_Derivatives.thy\#L410}
%\qquad(G\Rightarrow\fderiv{f}{e_1}{x}=\fderiv{f}{e_2}{x}) 
%&\Rightarrow \diffinv\, x\, f\, G\, \reals_{\geq 0}\, 0\, (e_1 = e_2),\label{eq:dinv-eq}\tag{dinv-eq}\\
%\isalink{https://github.com/isabelle-utp/Hybrid-Verification/blob/fa0d8409cffdc04b0b9769762124d360c0bdd07c/Framed_Derivatives.thy\#L424}
%\qquad(G\Rightarrow\fderiv{f}{e_1}{x}\leq\fderiv{f}{e_2}{x}) 
%&\Rightarrow \diffinv\, x\, f\, G\, \reals_{\geq 0}\, 0\, (e_1 \leq e_2),\label{eq:dinv-leq}\tag{dinv-leq}\\
%\isalink{https://github.com/isabelle-utp/Hybrid-Verification/blob/fa0d8409cffdc04b0b9769762124d360c0bdd07c/Framed_Derivatives.thy\#L438}
%\qquad(G\Rightarrow\fderiv{f}{e_1}{x}\leq\fderiv{f}{e_2}{x}) 
%&\Rightarrow \diffinv\, x\, f\, G\, \reals_{\geq 0}\, 0\, (e_1 <e_2).\label{eq:dinv-less}\tag{dinv-less}
%\end{align*}
The rule (\ref{eq:dinv-eq}) asserts that showing that an equality is an invariant reduces to showing that both sides of the equality change at the same rate over time. Similarly, the rules (\ref{eq:dinv-leq}) and (\ref{eq:dinv-less}) state that showing that inequalities are invariants requires showing that the rates of change on both sides preserve or augment the initial difference. 

From the users' perspective, $\fderiv{f}{e}{\lens}$ operates as the 
derivative of expression $e$ with respect to the variables $x$ according 
to the system of ODEs $f$. Well-known laws hold.\hfill\isalink{https://github.com/isabelle-utp/Hybrid-Verification/blob/7aa6a58e057586085f9eb6d646d523c6b7c85cb5/Framed_Derivatives.thy\#L155}\vspace{-2ex}

\begin{align}
    \fderiv{f}{k}{\lens} &= 0 & \text{if } x \unrest k, \label{eq:const} \\
    \fderiv{f}{\lens}{y} &= 0 &\text{if } x \lindep y, \label{eq:disc-var} \\
    \fderiv{f}{\lens}{X} &= \substlk{f}~x & \text{if } x\in X \text { and } \lget_{x \lquot X} \text{ is a bounded linear operator}, \label{eq:cont-var} \\
    \fderiv{f}{(e_1 + e_2)}{\lens} & \omit\rlap{$\hspace{.65ex}= (\fderiv{f}{e_1}{\lens}) + (\fderiv{f}{e_2}{\lens}),$}\label{eq:plus} \\
    \fderiv{f}{(e_1 \cdot e_2)}{\lens} &\omit\rlap{$\hspace{.65ex}= (e_1 \cdot \fderiv{f}{e_2}{\lens}) + (\fderiv{f}{e_1}{\lens} \cdot e_2),$}\\
    \fderiv{f}{e^n}{\lens} &\omit\rlap{$\hspace{.65ex}=n \cdot (\fderiv{f}{e}{\lens}) \cdot e^{(n-1)},$} \label{eq:power} \\
    \fderiv{f}{\ln(e)}{\lens} &= (\fderiv{f}{e}{\lens}) / e & \text{if } e > 0. \label{eq:ln}
\end{align}

In summary, users can read~(\ref{eq:const}) and~(\ref{eq:disc-var}) as 
 stating that the derivative of constants or variables
 outside the frame of differentiation are $0$. Law~(\ref{eq:cont-var}) says
 that the derivative of a variable inside the frame is dictated by the ODE, 
 and thus, users simply need to substitute according to $f$. The remaining 
 laws are well-known differentiation properties such as linearity of 
 derivatives or the Leibniz rule.\vspace{1ex}

 \begin{example} First, consider the various Fr\'echet derivatives of the expression $z^2$. Using the ODE $z'=1$, the resulting expression is 
 \begin{equation*}
 \fderiv{[z \smapsto 1]}{z^2}{z} = 2\cdot(\fderiv{[z \smapsto 1]}{z}{z})\cdot z
 = 2\cdot 1\cdot z=2z.
 \end{equation*}
 However, differentiating with respect to a different
 variable yields 
 \begin{equation*}
 \fderiv{[z \smapsto 1]}{z^2}{y} 
 = 2\cdot(\fderiv{[z \smapsto 1]}{z}{y})\cdot 
z= 2\cdot 0\cdot z=0,
\end{equation*}
assuming $y\lindep z$. Finally, the ODE changes
the final result 
\begin{equation*}
\fderiv{[z \smapsto 2]}{z^2}{z} = 
2\cdot(\fderiv{[z \smapsto 2]}{z}{z})\cdot z= 2\cdot 2\cdot z=4z. 
\end{equation*}
Therefore, certifying invariants for evolution commands reduces to computing framed Fr\'echet derivatives and comparing the results which
are often easily computable by rewriting. For instance, to certify that 
$z > 0$ is an invariant of $z' = z^2$, it suffices to check 
\begin{equation*}
0 =\fderiv{[z\leadsto z^2]}{0}{z}\leq \fderiv{[z\leadsto z^2]}{z}{z}
=z^2
\end{equation*} 
by rule~(\ref{eq:dinv-less}). We can now formally culminate the proof in Example~\ref{ex:diffinduct}. By rules~(\ref{eq:dinv-less}) and~(\ref{eq:dinv-leq}) respectively, the assertions $x>c$ and $y\geq x$ are invariants since
\begin{equation*}
1 =\fderiv{[x\leadsto 1,y\leadsto 2]}{x}{x\lplus y}\geq \fderiv{[x\leadsto 1,y\leadsto 2]}{c}{x\lplus y}
=0
\text{ and }
2 =\fderiv{[x\leadsto 1,y\leadsto 2]}{y}{x\lplus y}\geq \fderiv{[x\leadsto 1,y\leadsto 2]}{x}{x\lplus y}
=1.
\end{equation*}
Thus, this example and Example~\ref{ex:diffinduct} illustrate how to certify invariants through \dL's differential induction method by applying our semantic
framed (Fr\'echet) differentiation.\qed
\end{example}

%%%%%%%%%%%%% DIFF GHOSTS %%%%%%%%%%%%%%
\subsection{Ghosts and Darboux rules}\label{subsec:darboux}
Differential induction does not suffice to prove all 
invariant certifications~\cite{Platzer12, PlatzerT18}. For instance, 
applying rule~(\ref{eq:dinv-less}) to the dynamics $x' =-x$ of 
Examples~\ref{ex:odes}-\ref{ex:fddia} to show invariance
of $x>0$ does not lead to a concluding proof state.
Indeed, $\fderiv{[x\leadsto -x]}{x}{x}=-x$ is not necessarily greater or
equal to $0=\fderiv{[x\leadsto -x]}{0}{x}$. For those cases, differential
dynamic logic \dL includes the \emph{differential ghost}~\cite{PlatzerT18}
rule which asserts the correctness of an evolution command given the 
correctness of a higher-dimensional but equivalent system of ODEs. Here
we generalise our previous formalisation of this rule~\cite{FosterMGS21}
and use it to derive \dL's three Darboux rules~\cite{PlatzerT18}. 
Concretely, we formalise and prove soundness of the rules:

%\begin{center}
%\begin{tabular}{l c r}
%(\customtag{dG}{eq:dG})
%& $\frac{\hoare{P}{\{x' = f, y' = A \cdot y + b\ \&\ G\}}{Q}
%    }{\hoare{\exists v.\, \lsubst{P}{v}{y}}{\{x' = f\ \&\ G\}}{\exists v.\, \lsubst{Q}{v}{y}}}$                         
%&\isalink{https://github.com/isabelle-utp/Hybrid-Verification/blob/2f79ded6abbc8efc9101b226a1577651ba29dd90/Hybrid\_Programs/Diff\_Dyn\_Logic.thy\#L212}\\
%(\customtag{dbx-eq}{eq:dbx-eq})
%& $\frac{\fderiv{f(y\leadsto -c\cdot y)}{e}{\lens+_L y}=c\cdot e
%    }{\hoare{e= 0}{\{x' = f\ \&\ G\}}{e= 0}}$                         
%&\isalink{https://github.com/isabelle-utp/Hybrid-Verification/blob/2f79ded6abbc8efc9101b226a1577651ba29dd90/Hybrid\_Programs/Diff\_Dyn\_Logic.thy\#L625}\\
%(\customtag{dbx-ge}{eq:dbx-ge})
%& $\frac{\fderiv{f(y\leadsto -c\cdot y)}{e}{\lens+_L y}\geq c\cdot e
%    }{\hoare{e\mathrel{\rotatebox[origin=c]{180}{$\propto$}} 0}{\{x' = f\ \&\ G\}}{e\mathrel{\rotatebox[origin=c]{180}{$\propto$}} 0}}$                         
%&\isalink{https://github.com/isabelle-utp/Hybrid-Verification/blob/2f79ded6abbc8efc9101b226a1577651ba29dd90/Hybrid\_Programs/Diff\_Dyn\_Logic.thy\#L521}\\
%\end{tabular}
%\end{center}

\leqnomode
\begin{minipage}{.85\linewidth}
\begin{equation}
\frac{\hoare{P}{\{x' = f, y' = A \cdot y + b\ \&\ G\}}{Q}
    }{\hoare{\exists v.\, \lsubst{P}{v}{y}}{\{x' = f\ \&\ G\}}{\exists v.\, \lsubst{Q}{v}{y}}},
\hspace{1.5em}\isalink{https://github.com/isabelle-utp/Hybrid-Verification/blob/2f79ded6abbc8efc9101b226a1577651ba29dd90/Hybrid\_Programs/Diff\_Dyn\_Logic.thy\#L212}
\label{eq:dG}\tag{dG}
\end{equation}
\end{minipage}\\

\begin{minipage}{.85\linewidth}
\begin{equation}
\frac{\fderiv{f(y\leadsto -c\cdot y)}{e}{\lens+_L y}=c\cdot e
    }{\hoare{e= 0}{\{x' = f\ \&\ G\}}{e= 0}},
\hspace{2.65em}\isalink{https://github.com/isabelle-utp/Hybrid-Verification/blob/2f79ded6abbc8efc9101b226a1577651ba29dd90/Hybrid\_Programs/Diff\_Dyn\_Logic.thy\#L625}
\label{eq:dbx-eq}\tag{dbx-eq}
\end{equation}
\end{minipage}\\

\begin{minipage}{.85\linewidth}
\begin{equation}
\frac{\fderiv{f(y\leadsto -c\cdot y)}{e}{\lens+_L y}\geq c\cdot e
    }{\hoare{e\mathrel{\rotatebox[origin=c]{180}{$\propto$}} 0}{\{x' = f\ \&\ G\}}{e\mathrel{\rotatebox[origin=c]{180}{$\propto$}} 0}},
\hspace{2.8em}\isalink{https://github.com/isabelle-utp/Hybrid-Verification/blob/2f79ded6abbc8efc9101b226a1577651ba29dd90/Hybrid\_Programs/Diff\_Dyn\_Logic.thy\#L521}
\label{eq:dbx-ge}\tag{dbx-ge}
\end{equation}
\end{minipage}\\
\reqnomode

\vspace{1ex}

\noindent where $A$ is a square matrix, $b$ is a vector,
$\mathrel{\rotatebox[origin=c]{180}{$\propto$}}\in\{>,\geq\}$, $x$ and 
$y$ are independent variables $y\!\lindep\!x$, and $y$ does not appear
in $G$ or $f$: $y \unrest (G, f)$. In contrast with their usual presentation, 
our semantic formalisation of the Darboux rules also requires the 
existence of a third independent variable $z$ such that $z \unrest (G, f)$ 
because our proof requires two applications of the~(\ref{eq:dG}) 
rule. More work is needed to provide an alternative proof
without these conditions, and to generalise $A$ and $b$
in~(\ref{eq:dG}) to be functions on $x$ that do not mention $y$. The
use of matrices in the~(\ref{eq:dG}) rule was possible due to 
our previous work on linear systems~\cite{FosterGC20,Munive20} but 
further generalisations are possible in terms of bounded linear operators.

\vspace{1ex}
% example ghosts 
\begin{example}\label{ex:ghost}
Below, we provide an alternative (\ref{eq:dbx-ge})-based Hoare-style 
proof that $I\Leftrightarrow (g\geq 0)$ is an invariant for 
$\texttt{blood\_sugar}= \LOOP{(\texttt{ctrl}\seqcomp\texttt{dyn})}$
from Examples~\ref{ex:hp}-\ref{ex:fddia}.
\begin{align*}
&\hoare{I}{\INV{\LOOP{(\texttt{ctrl}\seqcomp\texttt{dyn})}}{I}}{I} \\
& \Leftarrow \hoare{I}{\texttt{ctrl}\seqcomp\texttt{dyn}}{I}\\
& \Leftarrow \hoare{I}{g:=g_M}{I}\land \hoare{I}{g'=-g}{I}\\
& \Leftarrow (g\geq 0)\preceq(g_M\geq 0)\land\hoare{g_M\geq 0}{g:=g_M}{g\geq 0}\land \fderiv{[g\leadsto -g,y\leadsto - 1\cdot y]}{g}{g+_L y}\geq -1\cdot g\\
& \Leftarrow \top\land\top\land -g\geq - g \Leftarrow \top.
\end{align*}
The second implication follows by~(\ref{eq:h-seq}), the third one
by~(\ref{eq:h-cons}) and~(\ref{eq:dbx-ge}), and the last one is true 
by~(\ref{eq:h-assign}), definitions, and the framed derivative rules. This
concludes the proof. As previously noted, differential induction cannot 
certify that $I$ is an invariant of $g' = -g$. Although its invariance could
be verified with the flow as in Example~\ref{ex:vcg}, it is not always 
straightforward to find the solution to a system of ODEs (see 
Sections~\ref{sec:eval} and~\ref{sec:ex}). Hence, in \dL-style reasoning, one sometimes needs to embed the ODEs
into a higher-dimensional space to prove invariance. Despite extant Isabelle’s HOL-based proof strategies to certify this 
example~\cite{Munive21}, our formalisation of the Darboux rules expands our 
pool of methods to tackle similar problems in the style of \dL.\qed
\end{example}

\section{Reasoning Components}\label{sec:tactics}

Here, we describe our recently improved support for proof 
automation in our verification framework. Specifically, we discuss the 
proof methods we have developed using Isabelle's Eisbach 
tool~\cite{MatichukMW16} and the underlying formalisations of
mathematical concepts required to make the automation effective.
That is, our methods not only employ the proof rules 
introduced in Sections~\ref{subsec:vcg} and~\ref{sec:frames}, but
we also add lemmas and formalisations in this section that aid in
making previously described procedures hidden from the user. Thus, 
our methods often discharge any side conditions 
generated in the verification process.

% VCG is automated by application of the simplifier (\skey{simp}), and Isabelle's classical reasoner, as provided in methods like \skey{blast} and \skey{auto}.
As noted in Example~\ref{ex:hp}, verification problems for hybrid 
programs often follow the pattern
%$\INV {\LOOP (\texttt{ctrl} \,\seqcomp\, \{\vec{x}' = e \,|\, G\})}{I}$ 
$\INV {\LOOP (\texttt{ctrl} \,\seqcomp\, \texttt{dyn})}{I}$. That is, 
they are an iteration of a discrete controller
intervening in a continuous dynamical system. %as seen in Example~\ref{ex:hp}. 
We can therefore apply invariant reasoning with the laws~\eqref{eq:h-loopi} and~\eqref{eq:h-whilei}, meaning that often the main task requires verifying
%$\hoare{I}{\texttt{ctrl} \,;\, \{\vec{x}' = e \,|\, G\}}{I}$.
$\hoare{I}{\texttt{ctrl} \,;\, \texttt{dyn}}{I}$. As foretold at the end of
Section~\ref{subsec:hoare}, two workflows can be applied to verify this 
Hoare triple. If the system of ODEs is solvable, then
we can insert the flow using law \eqref{eq:wlp-flow}, and
VCG becomes purely equational using the forward
box laws (Section~\ref{subsec:fdbox}). This approach is implemented in 
our proof method \isa{wlp{\isacharunderscore}full} 
(described below in Section~\ref{subsec:auto-flow-vcg}). 
Alternatively, if a solution is not available, we can find a suitable invariant $I$
and use our differential induction proof method \isa{dInduct}, 
and its variants~(Section~\ref{subsec:auto-dinduct}), to verify 
$\hoare{I}{\texttt{dyn}}{I}$. 
%$\hoare{I}{\{\vec{x}' = e \,|\, G\}}{I}$. 
In the remainder, we describe the proof methods and formalisations that 
support each of these workflows. Specifically, for 
\isa{wlp{\isacharunderscore}simp} we
describe our automation of the certification of differentiation, Lipschitz continuity, uniqueness of solutions, and the integration of computer algebra
systems (CASs) into IsaVODEs. On the \isa{dInduct} side, we provide proof methods 
for automating differential induction, weakening, and differential ghosts.
Finally, we lay the foundations for automating the proofs of obligations that
depend on the analysis of real-valued functions. We do this by formalising 
and proving well-known derivative test theorems ubiquitous in calculus. 
% \sfin{We need some introductory text for this section. Mention the use of Eisbach.}

% Here we describe various tactics that increase VCG automation within our framework. In 
% particular, we describe our automated method for certifying that a given function is a 
% derivative of another, a method for automatically checking that a function
% is Lipschitz continuous, a tactic for automatically certifying that a given solution is the flow 
% of an ODE, and tactics automating 
% $\wlp$-reasoning for hybrid systems. % This provides a complete high-level tactic language for reasoning about CPS model.
%\subsection{Verification Condition Generation}

%%%%%%%%%%%%% DERIVATIVES %%%%%%%%%%%%%%
\subsection{Automatic certification of differentiation}\label{subsec:auto-deriv}
Our tactic \isa{vderiv} for automatically discharging statements of the 
form $f' = g$ for functions $f,g:\reals\to\csrc$ has been described 
before~\cite{MuniveS22} but it is a component of other tactics and we have 
extended its capabilities here. Essentially, by the chain 
rule of differential calculus, derivative laws often require further 
certifications of simpler derivatives. A recursive procedure emerges 
where the certification of $f'\, t= g\, t$ with $f= f_1\circ\cdots\circ f_n$ 
is determined by a list of proven derivative rules 
$(f_i\, (h\, t))'=f_i'(h\, t)\cdot (h'\, t)$ for $i\in\{1,\dots,n\}$ together
with $t'=1$. The tactic \isa{vderiv} recursively applies these rules until 
it determines that 
\begin{equation*}
f'\, t=(f_n'\, t)\cdot\prod_{i=1}^{n-1}f_i'((f_{i+1}\circ\cdots\circ f_n)\, t).
\end{equation*}
Then, it uses Isabelle's support for higher-order unification to try to 
show that
\begin{equation*}
g\, t = (f_n'\, t)\cdot\prod_{i=1}^{n-1}f_i'((f_{i+1}\circ\cdots\circ f_n)\, t).
\end{equation*}
For this work, we have added derivative rules for the real-valued 
exponentiation $\exp(-)$, the square root $\sqrt{-}$, the tangent $\tan(-)$ and cotangent $\cot(-)$ trigonometric functions, and vectors' inner products $*_R$ and norms $\lVert-\lVert$. The tactic \isa{vderiv} is an integral part of those described below, and it is tacitly used in our example of Section~\ref{sec:rotdyn}.

%%%%%%%%%%%%% LIPSCHITZ %%%%%%%%%%%%%%%
\subsection{Automatic certification of Lipschitz continuity}\label{subsec:lipschitz}
As evidenced in Examples~\ref{ex:odes}-\ref{ex:fddia}, verification of 
hybrid systems might depend on knowing that there is a unique solution 
for a system of differential equations. In practice, certifying the existence 
and uniqueness of solutions with a general-purpose prover has required 
finding the Lipschitz-continuity constant~\cite{FosterMGS21,MuniveS22}. If it exists, then by the Picard-Lindel\"of theorem (see Section~\ref{subsec:dynsys}), there is an interval around the initial time where solutions to the system of ODEs are unique.
Alternatively, a domain-specific prover can restrict the specification 
language to a fragment where uniqueness is guaranteed~\cite{Platzer10} 
albeit limiting the space of verifiable dynamics.
Here, we provide the foundation for allowing a general-purpose prover to automatically certify the uniqueness of solutions to IVPs. Namely, we 
formalise the well-known fact that continuously differentiable ($C^1$) 
functions are Lipschitz continuous. The statement of the Isabelle lemma
is shown below.

\begin{isabellebody}
\isanewline
\isacommand{lemma}\isamarkupfalse%
\ c{\isadigit{1}}{\isacharunderscore}{\kern0pt}local{\isacharunderscore}{\kern0pt}lipschitz{\isacharcolon}{\kern0pt}\hfill\isalink{https://github.com/isabelle-utp/Hybrid-Verification/blob/0cd56918e956f0268d2d7ddc43c20df3cde18b89/HS_Preliminaries.thy\#L771} \isanewline
\ \ \isakeyword{fixes}\ f{\isacharcolon}{\kern0pt}{\isacharcolon}{\kern0pt}{\isachardoublequoteopen}real\ {\isasymRightarrow}\ {\isacharparenleft}{\kern0pt}{\isacharprime}{\kern0pt}a{\isacharcolon}{\kern0pt}{\isacharcolon}{\kern0pt}{\isacharbraceleft}{\kern0pt}banach{\isacharcomma}{\kern0pt}perfect{\isacharunderscore}{\kern0pt}space{\isacharbraceright}{\kern0pt}{\isacharparenright}{\kern0pt}\ {\isasymRightarrow}\ {\isacharprime}{\kern0pt}a{\isachardoublequoteclose}\isanewline
\ \ \isakeyword{assumes}\ {\isachardoublequoteopen}open\ S{\isachardoublequoteclose}\ \isakeyword{and}\ {\isachardoublequoteopen}open\ T{\isachardoublequoteclose}\isanewline
\ \ \ \ \isakeyword{and}\ c{\isadigit{1}}hyp{\isacharcolon}{\kern0pt}\ {\isachardoublequoteopen}{\isasymforall}{\isasymtau}\ {\isasymin}\ T{\isachardot}{\kern0pt}\ {\isasymforall}s\ {\isasymin}\ S{\isachardot}{\kern0pt}\ D\ {\isacharparenleft}{\kern0pt}f\ {\isasymtau}{\isacharparenright}{\kern0pt}\ {\isasymmapsto}\ {\isasymD}\ {\isacharparenleft}{\kern0pt}at\ s\ within\ S{\isacharparenright}{\kern0pt}{\isachardoublequoteclose}\isanewline
\ \ \ \ \isakeyword{and}\ {\isachardoublequoteopen}continuous{\isacharunderscore}{\kern0pt}on\ S\ {\isasymD}{\isachardoublequoteclose}\isanewline
\ \ \isakeyword{shows}\ {\isachardoublequoteopen}local{\isacharunderscore}{\kern0pt}lipschitz\ T\ S\ f{\isachardoublequoteclose}\isanewline
\ \ $\langle proof\rangle$\isanewline
\end{isabellebody}

\noindent Then, a procedure emerges when applying~(\ref{eq:wlp-flow}) 
or~(\ref{eq:fdia-flow}):
\begin{enumerate}
    \item Users try to rewrite a correctness specification that includes an evolution command. %via~(\ref{eq:wlp-flow}) or~(\ref{eq:fdia-flow}). 
    That is they try to rewrite $\fdbox{(x' = f\, \&\, G)_U^{t_0}}{Q}\, s$ or 
    $\fddia{(x' = f\, \&\, G)_U^{t_0}}{Q}\, s$.
    \item In order to guarantee that~(\ref{eq:wlp-flow}) 
or~(\ref{eq:fdia-flow}) are applicable, users need to show that there is a flow $\varphi$ for the ODEs $x'=f$. This is formalised in Isabelle/HOL as the proof obligation \isa{local{\isacharunderscore}flow{\isacharunderscore}on\ f\ x\ T\ S\ {\isasymphi}}.
    \item Unfolding this predicate's definition yields the obligation 
\isa{local{\isacharunderscore}lipschitz\ T\ S\ f}
to show that the vector field $f$ is Lipschitz continuous on $T$.
    \item Then, users can apply our lemma 
\isa{c{\isadigit{1}}{\isacharunderscore}{\kern0pt}local{\isacharunderscore}{\kern0pt}lipschitz} above which
requires them to show that $(1)$ $T$ and $S$ are open sets and that 
$(2)$ the derivative of $f\, \tau$ is some function \isa{{\isasymD}}
that $(3)$ is continuous on $S$. Users can either supply \isa{{\isasymD}} or let Isabelle reconstruct it through its support for higher-order unification later in the proof.
    \item Users discharge the remaining proof obligations: openness, continuity, differentiability, and original obligation with $\fdbox{(x' = f\, \&\, G)_U^{t_0}}{Q}\, s$ rewritten by~(\ref{eq:wlp-flow}).
\end{enumerate}\vspace{1ex}

\begin{example} We illustrate the procedure for certifying the uniqueness of solutions in Isabelle below by formalising part of the argument of Example~\ref{ex:vcg}. Recall the definitions of the control and dynamics of the problem.

\begin{isabellebody}\isanewline
\isacommand{abbreviation}\isamarkupfalse%
\ {\isachardoublequoteopen}ctrl\ {\isasymequiv}\ IF\ g\ {\isasymle}\ g\isactrlsub m\ THEN\ g\ {\isacharcolon}{\kern0pt}{\isacharcolon}{\kern0pt}{\isacharequal}{\kern0pt}\ g\isactrlsub M\ ELSE\ skip{\isachardoublequoteclose}\isanewline
\isanewline
\isacommand{abbreviation}\isamarkupfalse%
\ {\isachardoublequoteopen}dyn\ {\isasymequiv}\ {\isacharbraceleft}{\kern0pt}g{\isacharbackquote}{\kern0pt}\ {\isacharequal}{\kern0pt}\ {\isacharminus}{\kern0pt}g{\isacharbraceright}{\kern0pt}{\isachardoublequoteclose}\isanewline
\isanewline
\isacommand{abbreviation}\isamarkupfalse%
\ {\isachardoublequoteopen}flow\ {\isasymtau}\ {\isasymequiv}\ {\isacharbrackleft}{\kern0pt}g\ {\isasymleadsto}\ g\ {\isacharasterisk}{\kern0pt}\ exp\ {\isacharparenleft}{\kern0pt}{\isacharminus}{\kern0pt}\ {\isasymtau}{\isacharparenright}{\kern0pt}{\isacharbrackright}{\kern0pt}{\isachardoublequoteclose}\isanewline
\isanewline
\isacommand{abbreviation}\isamarkupfalse%
\ {\isachardoublequoteopen}blood{\isacharunderscore}{\kern0pt}sugar\ {\isasymequiv}\ LOOP\ {\isacharparenleft}{\kern0pt}ctrl{\isacharsemicolon}{\kern0pt}\ dyn{\isacharparenright}{\kern0pt}\ INV\ {\isacharparenleft}{\kern0pt}g\ {\isasymge}\ {\isadigit{0}}{\isacharparenright}{\kern0pt}{\isachardoublequoteclose}
\isanewline
\end{isabellebody}

The procedure appears in a partial apply-style Isabelle proof below. We label each procedure step with its number and add Isabelle outputs after each \isa{\isacommand{apply}} command.

\begin{isabellebody}\isanewline
\isacommand{lemma}\isamarkupfalse%
\ {\isachardoublequoteopen}\isactrlbold {\isacharbraceleft}{\kern0pt}g\ {\isasymge}\ {\isadigit{0}}\isactrlbold {\isacharbraceright}{\kern0pt}\ blood{\isacharunderscore}{\kern0pt}sugar\ \isactrlbold {\isacharbraceleft}{\kern0pt}g\ {\isasymge}\ {\isadigit{0}}\isactrlbold {\isacharbraceright}{\kern0pt}{\isachardoublequoteclose}\isanewline
% step 0
\ \ \isacommand{apply}\isamarkupfalse%
\ {\isacharparenleft}{\kern0pt}wlp{\isacharunderscore}{\kern0pt}simp{\isacharparenright}{\kern0pt}\ %
\ \ \hspace{7.5em}\isamarkupcmt{the forward box $\fdbox{g'=-g}{0\leq g}$ appears}\isanewline
% step 1
\ \ \isacommand{apply}\isamarkupfalse%
\ {\isacharparenleft}{\kern0pt}subst\ fbox{\isacharunderscore}{\kern0pt}solve{\isacharbrackleft}{\kern0pt}\isakeyword{where}\ {\isasymphi}{\isacharequal}{\kern0pt}flow{\isacharbrackright}{\kern0pt}{\isacharparenright}{\kern0pt}
\ \ \hspace{12.65em}\isamarkupcmt{Step (1)}\isanewline
% step 2
\ \ \texttt{Isabelle's output}: local{\isacharunderscore}{\kern0pt}flow{\isacharunderscore}{\kern0pt}on\ {\isacharbrackleft}{\kern0pt}g\ {\isasymleadsto}\ g{\isacharbrackright}\ g\ $\reals$\ $\reals$\ flow
\ \ \hspace{5.15em}\isamarkupcmt{Step (2)}\isanewline
% step 3
\ \ \ \ \ \isacommand{apply}\isamarkupfalse%
\ {\isacharparenleft}{\kern0pt}{\isacharparenleft}{\kern0pt}clarsimp\ simp{\isacharcolon}{\kern0pt}\ local{\isacharunderscore}{\kern0pt}flow{\isacharunderscore}{\kern0pt}on{\isacharunderscore}{\kern0pt}def{\isacharparenright}{\kern0pt}{\isacharquery}{\kern0pt}{\isacharcomma}{\kern0pt}\ unfold{\isacharunderscore}{\kern0pt}locales{\isacharsemicolon}{\kern0pt}\ clarsimp{\isacharquery}{\kern0pt}{\isacharparenright}{\kern0pt}\isanewline
\ \ \ \ \ \texttt{Isabelle's output}: local{\isacharunderscore}{\kern0pt}lipschitz\ $\reals$\ $\reals$\ ({\isasymlambda}t{\isachardot}{\kern0pt}\ {\isacharbrackleft}g\ {\isasymleadsto}\ g{\isacharbrackright}{\kern0pt}{\isasymdown}\isactrlsub S\isactrlsub u\isactrlsub b\isactrlsub s\isactrlsub t\isactrlbsub g\isactrlesub)\ \ \isamarkupcmt{Step (3)}\isanewline
% step 4
\ \ \ \ \ \ \ \isacommand{apply}\isamarkupfalse%
\ {\isacharparenleft}{\kern0pt}rule\ c{\isadigit{1}}{\isacharunderscore}{\kern0pt}local{\isacharunderscore}{\kern0pt}lipschitz{\isacharparenright}{\kern0pt}
\ \ \hspace{13.9em}\isamarkupcmt{Step (4)}\isanewline
\ \ \ \ \ \ \ \texttt{Isabelle's outputs}: open\ $\reals$,\ open\ $\reals$,\ continuous{\isacharunderscore}on\ $\reals$\ {\isasymD},$\dots$
\isanewline
\ \ $\langle proof\rangle$\hspace{31em}\isamarkupcmt{Step (5)}\isanewline
\end{isabellebody}

\noindent In practice, certifying openness of $\reals$ or 
intervals $(a,b)=\{x\mid a < x < b\}$ with $a\leq b$ is automatic thanks to 
Isabelle's simplifier. Finding derivatives or checking continuity is not always as 
straightforward but simple linear combinations are automatically certifiable. We have bundled the procedure for certifying the uniqueness of solutions to IVPs in an Isabelle tactic \isa{local{\isacharunderscore}flow{\isacharunderscore}on{\isacharunderscore}auto}
described in Section~\ref{subsec:auto-flow} but below we focus on automating the certification of Lipschitz continuity. \qed
%For completeness, we show the same correctness specification below but proved in two lines with our tactics automating the aforementioned procedure.
%
%\begin{isabellebody}\isanewline
%\isacommand{lemma}\isamarkupfalse%
%\ {\isachardoublequoteopen}\isactrlbold {\isacharbraceleft}{\kern0pt}g\ {\isasymge}\ {\isadigit{0}}\isactrlbold {\isacharbraceright}{\kern0pt}\ blood{\isacharunderscore}{\kern0pt}sugar\ \isactrlbold {\isacharbraceleft}{\kern0pt}g\ {\isasymge}\ {\isadigit{0}}\isactrlbold {\isacharbraceright}{\kern0pt}{\isachardoublequoteclose}\isanewline
%\ \ \isacommand{apply}\isamarkupfalse%
%\ {\isacharparenleft}{\kern0pt}wlp{\isacharunderscore}{\kern0pt}expr{\isacharunderscore}{\kern0pt}solve\ {\isachardoublequoteopen}flow{\isachardoublequoteclose}{\isacharparenright}{\kern0pt}\isanewline
%\ \ \isacommand{using}\isamarkupfalse%
%\ ge{\isacharunderscore}{\kern0pt}gm\ \isacommand{by}\isamarkupfalse%
%\ force\isanewline
%\end{isabellebody}
\end{example}\vspace{1ex}

In Example~\ref{ex:odes}, we use the fact that continuously differentiable
($C^1$) functions are Lipschitz continuous to argue that a system of 
ODEs $f$ has unique solutions by the Picard-Lindel\"of theorem. 
Thus, formalisation of the rule \isa{c{\isadigit{1}}{\isacharunderscore}{\kern0pt}local{\isacharunderscore}{\kern0pt}lipschitz} is the first step 
towards automating the certification of the uniqueness of solutions to IVPs 
and a crucial step in hybrid system verification (via flows) in general-purpose proof assistants. Next, we provide tactics
\isa{c{\isadigit{1}}{\isacharunderscore}{\kern0pt}lipschitz}
and \isa{c{\isadigit{1}}{\isacharunderscore}{\kern0pt}lipschitzI}
to make certification of Lipschitz continuity as seamless in VCG as 
in pen and paper proofs, like that of Example~\ref{ex:odes}. 
Our tactics automate an initial application, in a backward 
reasoning style, of our lemma 
\isa{c{\isadigit{1}}{\isacharunderscore}{\kern0pt}local{\isacharunderscore}{\kern0pt}lipschitz}.
The tactics discharge  the emerging proof obligations by replicating the 
behaviour of \isa{vderiv} but using rules for continuity and Fr\'echet 
differentiability available in Isabelle's HOL-Analysis library. 
The difference between both tactics is that 
\isa{c{\isadigit{1}}{\isacharunderscore}{\kern0pt}lipschitzI}
allows users more control by enabling them to supply the derivative 
\isa{\isasymD}. We exemplify \isa{c{\isadigit{1}}{\isacharunderscore}{\kern0pt}lipschitz} in VCG below:

\begin{isabellebody}
\isanewline\isacommand{lemma}\isamarkupfalse%
\ {\isachardoublequoteopen}vwb{\isacharunderscore}{\kern0pt}lens\ x\ {\isasymLongrightarrow}\ vwb{\isacharunderscore}{\kern0pt}lens\ y\ {\isasymLongrightarrow}\ vwb{\isacharunderscore}{\kern0pt}lens\ z\isanewline
\ \ {\isasymLongrightarrow}\ x\ {\isasymbowtie}\ y\ {\isasymLongrightarrow}\ x\ {\isasymbowtie}\ z\ {\isasymLongrightarrow}\ y\ {\isasymbowtie}\ z
\isanewline\ \ {\isasymLongrightarrow}\ local{\isacharunderscore}{\kern0pt}lipschitz\ UNIV\ UNIV\ {\isacharparenleft}{\kern0pt}{\isasymlambda}t{\isacharcolon}{\kern0pt}{\isacharcolon}{\kern0pt}real{\isachardot}{\kern0pt}\ {\isacharbrackleft}{\kern0pt}x\ {\isasymleadsto}\ {\isachardollar}{\kern0pt}y{\isacharcomma}{\kern0pt}\ y\ {\isasymleadsto}\ {\isachardollar}{\kern0pt}z{\isacharbrackright}{\kern0pt}\ {\isasymdown}\isactrlsub S\isactrlsub u\isactrlsub b\isactrlsub s\isactrlsub t\isactrlbsub x\ {\isacharplus}{\kern0pt}\isactrlsub L\ y\isactrlesub \ s{\isacharparenright}{\kern0pt}{\isachardoublequoteclose}\isanewline\ \ \isacommand{by}\isamarkupfalse%
\ c{\isadigit{1}}{\isacharunderscore}{\kern0pt}lipschitz%
\isanewline

\isacommand{lemma}\isamarkupfalse%
\ {\isachardoublequoteopen}vwb{\isacharunderscore}{\kern0pt}lens\ x\ {\isasymLongrightarrow}\ local{\isacharunderscore}{\kern0pt}lipschitz\ UNIV\ UNIV\ {\isacharparenleft}{\kern0pt}{\isasymlambda}t{\isachardot}{\kern0pt}\ {\isacharbrackleft}{\kern0pt}x\ {\isasymleadsto}\ {\isadigit{1}}\ {\isacharminus}{\kern0pt}\ {\isachardollar}{\kern0pt}x{\isacharbrackright}{\kern0pt}\ {\isasymdown}\isactrlsub S\isactrlsub u\isactrlsub b\isactrlsub s\isactrlsub t\isactrlbsub x\isactrlesub \ s{\isacharparenright}{\kern0pt}{\isachardoublequoteclose}\isanewline
\ \ \isacommand{by}\isamarkupfalse%
\ c{\isadigit{1}}{\isacharunderscore}{\kern0pt}lipschitz%
\isanewline

\isacommand{lemma}\isamarkupfalse%
\ {\isachardoublequoteopen}vwb{\isacharunderscore}{\kern0pt}lens\ x\ {\isasymLongrightarrow}\ x\ {\isasymbowtie}\ y\ 
\isanewline\ \ {\isasymLongrightarrow}\ local{\isacharunderscore}{\kern0pt}lipschitz\ UNIV\ UNIV\ {\isacharparenleft}{\kern0pt}{\isasymlambda}t{\isacharcolon}{\kern0pt}{\isacharcolon}{\kern0pt}real{\isachardot}{\kern0pt}\ {\isacharbrackleft}{\kern0pt}x\ {\isasymleadsto}\ {\isacharminus}{\kern0pt}\ {\isacharparenleft}{\kern0pt}{\isachardollar}{\kern0pt}y\ {\isacharasterisk}{\kern0pt}\ {\isachardollar}{\kern0pt}x{\isacharparenright}{\kern0pt}{\isacharbrackright}{\kern0pt}\ {\isasymdown}\isactrlsub S\isactrlsub u\isactrlsub b\isactrlsub s\isactrlsub t\isactrlbsub x\isactrlesub \ s{\isacharparenright}{\kern0pt}{\isachardoublequoteclose}\isanewline\ \ \isacommand{by}\isamarkupfalse%
\ c{\isadigit{1}}{\isacharunderscore}{\kern0pt}lipschitz%
\isanewline
\end{isabellebody}

In applications, we use our tactics on framed substitutions, e.g.
\isa{{\isacharbrackleft}{\kern0pt}x\ {\isasymleadsto}\ {\isadigit{1}}\ {\isacharminus}{\kern0pt}\ {\isachardollar}{\kern0pt}x{\isacharbrackright}{\kern0pt}\ {\isasymdown}\isactrlsub S\isactrlsub u\isactrlsub b\isactrlsub s\isactrlsub t\isactrlbsub x\isactrlesub}, that represent systems of ODEs $f$. 
VCG also requires assumptions of lens-independence and satisfaction of 
lens laws due to our use of shallow expressions and frames. 
These are automatically provided by our \isa{\isacommand{dataspace}} 
Isabelle command and taken into account in our 
\isa{c{\isadigit{1}}{\isacharunderscore}{\kern0pt}lipschitz} tactics. The tactic \isa{c{\isadigit{1}}{\isacharunderscore}{\kern0pt}lipschitz} is indirectly used in the verification problem in Section~\ref{sec:rotdyn}. Both 
tactics automatically discharge proof obligations where the ODEs' (the vector field $f$) form a linear system of ODEs. This already yields polynomial 
and transcendental functions as solutions $\varphi$ to these systems. 
%Even if the ODEs' right-hand-sides do not have a linear combination of variables in the system of ODEs, we can often provide an equivalent evolution command that does. For instance, although \isa{c1{\isacharunderscore}lipschitz} does not automatically certify Lipschitz continuity for $x'=x^2$, it does so for the \jhsb{equivalent}{Double check} evolution command $(x'=y,\ y'=2xy \, \&\, y=x^2)$. 
Accordingly, users can employ 
our tool support for linear systems of ODEs~\cite{Munive20}. We leave the automation of Lipschitz continuity certification for non-linear systems as future work. Alternative
proof strategies like differential induction (see Section~\ref{sec:ex}) are 
also available for cases when these tactics fail.

%%%%%%%%%%%%% UNIQUENESS %%%%%%%%%%%%%%
\subsection{Automatic certification of the flow}\label{subsec:auto-flow}
In Section~\ref{subsec:lipschitz}, we describe a procedure to verify 
the partial correctness of an evolution command by supplying the flow 
to its system of ODEs. Here, we automate the flow certification part of this
procedure in a tactic 
\isa{local{\isacharunderscore}flow{\isacharunderscore}on{\isacharunderscore}auto}.
This proof method calls our two previously described tactics 
\isa{c1{\isacharunderscore}lipschitz} and \isa{vderiv}. The first one 
discharges the Lipschitz continuity requirement from the Picard-Lindel\"of
theorem to guarantee uniqueness of solutions to the associated IVP. The
tactic \isa{vderiv} certifies that the supplied flow $\varphi$ is a solution to 
the system of ODEs. The following lines exemplify our tactic usage in 
hybrid systems verification tasks. Our tactic is robust as it automatically 
discharges frequently occurring proof obligations. Notice that for each 
Isabelle lemma below, there is a corresponding (local) Lipschitz
continuity example from Section~\ref{subsec:lipschitz}.

\begin{isabellebody}
\isanewline
\isacommand{lemma}\isamarkupfalse%
\ {\isachardoublequoteopen}vwb{\isacharunderscore}{\kern0pt}lens\ x\ {\isasymLongrightarrow}\ vwb{\isacharunderscore}{\kern0pt}lens\ y\ {\isasymLongrightarrow}\ vwb{\isacharunderscore}{\kern0pt}lens\ z\isanewline
\ \ {\isasymLongrightarrow}\ x\ {\isasymbowtie}\ y\ {\isasymLongrightarrow}\ x\ {\isasymbowtie}\ z\ {\isasymLongrightarrow}\ y\ {\isasymbowtie}\ z
\isanewline\ \ {\isasymLongrightarrow}\ local{\isacharunderscore}{\kern0pt}flow{\isacharunderscore}{\kern0pt}on\ {\isacharbrackleft}{\kern0pt}x\ {\isasymleadsto}\ {\isachardollar}{\kern0pt}y{\isacharcomma}{\kern0pt}\ y\ {\isasymleadsto}\ {\isachardollar}{\kern0pt}z{\isacharbrackright}{\kern0pt}\ {\isacharparenleft}{\kern0pt}x\ {\isacharplus}{\kern0pt}\isactrlsub L\ y{\isacharparenright}{\kern0pt}\ UNIV\ UNIV\ 
\isanewline\ \ {\isacharparenleft}{\kern0pt}{\isasymlambda}t{\isachardot}{\kern0pt}\ {\isacharbrackleft}{\kern0pt}x\ {\isasymleadsto}\ {\isachardollar}{\kern0pt}z\ {\isacharasterisk}{\kern0pt}\ t\isactrlsup {\isadigit{2}}\ {\isacharslash}{\kern0pt}\ {\isadigit{2}}\ {\isacharplus}{\kern0pt}\ {\isachardollar}{\kern0pt}y\ {\isacharasterisk}{\kern0pt}\ t\ {\isacharplus}{\kern0pt}\ {\isachardollar}{\kern0pt}x{\isacharcomma}{\kern0pt}\ y\ {\isasymleadsto}\ {\isachardollar}{\kern0pt}z\ {\isacharasterisk}{\kern0pt}\ t\ {\isacharplus}{\kern0pt}\ {\isachardollar}{\kern0pt}y{\isacharbrackright}{\kern0pt}{\isacharparenright}{\kern0pt}{\isachardoublequoteclose}
\isanewline\ \ \isacommand{by}\isamarkupfalse%
\ local{\isacharunderscore}{\kern0pt}flow{\isacharunderscore}{\kern0pt}on{\isacharunderscore}{\kern0pt}auto
\isanewline

\isacommand{lemma}\isamarkupfalse%
\ {\isachardoublequoteopen}vwb{\isacharunderscore}{\kern0pt}lens\ x\ {\isasymLongrightarrow}\ local{\isacharunderscore}{\kern0pt}flow{\isacharunderscore}{\kern0pt}on\ {\isacharbrackleft}{\kern0pt}x\ {\isasymleadsto}\ {\isacharminus}{\kern0pt}\ {\isachardollar}{\kern0pt}x\ {\isacharplus}{\kern0pt}\ {\isadigit{1}}{\isacharbrackright}{\kern0pt}\ x\ UNIV\ UNIV
\isanewline\ \ {\isacharparenleft}{\kern0pt}{\isasymlambda}t{\isachardot}{\kern0pt}\ {\isacharbrackleft}{\kern0pt}x\ {\isasymleadsto}\ {\isadigit{1}}\ {\isacharminus}{\kern0pt}\ exp\ {\isacharparenleft}{\kern0pt}{\isacharminus}{\kern0pt}\ t{\isacharparenright}{\kern0pt}\ {\isacharplus}{\kern0pt}\ {\isachardollar}{\kern0pt}x\ {\isacharasterisk}{\kern0pt}\ exp\ {\isacharparenleft}{\kern0pt}{\isacharminus}{\kern0pt}\ t{\isacharparenright}{\kern0pt}{\isacharbrackright}{\kern0pt}{\isacharparenright}{\kern0pt}{\isachardoublequoteclose}
\isanewline\ \ \isacommand{by}\isamarkupfalse%
\ local{\isacharunderscore}{\kern0pt}flow{\isacharunderscore}{\kern0pt}on{\isacharunderscore}{\kern0pt}auto
\isanewline

\isacommand{lemma}\isamarkupfalse%
\ {\isachardoublequoteopen}vwb{\isacharunderscore}{\kern0pt}lens\ {\isacharparenleft}{\kern0pt}x{\isacharcolon}{\kern0pt}{\isacharcolon}{\kern0pt}real\ {\isasymLongrightarrow}\ {\isacharprime}{\kern0pt}s{\isacharparenright}{\kern0pt}\ {\isasymLongrightarrow}\ x\ {\isasymbowtie}\ y\ 
\isanewline\ \ {\isasymLongrightarrow}\ local{\isacharunderscore}{\kern0pt}flow{\isacharunderscore}{\kern0pt}on\ {\isacharbrackleft}{\kern0pt}x\ {\isasymleadsto}\ {\isacharminus}{\kern0pt}\ {\isachardollar}{\kern0pt}y\ {\isacharasterisk}{\kern0pt}\ {\isachardollar}{\kern0pt}x{\isacharbrackright}{\kern0pt}\ x\ UNIV\ UNIV
\isanewline\ \ {\isacharparenleft}{\kern0pt}{\isasymlambda}t{\isachardot}{\kern0pt}\ {\isacharbrackleft}{\kern0pt}x\ {\isasymleadsto}\ {\isachardollar}{\kern0pt}x\ {\isacharasterisk}{\kern0pt}\ exp\ {\isacharparenleft}{\kern0pt}{\isacharminus}{\kern0pt}\ t\ {\isacharasterisk}{\kern0pt}\ {\isachardollar}{\kern0pt}y{\isacharparenright}{\kern0pt}{\isacharbrackright}{\kern0pt}{\isacharparenright}{\kern0pt}{\isachardoublequoteclose}
\isanewline\ \ \isacommand{by}\isamarkupfalse%
\ local{\isacharunderscore}{\kern0pt}flow{\isacharunderscore}{\kern0pt}on{\isacharunderscore}{\kern0pt}auto
\isanewline
\end{isabellebody}

Crucially, \isa{local{\isacharunderscore}flow{\isacharunderscore}on{\isacharunderscore}auto}
certifies the exponential and trigonometric solutions 
required for these ODEs as evidenced by its successful 
application on the examples above and in the verification problem of Section~\ref{sec:rotdyn}. 

%%%%%%%%%%%%%%%% VCG %%%%%%%%%%%%%%%%
\subsection{Automatic VCG with flows}\label{subsec:auto-flow-vcg}
In addition to our derivative, Lipschitz continuity, and flow automatic
certifications, we have added various tactics for verification condition 
generation. The simplest one is \isa{wlp{\isacharunderscore}simp}. It 
merely calls Isabelle's simplifier adding Section~\ref{sec:prelim}'s $\wlp$ 
equational laws as rewriting rules except those for finite iterations and 
while loops which it initially tries to remove with invariant 
reasoning via~(\ref{eq:h-loopi}) and~(\ref{eq:h-whilei}). The tactic assumes
that the hybrid program has the standard shape 
$\INV {\LOOP (\texttt{ctrl};\texttt{dyn})}{I}$ iterating a discrete controller
intervening in a continuous dynamical system as seen in 
Example~\ref{ex:hp}. From this initial proof method, we create two 
tactics for supplying flows: the tactic
\isa{wlp{\isacharunderscore}flow} takes as input a 
previously proven flow-certification theorem asserting our predicate 
\isa{local{\isacharunderscore}flow{\isacharunderscore}on}. 
Alternatively, users can try to make this certification automatic
with the tactic \isa{wlp{\isacharunderscore}solve} which 
requires as input a candidate solution $\varphi$. It calls 
\isa{local{\isacharunderscore}flow{\isacharunderscore}on{\isacharunderscore}auto}
after applying \isa{wlp{\isacharunderscore}simp} and
trying~(\ref{eq:wlp-flow}) with the input $\varphi$. The intention
is that both \isa{wlp{\isacharunderscore}flow} and
\isa{wlp{\isacharunderscore}solve} leave only proof obligations
that require reasoning of first-order logic of real numbers. 
Complementary tactics \isa{wlp{\isacharunderscore}full} and
\isa{wlp{\isacharunderscore}expr{\isacharunderscore}solve}
try to discharge these remaining proof obligations 
automatically leaving raw Isabelle terms in the proof obligations 
without our syntax translations. We have also supplied simple tactics
for algebraic reasoning more interactively. Specifically, 
we have provided a tactic for distributing factors over additions
and a tactic for simplifying powers in multi-variable monomials. See Section~\ref{sec:rotdyn} for the application of these in a verification problem.

%%%%%%%%%%%%%% CAS INTEGRATION %%%%%%%%%%%%%%
\subsection{Solutions from Computer Algebra Systems}\label{sec:cas}
% TODO: distil section 7.3 into a couple of paragraphs
% TODO: merge 7.1 and 7.2

We have integrated two Computer Algebra Systems (CASs), namely SageMath and the Wolfram Engine, into Isabelle/HOL to supply symbolic solutions to ODEs. The user can make use of the integration via the  \isakwmaj{\textbf{find{\isacharunderscore}local{\isacharunderscore}flow}} command, which supplies a solution to the first ODE it finds within the current subgoal.

Below, we show an application of our integration and its corresponding output in Isabelle. Users can click the greyed-out area to automatically insert the suggestion. 

\begin{isabellebody}\isanewline
\isacommand{lemma}\ \isamarkupfalse%
local{\isacharunderscore}flow{\isacharunderscore}example{\isacharcolon}\isanewline
\ \ {\isachardoublequoteopen}\isactrlbold {\isacharbraceleft}{\kern0pt}x\ {\isachargreater}{\kern0pt}\ {\isadigit{0}}\isactrlbold {\isacharbraceright}{\kern0pt}\ x\ {\isacharcolon}{\isacharcolon}{\isacharequal}\ {\isadigit{1}}{\isacharsemicolon}\ {\isacharbraceleft}x{\isasymacute}\ {\isacharequal}\ {\isadigit{1}}{\isacharbraceright} \isactrlbold {\isacharbraceleft}{\kern0pt}x {\isachargreater}{\kern0pt}\ {\isadigit{1}}\isactrlbold {\isacharbraceright}{\kern0pt}{\isachardoublequoteclose}\isanewline
\ \ \isakwmaj{find{\isacharunderscore}local{\isacharunderscore}flow}\isanewline
\texttt{Output}:\isanewline
\ \ \texttt{Calling Wolfram...}\isanewline
\ \ {\isasymlambda}t{\isachardot}\ {\isacharbrackleft}x\ {\isasymleadsto}\ t\ {\isacharplus}\ {\isachardollar}x{\isacharbrackright}\isanewline
\ \ \texttt{try this:}\ {\setlength{\fboxsep}{0.1 pt}\colorbox{lightgray}{apply (wlp{\isacharunderscore}solve {\isachardoublequoteopen}{\isasymlambda}t{\isachardot}\ {\isacharbrackleft}x\ {\isasymleadsto}\ t\ {\isacharplus}\ {\isachardollar}x{\isacharbrackright}{\isachardoublequoteclose})}}
\isanewline
\end{isabellebody}
\noindent
In this case, our plugin requests a solution to the simple ODE $x' = 1$ from the Wolfram Engine. The solution is
expressed as a $\lambda$-abstraction, which takes the current time as input. In the body, a substitution is constructed
which gives the value for each continuous variable at time $t$. In this case, the solution is simply $x\, t = t + x\, 0$, which
is represented by the substitution $[x \leadsto t + x]$.

Our integration follows the following steps in order to produce a solution:

\begin{enumerate}
    \item Retrieve an Isabelle term describing the system of ODEs.
    \item Convert the term into an intermediate representation.
    \item Use one of the CAS backends to solve the ODE.
    \item Convert the solution to an Isabelle term.
    \item Certify the solution using the \texttt{wlp\_solve} Isabelle tactic.
\end{enumerate}

\noindent We consider each of these stages below.

\paragraph{Intermediate representation}

\noindent To have an extensible and modular interface, we implement $(1)$ an intermediate data structure for representing ODEs and their solutions, and $(2)$ procedures for translating back and forth between Isabelle and our intermediate representation. This allows us to capture the structure of the ODEs without the added overhead of Isabelle syntax. It also gives us a unified interface between Isabelle and any CAS.

Our plugin assumes that a system of ODEs as a substitution $[x \leadsto e, y \leadsto f, \cdots]$, as described in Section~\ref{subsec:dynevol}. Each expression ($e, f, \cdots$) gives the derivative expression for each corresponding variable, potentially in terms of other variables. Naturally, for an ODE, these expressions are formed using only arithmetic operators and mathematical functions. We therefore derive the following constrained grammar for arithmetic expressions:\vspace{1ex}

%\begin{minted}{SML}
\begin{alltt}
datatype AExp = NConst of string | UOp of string * AExp |
            BOp of string * AExp * AExp | NNat of int | NInt of int |
            NReal of real | CVar of string | SVar of string | IVar
\end{alltt}
%\end{minted}
\vspace{1ex}

\noindent\texttt{NConst} represents numeric constants, such as $e$ or $\pi$. \texttt{UOp} is for unary operations and \texttt{BOp} is for binary operations. Both of these take a name, and the number of required parameter expressions. The name corresponds to the internal name for the operator in Isabelle/HOL. The operator ``+'', for example, has the name \texttt{Groups.plus\_class.plus}. We then have three constructors for numeric constants: \texttt{NNat} for naturals, \texttt{NInt} for integers and \texttt{NReal} for reals. We have three characterisations of variables: \texttt{CVar} are arbitrary but fixed variables, \texttt{SVar} are mutable (state) variables, and \texttt{IVar} represents the independent variable of our system, usually time.

A system of ODEs is then modelled simply as a finite map from variable names to derivative expressions. Converting back and forth between Isabelle terms (i.e. elements of the \texttt{term} datatype in Isabelle/ML) and the \texttt{AExp} datatype consists of recursing through the structure of the term, using a lookup table to translate each named arithmetic function. % We implement two ways of representing SODEs in Isabelle: tuples and substitutions.

Integrating a CAS into IsaVODEs consists of three separate components: a translation from the intermediate representation to the CAS input format, an interface with the CAS to obtain a solution, and a translation from the solution back into the required format.

\paragraph{Wolfram Engine}
The Wolfram Engine is the CAS behind WolframAlpha and Mathematica~\cite{wolfram_engine}. It is one of the leading CASs on the market, with powerful features for solving differential equations, amongst many other applications. The Wolfram language provides the DSolve function which produces solutions to various kinds of differential equations.

The basic building block for this integration is a representation of Wolfram expressions as an ML datatype. In the Wolfram language, everything is an expression~\cite{wolfram_expression}, so this is sufficient for a complete interface. We represent these expressions as follows:

%\sfin{Can we point to a reference for where this expression type comes from?}

\begin{alltt}
\textbf{datatype} expr = Int \textbf{of} int
              | Real \textbf{of} real
              | Id  \textbf{of} string
              | Fun \textbf{of} string * expr list
              | CurryFun \textbf{of} string * expr list list
\end{alltt}

\noindent\texttt{Real} and \texttt{Int} represent real and integer numbers respectively. \texttt{Id} represented an identifiers,
and \texttt{Fun} represents functions with only one list of arguments, which are distinguished from curried functions
\texttt{CurryFun} with multiple sets of arguments.

Our approach for translation from \texttt{AExp} to \texttt{expr} is the following:
\begin{enumerate}
    \item Generate an alphabetically ordered variable mapping to avoid name clashes and ease solution reconstruction.
    \item Traverse the expression tree and translate each term to Wolfram.
    \item Wrap in the Wolfram \texttt{DSolve} function, supplying the state and independent variables along with our ODE. 
\end{enumerate}

Once we have a well-formed Wolfram expression, we use the \texttt{wolframscript} command-line interface to obtain a
solution in the form of a list of Wolfram \texttt{Rule} expressions, which are isomorphic to our substitutions. We can
parse the result back into the \texttt{expr} type, and translate this into the \texttt{AExp} type using a translation
table for function names and the variable name mapping we constructed.

% \cpin{Check that SageMath integration works}

\paragraph{SageMath}
SageMath~\cite{sagemath} is an open-source competitor to the Wolfram Engine. It is accessed via calls to a Python API. It integrates several open-source CASs to provide its functionality, choosing in each case the best implementation for a particular symbolic computation. This makes SageMath an ideal target for integration with Isabelle.

The translation process is similar to that of our Wolfram Engine integration, but we also apply some preprocessing to the ODEs in order to make the solving more efficient. Since the CASs integrated with SageMath are better at solving smaller ODEs, we can improve their performance by rewriting the following kinds of ODEs:

\begin{enumerate}
\item SODEs formed from a higher order ODE, where an extra variable has been introduced in order to make all derivatives first order, can be recast into their higher-order form. For example, the ODE \(x' = 2x+y, y'=x\) can be rewritten as \(x'' = 2x + x'\).  
\item Independent components of an ODE can be solved independently by the CAS. For example, in the ODE \(x' = x, t' = 1\), x and t are independent and so they can be solved independently.
\end{enumerate}

%%%%%%%%%%%%% DIFFERENTIAL INDUCTION %%%%%%%%%%%%%%
\subsection{Automatic differential invariants}\label{subsec:auto-dinduct}

% \sfin{Need narrative on \texttt{dWeaken} and \texttt{dGhost}}

%\sfin{Where is the differential weakening law?}

We have developed three main proof methods for automating differential invariant proofs: \isa{dWeaken}, \isa{dInduct},
and \isa{dGhost}, which implement differential weakening, induction, and ghosts, respectively~\cite{Platzer18, MuniveS22}. The first, \isa{dWeaken},
applies the differential weakening law to prove $\hoaretriple{P}{\{x' = f ~|~ G \}}{Q}$, when $G \implies Q$. Proof of the implication is attempted via the \isa{auto} proof method.

Application of differential induction (see \S\ref{subsec:dinv}) is automated by the \isa{dInduct} proof method. To prove goals 
$\hoaretriple{I}{\{x' = f ~|~ G \}}{I}$, it (1) applies
% $\hoare{I}{(x' = f\, \&\, G)_U^{t_0}}{I}$, it (1) applies
% $\hoaretriple{I}{\{\vec{x}' = e ~|~ G \}}{I}$, it (1) applies
rules~\eqref{eq:dinv-eq}-\eqref{eq:dinv-less} to produce a framed derivative expression, and (2) calculates derivatives
using the laws~\eqref{eq:const}-\eqref{eq:ln}, substitution laws, and basic simplification laws. This yields derivative-free
equality or inequality predicates. \isa{dInduct} uses only the simplifier for calculating invariants, and so it is
both efficient and yields readable VCs.

For cases requiring deduction to solve the VCs, we have implemented \isa{dInduct\_auto}, which applies \skey{expr-auto}
after \isa{dInduct}, plus further simplification lemmas from \textsf{HOL-Analysis}. Ultimately such heuristics should be
based on decision procedures~\cite{Holzl2009-Approximate,Blanchette2016Hammers,Li2017-Poly,Cordwell2021BKR}, as oracles
or as verified
components. % \hfill\isalink{https://github.com/isabelle-utp/Hybrid-Verification/blob/7baedd092dff0182336ef0bb6251fd8beff6a1cc/HS_Lie_Derivatives.thy\#L374}

% As mentioned in Section~\ref{sec:vcg}, certification of differential invariants is an alternative approach to supplying solutions. We have previously made tactics that automate the 
% differential induction process for invariant certification. Our tactic \isa{dInduct} and its variants~\cite{FosterMGS21} use our framed
% Fr\'echet derivative rules while

Should differential induction fail, our \isa{diff{\isacharunderscore}inv{\isacharunderscore}on} tactics use the vector
derivative laws from Isabelle's ODEs library~\cite{afp:odes,MuniveS22}. The latter is more interactive as users need to
provide the derivatives of both sides of the (in)equality.  Yet, this provides more fine-grained control and makes the
tactic more likely to succeed.

While \isa{dInduct\_auto} suffices for simpler examples, differential induction must often be combined with weakening
and cut rules. These rules have been explained elsewhere~\cite{Platzer18, MuniveS22}. This process is automated by a search-based proof method called \isa{dInduct\_mega}. The following steps
are executed iteratively until all goals are proved or no rule applies: (1) try any fact labelled with attribute
\texttt{facts}, (2) try differential weakening to prove the goal, (3) try differential cut to split it into two
differential invariants, (4) try \isa{dInduct\_auto}. The rules are applied using backtracking so that if one rule
fails, another one is tried.

Method \isa{dInduct\_mega} is applied by the method \skey{dInv}, which we applied in \S\ref{sec:planar-flight}. The
\isa{dInv} method allows us to prove a goal of the form $\hoaretriple{P}{\{x' = f ~|~ G \}}{Q}$ by supplying an
assertion $I$, and proving that this is an invariant of ${\{x' = f ~|~ G \}}$ and also that $P \implies I$. This
method also requires us to apply differential cut and weakening laws.

%\sfin{I think we should update dGhost to apply J's updated rule, but only after submission.}

Finally, the differential ghost law \eqref{eq:dG} is automated through the \isa{dGhost} proof method.
% It requires three parameters: a lens $y$ that acts as the differential ghost variable, 
We have complemented these tactics with sound rules from differential dynamic logic. Together with the tactics described
in this Section, they enable users to seamlessly prove properties about hybrid systems in the style of \dL. Yet, users
of the general-purpose prover can also use the interactive style provided by Isabelle's Isar scripting
language. See Section~\ref{sec:longsol} for an example where our tactics largely automate a proof of invariance.

% Reintroduce dInduct-mega

%%%%%%%%%%%%% DERIV TESTS %%%%%%%%%%%%%%
\subsection{Derivative tests}\label{subsec:deriv-tests}
To culminate this section, we lay the foundations for further automating
the verification process in our framework. 
Both \isa{wlp{\isacharunderscore}full} and \isa{dInduct}
methods cannot certify complex real arithmetic proof obligations after discharging
any derivative-related intermediate steps. In many cases, this requires
determining the local minima and maxima of expressions and using them for further argumentation. A well-known application of differentiation is the 
analysis of real-valued functions: determining their local minima and 
maxima or where they are increasing or decreasing. Optimisation problems 
in physics and engineering frequently require this kind of analysis. 
Behind it, there are 
two key results generally called the \emph{first} and \emph{second 
derivative tests}. We formalise and prove these basic theorems in 
Isabelle/HOL, setting the foundation for automatic certification of the
analysis of real-valued functions (see Section~\ref{sec:rocket}).
We start by defining increasing/decreasing functions and local extrema:

\begin{isabellebody}
\isanewline
\isacommand{definition}\isamarkupfalse%
\ {\isachardoublequoteopen}increasing{\isacharunderscore}{\kern0pt}on\ T\ f\ {\isasymlongleftrightarrow}\ {\isacharparenleft}{\kern0pt}{\isasymforall}x{\isasymin}T{\isachardot}{\kern0pt}\ {\isasymforall}y{\isasymin}T{\isachardot}{\kern0pt}\ x\ {\isasymle}\ y\ {\isasymlongrightarrow}\ f\ x\ {\isasymle}\ f\ y{\isacharparenright}{\kern0pt}{\isachardoublequoteclose}

\isanewline
\isacommand{definition}\isamarkupfalse%
\ {\isachardoublequoteopen}decreasing{\isacharunderscore}{\kern0pt}on\ T\ f\ {\isasymlongleftrightarrow}\ {\isacharparenleft}{\kern0pt}{\isasymforall}x{\isasymin}T{\isachardot}{\kern0pt}\ {\isasymforall}y{\isasymin}T{\isachardot}{\kern0pt}\ x\ {\isasymle}\ y\ {\isasymlongrightarrow}\ f\ y\ {\isasymle}\ f\ x{\isacharparenright}{\kern0pt}{\isachardoublequoteclose}

\isanewline
\isacommand{definition}\isamarkupfalse%
\ {\isachardoublequoteopen}strict{\isacharunderscore}{\kern0pt}increasing{\isacharunderscore}{\kern0pt}on\ T\ f\ {\isasymlongleftrightarrow}\ {\isacharparenleft}{\kern0pt}{\isasymforall}x{\isasymin}T{\isachardot}{\kern0pt}\ {\isasymforall}y{\isasymin}T{\isachardot}{\kern0pt}\ x\ {\isacharless}{\kern0pt}\ y\ {\isasymlongrightarrow}\ f\ x\ {\isacharless}{\kern0pt}\ f\ y{\isacharparenright}{\kern0pt}{\isachardoublequoteclose}

\isanewline
\isacommand{definition}\isamarkupfalse%
\ {\isachardoublequoteopen}strict{\isacharunderscore}{\kern0pt}decreasing{\isacharunderscore}{\kern0pt}on\ T\ f\ {\isasymlongleftrightarrow}\ {\isacharparenleft}{\kern0pt}{\isasymforall}x{\isasymin}T{\isachardot}{\kern0pt}\ {\isasymforall}y{\isasymin}T{\isachardot}{\kern0pt}\ x\ {\isacharless}{\kern0pt}\ y\ {\isasymlongrightarrow}\ f\ y\ {\isacharless}{\kern0pt}\ f\ x{\isacharparenright}{\kern0pt}{\isachardoublequoteclose}

\isanewline
\isacommand{definition}\isamarkupfalse%
\ {\isachardoublequoteopen}local{\isacharunderscore}{\kern0pt}maximum{\isacharunderscore}{\kern0pt}at\ T\ f\ x\ {\isasymlongleftrightarrow}\ {\isacharparenleft}{\kern0pt}{\isasymforall}y{\isasymin}T{\isachardot}{\kern0pt}\ f\ y\ {\isasymle}\ f\ x{\isacharparenright}{\kern0pt}{\isachardoublequoteclose}

\isanewline
\isacommand{definition}\isamarkupfalse%
\ {\isachardoublequoteopen}local{\isacharunderscore}{\kern0pt}minimum{\isacharunderscore}{\kern0pt}at\ T\ f\ x\ {\isasymlongleftrightarrow}\ {\isacharparenleft}{\kern0pt}{\isasymforall}y{\isasymin}T{\isachardot}{\kern0pt}\ f\ y\ {\isasymge}\ f\ x{\isacharparenright}{\kern0pt}{\isachardoublequoteclose}
\isanewline
\end{isabellebody}

We also prove simple consequences of these definitions on closed intervals 
$[a,b]$ denoted in Isabelle as 
\isa{{\isacharbraceleft}{\kern0pt}a{\isachardot}{\kern0pt}{\isachardot}{\kern0pt}b{\isacharbraceright}}.
For instance, we prove the transitivity of the increasing and decreasing
properties over consecutive intervals and their relationship to local 
extrema. We exemplify some of these results below and refer to
our repository for all proved properties.\hfill\isalink{https://github.com/isabelle-utp/Hybrid-Verification/blob/7aa6a58e057586085f9eb6d646d523c6b7c85cb5/HS_Preliminaries.thy\#L595}

\begin{isabellebody}
\isanewline
\isanewline
\isacommand{lemma}\isamarkupfalse%
\ increasing{\isacharunderscore}{\kern0pt}on{\isacharunderscore}{\kern0pt}trans{\isacharcolon}{\kern0pt}\isanewline
\ \ \isakeyword{fixes}\ f\ {\isacharcolon}{\kern0pt}{\isacharcolon}{\kern0pt}\ {\isachardoublequoteopen}{\isacharprime}{\kern0pt}a{\isacharcolon}{\kern0pt}{\isacharcolon}{\kern0pt}linorder\ {\isasymRightarrow}\ {\isacharprime}{\kern0pt}b{\isacharcolon}{\kern0pt}{\isacharcolon}{\kern0pt}preorder{\isachardoublequoteclose}\isanewline
\ \ \isakeyword{assumes} {\isachardoublequoteopen}a\ {\isasymle}\ b{\isachardoublequoteclose}\ \isakeyword{and}\ {\isachardoublequoteopen}b\ {\isasymle}\ c{\isachardoublequoteclose}\isanewline
\ \ \ \ \isakeyword{and} {\isachardoublequoteopen}increasing{\isacharunderscore}{\kern0pt}on\ {\isacharbraceleft}{\kern0pt}a{\isachardot}{\kern0pt}{\isachardot}{\kern0pt}b{\isacharbraceright}{\kern0pt}\ f{\isachardoublequoteclose}\ \isakeyword{and}\ {\isachardoublequoteopen}increasing{\isacharunderscore}{\kern0pt}on\ {\isacharbraceleft}{\kern0pt}b{\isachardot}{\kern0pt}{\isachardot}{\kern0pt}c{\isacharbraceright}{\kern0pt}\ f{\isachardoublequoteclose}\isanewline
\ \ \isakeyword{shows}\ {\isachardoublequoteopen}increasing{\isacharunderscore}{\kern0pt}on\ {\isacharbraceleft}{\kern0pt}a{\isachardot}{\kern0pt}{\isachardot}{\kern0pt}c{\isacharbraceright}{\kern0pt}\ f{\isachardoublequoteclose}\isanewline
\ \ \isacommand{unfolding}\isamarkupfalse%
\ increasing{\isacharunderscore}{\kern0pt}on{\isacharunderscore}{\kern0pt}def\ \ \isanewline
\ \ \isacommand{by}\isamarkupfalse%
\ auto\ {\isacharparenleft}{\kern0pt}smt\ {\isacharparenleft}{\kern0pt}verit{\isacharcomma}{\kern0pt}\ best{\isacharparenright}{\kern0pt}\ intervalE\ nle{\isacharunderscore}{\kern0pt}le\ order{\isacharunderscore}{\kern0pt}trans{\isacharparenright}{\kern0pt}%

\isanewline
\isacommand{lemma}\isamarkupfalse%
\ increasing{\isacharunderscore}{\kern0pt}on{\isacharunderscore}{\kern0pt}local{\isacharunderscore}{\kern0pt}maximum{\isacharcolon}{\kern0pt}\isanewline
\ \ \isakeyword{fixes}\ f\ {\isacharcolon}{\kern0pt}{\isacharcolon}{\kern0pt}\ {\isachardoublequoteopen}{\isacharprime}{\kern0pt}a{\isacharcolon}{\kern0pt}{\isacharcolon}{\kern0pt}preorder\ {\isasymRightarrow}\ {\isacharprime}{\kern0pt}b{\isacharcolon}{\kern0pt}{\isacharcolon}{\kern0pt}preorder{\isachardoublequoteclose}\isanewline
\ \ \isakeyword{assumes} {\isachardoublequoteopen}a\ {\isasymle}\ b{\isachardoublequoteclose}\ \isakeyword{and} {\isachardoublequoteopen}increasing{\isacharunderscore}{\kern0pt}on\ {\isacharbraceleft}{\kern0pt}a{\isachardot}{\kern0pt}{\isachardot}{\kern0pt}b{\isacharbraceright}{\kern0pt}\ f{\isachardoublequoteclose}\isanewline
\ \ \isakeyword{shows} {\isachardoublequoteopen}local{\isacharunderscore}{\kern0pt}maximum{\isacharunderscore}{\kern0pt}at\ {\isacharbraceleft}{\kern0pt}a{\isachardot}{\kern0pt}{\isachardot}{\kern0pt}b{\isacharbraceright}{\kern0pt}\ f\ b{\isachardoublequoteclose}\ \isanewline
\ \ \isacommand{by}\isamarkupfalse%
\ {\isacharparenleft}{\kern0pt}auto\ simp{\isacharcolon}{\kern0pt}\ increasing{\isacharunderscore}{\kern0pt}on{\isacharunderscore}{\kern0pt}def\ local{\isacharunderscore}{\kern0pt}maximum{\isacharunderscore}{\kern0pt}at{\isacharunderscore}{\kern0pt}def{\isacharparenright}

\isanewline
\isacommand{lemma}\isamarkupfalse%
\ incr{\isacharunderscore}{\kern0pt}decr{\isacharunderscore}{\kern0pt}local{\isacharunderscore}{\kern0pt}maximum{\isacharcolon}{\kern0pt}\isanewline
\ \ \isakeyword{fixes}\ f\ {\isacharcolon}{\kern0pt}{\isacharcolon}{\kern0pt}\ {\isachardoublequoteopen}{\isacharprime}{\kern0pt}a{\isacharcolon}{\kern0pt}{\isacharcolon}{\kern0pt}linorder\ {\isasymRightarrow}\ {\isacharprime}{\kern0pt}b{\isacharcolon}{\kern0pt}{\isacharcolon}{\kern0pt}preorder{\isachardoublequoteclose}\isanewline
\ \ \isakeyword{assumes} {\isachardoublequoteopen}a\ {\isasymle}\ b{\isachardoublequoteclose}\ \isakeyword{and} {\isachardoublequoteopen}b\ {\isasymle}\ c{\isachardoublequoteclose}\isanewline
\ \ \ \ \isakeyword{and}\ {\isachardoublequoteopen}increasing{\isacharunderscore}{\kern0pt}on\ {\isacharbraceleft}{\kern0pt}a{\isachardot}{\kern0pt}{\isachardot}{\kern0pt}b{\isacharbraceright}{\kern0pt}\ f{\isachardoublequoteclose}\ \isakeyword{and}\ {\isachardoublequoteopen}decreasing{\isacharunderscore}{\kern0pt}on\ {\isacharbraceleft}{\kern0pt}b{\isachardot}{\kern0pt}{\isachardot}{\kern0pt}c{\isacharbraceright}{\kern0pt}\ f{\isachardoublequoteclose}\isanewline
\ \ \isakeyword{shows} {\isachardoublequoteopen}local{\isacharunderscore}{\kern0pt}maximum{\isacharunderscore}{\kern0pt}at\ {\isacharbraceleft}{\kern0pt}a{\isachardot}{\kern0pt}{\isachardot}{\kern0pt}c{\isacharbraceright}{\kern0pt}\ f\ b{\isachardoublequoteclose}\isanewline
\ \ \isacommand{unfolding}\isamarkupfalse%
\ increasing{\isacharunderscore}{\kern0pt}on{\isacharunderscore}{\kern0pt}def\ decreasing{\isacharunderscore}{\kern0pt}on{\isacharunderscore}{\kern0pt}def\ local{\isacharunderscore}{\kern0pt}maximum{\isacharunderscore}{\kern0pt}at{\isacharunderscore}{\kern0pt}def\isanewline
\ \ \isacommand{by}\isamarkupfalse%
\ auto\ {\isacharparenleft}{\kern0pt}metis\ intervalE\ linorder{\isacharunderscore}{\kern0pt}le{\isacharunderscore}{\kern0pt}cases{\isacharparenright}
\isanewline
\end{isabellebody}

Crucially, the proofs of all these results only require unfolding definitions 
and calling Isabelle's \emph{auto} proof method and 
\emph{Sledgehammer} tool~\cite{PaulsonB10} to discharge the last proof obligation. We then use our definitions of increasing and 
decreasing functions to state and prove the first derivative test. It states that if the
derivative of a function is greater (resp. less) than $0$ over an interval 
$T$, then the function is increasing (resp. decreasing) on that interval. 
The simple 2-line proof above uses an intermediate lemma 
\isa{has{\isacharunderscore}{\kern0pt}vderiv{\isacharunderscore}{\kern0pt}mono{\isacharunderscore}{\kern0pt}test} hiding our full proof of this basic result.\hfill\isalink{https://github.com/isabelle-utp/Hybrid-Verification/blob/7aa6a58e057586085f9eb6d646d523c6b7c85cb5/HS_Preliminaries.thy\#L671}

\begin{isabellebody}
\isanewline
\isacommand{lemma}\isamarkupfalse%
\ first{\isacharunderscore}{\kern0pt}derivative{\isacharunderscore}{\kern0pt}test{\isacharcolon}{\kern0pt}\isanewline
\ \ \isakeyword{assumes}\ T{\isacharunderscore}{\kern0pt}hyp{\isacharcolon}{\kern0pt}\ {\isachardoublequoteopen}is{\isacharunderscore}{\kern0pt}interval\ T{\isachardoublequoteclose}\ \isanewline
\ \ \ \ \isakeyword{and}\ d{\isacharunderscore}{\kern0pt}hyp{\isacharcolon}{\kern0pt}\ {\isachardoublequoteopen}D\ f\ {\isacharequal}{\kern0pt}\ f{\isacharprime}{\kern0pt}\ on\ T{\isachardoublequoteclose}\isanewline
\ \ \isakeyword{shows}\ {\isachardoublequoteopen}{\isasymforall}x{\isasymin}T{\isachardot}{\kern0pt}\ {\isacharparenleft}{\kern0pt}{\isadigit{0}}{\isacharcolon}{\kern0pt}{\isacharcolon}{\kern0pt}real{\isacharparenright}{\kern0pt}\ {\isasymle}\ f{\isacharprime}{\kern0pt}\ x\ {\isasymLongrightarrow}\ increasing{\isacharunderscore}{\kern0pt}on\ T\ f{\isachardoublequoteclose}\isanewline
\ \ \ \ \isakeyword{and}\ {\isachardoublequoteopen}{\isasymforall}x{\isasymin}T{\isachardot}{\kern0pt}\ f{\isacharprime}{\kern0pt}\ x\ {\isasymle}\ {\isadigit{0}}\ {\isasymLongrightarrow}\ decreasing{\isacharunderscore}{\kern0pt}on\ T\ f{\isachardoublequoteclose}\isanewline
\ \ \isacommand{unfolding}\isamarkupfalse%
\ increasing{\isacharunderscore}{\kern0pt}on{\isacharunderscore}{\kern0pt}def\ decreasing{\isacharunderscore}{\kern0pt}on{\isacharunderscore}{\kern0pt}def\isanewline
\ \ \isacommand{using}\isamarkupfalse%
\ has{\isacharunderscore}{\kern0pt}vderiv{\isacharunderscore}{\kern0pt}mono{\isacharunderscore}{\kern0pt}test{\isacharbrackleft}{\kern0pt}OF\ assms{\isacharbrackright}{\kern0pt}\ \isacommand{by}\isamarkupfalse%
\ blast
\isanewline
\end{isabellebody}

For the second derivative test, we formalise a frequently used property
of continuous real-valued functions. Namely, that if a function maps a 
point $t$ to some value above (resp. below) a threshold $c$, then there
is an open set around $t$ filled with points mapped to values above 
(resp. below) the threshold. This result and similar ones in terms of 
open balls with fixed radius around $t$ appear in our formalisations. For 
these, we also complement Isabelle's library of topological concepts with 
the definition of \emph{neighbourhood} and some of its properties and 
characterisations.\hfill\isalink{https://github.com/isabelle-utp/Hybrid-Verification/blob/7aa6a58e057586085f9eb6d646d523c6b7c85cb5/HS_Preliminaries.thy\#L707}

\begin{isabellebody}
\isanewline
\isacommand{definition}\isamarkupfalse%
\ {\isachardoublequoteopen}neighbourhood\ N\ x\ {\isasymlongleftrightarrow}\ {\isacharparenleft}{\kern0pt}{\isasymexists}X{\isachardot}{\kern0pt}\ open\ X\ {\isasymand}\ x\ {\isasymin}\ X\ {\isasymand}\ X\ {\isasymsubseteq}\ N{\isacharparenright}{\kern0pt}{\isachardoublequoteclose}

\isanewline
\isacommand{lemma}\isamarkupfalse%
\ continuous{\isacharunderscore}{\kern0pt}on{\isacharunderscore}{\kern0pt}Ex{\isacharunderscore}{\kern0pt}open{\isacharunderscore}{\kern0pt}less{\isacharcolon}{\kern0pt}\isanewline
\ \ \isakeyword{fixes}\ f\ {\isacharcolon}{\kern0pt}{\isacharcolon}{\kern0pt}\ {\isachardoublequoteopen}{\isacharprime}{\kern0pt}a\ {\isacharcolon}{\kern0pt}{\isacharcolon}{\kern0pt}\ topological{\isacharunderscore}{\kern0pt}space\ {\isasymRightarrow}\ real{\isachardoublequoteclose}\isanewline
\ \ \isakeyword{assumes}\ {\isachardoublequoteopen}continuous{\isacharunderscore}{\kern0pt}on\ T\ f{\isachardoublequoteclose}\isanewline
\ \ \ \ \isakeyword{and}\ {\isachardoublequoteopen}neighbourhood\ T\ t{\isachardoublequoteclose}\isanewline
\ \ \isakeyword{shows}\ {\isachardoublequoteopen}f\ t\ {\isachargreater}{\kern0pt}\ c\ {\isasymLongrightarrow}\ {\isasymexists}X{\isachardot}{\kern0pt}\ open\ X\ {\isasymand}\ t\ {\isasymin}\ X\ {\isasymand}\ X\ {\isasymsubseteq}\ T\ {\isasymand}\ {\isacharparenleft}{\kern0pt}{\isasymforall}{\isasymtau}{\isasymin}X{\isachardot}{\kern0pt}\ f\ {\isasymtau}\ {\isachargreater}{\kern0pt}\ c{\isacharparenright}{\kern0pt}{\isachardoublequoteclose}\isanewline
\ \ \ \ \isakeyword{and}\ {\isachardoublequoteopen}f\ t\ {\isacharless}{\kern0pt}\ c\ {\isasymLongrightarrow}\ {\isasymexists}X{\isachardot}{\kern0pt}\ open\ X\ {\isasymand}\ t\ {\isasymin}\ X\ {\isasymand}\ X\ {\isasymsubseteq}\ T\ {\isasymand}\ {\isacharparenleft}{\kern0pt}{\isasymforall}{\isasymtau}{\isasymin}X{\isachardot}{\kern0pt}\ f\ {\isasymtau}\ {\isacharless}{\kern0pt}\ c{\isacharparenright}{\kern0pt}{\isachardoublequoteclose}\isanewline
\end{isabellebody}

Finally, the second derivative test states that if the derivative of a
real-valued function is $0$ at a point $t$, and its second derivative
is continuous and positive (resp. negative) at $t$, then the original
function has a local minimum (resp. maximum) at $t$.

\begin{isabellebody}
\isanewline
\isacommand{lemma}\isamarkupfalse%
\ second{\isacharunderscore}{\kern0pt}derivative{\isacharunderscore}{\kern0pt}test{\isacharcolon}{\kern0pt}\isanewline
\ \ \isakeyword{assumes}\ {\isachardoublequoteopen}continuous{\isacharunderscore}{\kern0pt}on\ T\ f{\isacharprime}{\kern0pt}{\isacharprime}{\kern0pt}{\isachardoublequoteclose}\isanewline
\ \ \ \ \isakeyword{and}\ {\isachardoublequoteopen}neighbourhood\ T\ t{\isachardoublequoteclose}\isanewline
\ \ \ \ \isakeyword{and}\ f{\isacharprime}{\kern0pt}{\isacharcolon}{\kern0pt}\ {\isachardoublequoteopen}D\ f\ {\isacharequal}{\kern0pt}\ f{\isacharprime}{\kern0pt}\ on\ T{\isachardoublequoteclose}\isanewline
\ \ \ \ \isakeyword{and}\ f{\isacharprime}{\kern0pt}{\isacharprime}{\kern0pt}{\isacharcolon}{\kern0pt}\ {\isachardoublequoteopen}D\ f{\isacharprime}{\kern0pt}\ {\isacharequal}{\kern0pt}\ f{\isacharprime}{\kern0pt}{\isacharprime}{\kern0pt}\ on\ T{\isachardoublequoteclose}\isanewline
\ \ \ \ \isakeyword{and}\ {\isachardoublequoteopen}f{\isacharprime}{\kern0pt}\ t\ {\isacharequal}{\kern0pt}\ {\isacharparenleft}{\kern0pt}{\isadigit{0}}\ {\isacharcolon}{\kern0pt}{\isacharcolon}{\kern0pt}\ real{\isacharparenright}{\kern0pt}{\isachardoublequoteclose}\isanewline
\ \ \isakeyword{shows}\ {\isachardoublequoteopen}f{\isacharprime}{\kern0pt}{\isacharprime}{\kern0pt}\ t {\isacharless} {\isadigit{0}}\isanewline
\ \ \ \ {\isasymLongrightarrow}\ {\isasymexists}a\ b{\isachardot}{\kern0pt}\ a{\isacharless}t\ {\isasymand}\ t{\isacharless}b\ {\isasymand}\ {\isacharbraceleft}{\kern0pt}a{\isacharminus}{\kern0pt}{\isacharminus}{\kern0pt}b{\isacharbraceright}{\kern0pt}\ {\isasymsubseteq}\ T\ {\isasymand}\ local{\isacharunderscore}{\kern0pt}maximum{\isacharunderscore}{\kern0pt}at\ {\isacharbraceleft}{\kern0pt}a{\isacharminus}{\kern0pt}{\isacharminus}{\kern0pt}b{\isacharbraceright}{\kern0pt}\ f\ t{\isachardoublequoteclose}\isanewline
\ \ \ \ \isakeyword{and}\ {\isachardoublequoteopen}f{\isacharprime}{\kern0pt}{\isacharprime}{\kern0pt}\ t {\isachargreater} {\isadigit{0}}\isanewline
\ \ \ \ {\isasymLongrightarrow}\ {\isasymexists}a\ b{\isachardot}{\kern0pt}\ a{\isacharless}t\ {\isasymand}\ t{\isacharless}b\ {\isasymand}\ {\isacharbraceleft}{\kern0pt}a{\isacharminus}{\kern0pt}{\isacharminus}{\kern0pt}b{\isacharbraceright}{\kern0pt}\ {\isasymsubseteq}\ T\ {\isasymand}\ local{\isacharunderscore}{\kern0pt}minimum{\isacharunderscore}{\kern0pt}at\ {\isacharbraceleft}{\kern0pt}a{\isacharminus}{\kern0pt}{\isacharminus}{\kern0pt}b{\isacharbraceright}{\kern0pt}\ f\ t{\isachardoublequoteclose}\isanewline
\ \ \isacommand{unfolding}\isamarkupfalse%
\ local{\isacharunderscore}{\kern0pt}maximum{\isacharunderscore}{\kern0pt}at{\isacharunderscore}{\kern0pt}def\ local{\isacharunderscore}{\kern0pt}minimum{\isacharunderscore}{\kern0pt}at{\isacharunderscore}{\kern0pt}def\isanewline
\ \ \isacommand{using}\isamarkupfalse%
\ has{\isacharunderscore}{\kern0pt}vderiv{\isacharunderscore}{\kern0pt}max{\isacharunderscore}{\kern0pt}test{\isacharbrackleft}{\kern0pt}OF\ assms{\isacharbrackright}{\kern0pt}\ has{\isacharunderscore}{\kern0pt}vderiv{\isacharunderscore}{\kern0pt}min{\isacharunderscore}{\kern0pt}test{\isacharbrackleft}{\kern0pt}OF\ assms{\isacharbrackright}{\kern0pt}\ \isanewline
\ \ \isacommand{by}\isamarkupfalse%
\ blast{\isacharplus}
\isanewline
\end{isabellebody}

As before, we provide the complete argument in the proof of our lemmas 
\isa{has{\isacharunderscore}{\kern0pt}vderiv{\isacharunderscore}{\kern0pt}max{\isacharunderscore}{\kern0pt}test} and
\isa{has{\isacharunderscore}{\kern0pt}vderiv{\isacharunderscore}{\kern0pt}min{\isacharunderscore}{\kern0pt}test}. We refer interested readers to our repository for
complete results.\hfill\isalink{https://github.com/isabelle-utp/Hybrid-Verification/blob/7aa6a58e057586085f9eb6d646d523c6b7c85cb5/HS_Preliminaries.thy\#L801}
Section~\ref{sec:ex} provides an example where we use these derivative
tests to reason about real-arithmetic properties and establish the progress of
a dynamical system. Beyond that, this subsection showcases the openness of our framework. Anyone can formalise well-known analysis concepts that provide background theory engineering to
increase proof automation or that generalise extant verification methods.

We have presented various tactics that increase proof automation for hybrid systems verification in our framework. The automation supports both, solution and invariant-based reasoning. The definition of the tactics and their testing covers more than 500 lines of Isabelle code. From these 500 lines, the tactics setup (definitions and lemmas) comprises approximately 200 lines. This number does not take into account our described formalisations. The tactics have simplified our verification experience by discharging certifications of the uniqueness of solutions or differential inductions. We have used them extensively in a set of 66 hybrid systems verification problems. See Section~\ref{sec:eval} for more details.

\section{Evaluation}\label{sec:eval}
For the evaluation of our verification framework, we have tackled 66 problems
of the Hybrid Systems Theorem Proving (HSTP) category from the 9th 
International Workshop on Applied Verification of Continuous and 
Hybrid Systems (ARCH22)~\cite{ARCH22} Friendly Competition. Of 
those, 5 are verifications of regular programs, 32 are verifications of 
continuous dynamics and 29 are verifications of hybrid programs. 
In a different classification, the first 9 problems test the tool's ability to 
handle the interactions between hybrid programs' constructors through
various orders of loops, tests, assignments and dynamics. 
The next 30 problems test the 
tool's ability to tackle different kinds of continuous dynamics: one 
evolution command after another, an evolution command with many 
variables at once, and dynamics that in \dL require invariants, ghosts, 
differential cuts, weakenings, or Darboux rules. 21 problems come from 
tutorials~\cite{Platzer12a, QueselMLAP16} (9 and 12 respectively) on 
how to model and prove hybrid systems in \dL. They include event and 
time-triggered controls for straight-line motion and two-dimensional 
curved motion. 3 hybrid programs come from a case study on the 
verification of the European train control system 
protocol~\cite{PlatzerQ09}. The remaining 3 involve linear and 
nonlinear dynamics.\hfill\isalink{https://github.com/isabelle-utp/Hybrid-Verification/blob/c6b893eb66ea9c91f116cd4145367f1b99735a23/Hybrid_Programs/Verification_Examples/ARCH2022_Examples.thy\#L31}

We fully proved in Isabelle 58 of the 66 problems. Of the remaining 8,
we proved 4 with our verification framework while leaving some 
arithmetic certifications to external computer algebra systems. We could
not find the proofs for 2 of the remaining 4 problems while the last 2 
require the generalisation of our Darboux rule or a different proof 
method. This performance is similar to state-of-the-art tools 
participating in the same competition~\cite{ARCH22} on those 
benchmark problems in the proof-interaction/scripted format. In general,
our first approach to solving all 66 problems was a combination of our 
$\wlp$'s tactics with a supplied solution. Yet, we only were able to solve
38 of them with this approach. The remaining 20 problems required us to
employ higher-order logic methods or \dL techniques like differential 
induction, ghosts, cuts, weakenings or Darboux rules. Specifically, 
our three proofs using Isabelle's higher-order logic and analysis methods
involved nonlinear dynamics which immediately required us to increase 
our interaction with the tool. A generalisation of our invariance certification
methods would alleviate this because as solutions to the systems of ODEs
become more complex, their certification is more difficult. This explains 
why invariant reasoning becomes prominent in the remaining 17 solved 
problems.

For 14 problems, we provided several proofs to exemplify our tool's 
diversity of methods. Between providing solutions and using differential
induction, neither method is comparatively easier to use than the other
in the 14 problems tested. Most of the time they end up with the same 
number of lines of code (LOC) per proof and whenever one has more 
LOC for a problem, a different problem favours equally the other 
method. Quantitatively speaking, we used the solution method 36 times
and the induction method 26 times. We used differential ghosts 4 times 
and our Darboux rule twice. %We suspect that 2 of our unsolved problems will be solved by providing solutions and 2 more will require the generalisation of our Darboux rule. 
In terms of LOC, the average
length of the statement of a problem is 3.72 lines with a median and mode
of 1 totalling 246 LOCs for all problem statements. The average number of
LOC of the problems' shortest proofs is 8.11 with a median of 3 and a
mode of 1 totalling 511 LOC for problems' proofs. The shortest
proof for a problem is 1 while the longest is 98 LOC. These figures do 
not take into account any additional lemmata about real numbers necessary for having fully certified proofs in Isabelle/HOL.
In general, 23 of the 62 (at least partially) solved problems require the 
assertion of real arithmetical facts for their full verification. Thus, automation
of first-order logic arithmetic for real numbers in general-purpose ITPs is highly desired for the scalability of end-to-end verification 
within them. Otherwise, trust in computer algebra tools will be necessary.

Our additions of tactics to the verification framework 
have highly reduced the number of LOC to verify a problem as evidenced by the fact that 48 benchmark
problems are solved with a single call to one of our tactics. Our addition
of the fact that $C^1$-functions are Lipschitz continuous has largely
contributed to the success of this automation: 18 of our proofs call our tactic
\isa{wlp{\isacharunderscore}solve} and 9 use 
\isa{local{\isacharunderscore}flow{\isacharunderscore}on{\isacharunderscore}auto}.
Finally, thanks to our shallow expressions and our nondeterministic 
assignments, our formalisation of the benchmark problems is now fully 
faithful to the ARCH22 competition-required \dL syntax.

\section{Examples}\label{sec:ex}
In this section, we showcase the benefits of our contributions by applying them
in some of the ARCH22 competition benchmarks and examples of our own.

%%%%%%%%%%%%%% VCG-FLOW %%%%%%%%%%%%%%%
% Example with full workflow: suggested solution and using automated tactics
\subsection{Rotational dynamics 3.}\label{sec:rotdyn} Our first problem illustrates the 
integration of all our features for ODEs. It describes the preservation
of $I \iff d_1^2 + d_2^2 = w^2\cdot p^2\land d_1=-w\cdot x_2\land d_2=w\cdot x_1$
for the ODEs $x_1'=d_1, x_ 2'=d_2, d_1'=-w\cdot d_2, d_2'=w\cdot d_1$.
Observe that the relationship between $d_1$ and $d_2$ in the system of ODEs
is similar to that of scaled sine and cosine functions. Moreover, the invariant states 
a Pythagorean relation among them. Thus, we can expect the flow to involve 
trigonometric functions and the problem to be solved with differential invariants
in \dL due to its previously limited capabilities to explicitly state these 
functions~\cite{GallicchioTMP22}. Indeed, differential induction in Isabelle/HOL
can prove this example in one line:

\begin{isabellebody}\isanewline
\isacommand{lemma}\isamarkupfalse%
\ {\isachardoublequoteopen}{\isacharparenleft}{\kern0pt}d{\isadigit{1}}\isactrlsup {\isadigit{2}}\ {\isacharplus}{\kern0pt}\ d{\isadigit{2}}\isactrlsup {\isadigit{2}}\ {\isacharequal}{\kern0pt}\ w\isactrlsup {\isadigit{2}}\ {\isacharasterisk}{\kern0pt}\ p\isactrlsup {\isadigit{2}}\ {\isasymand}\ d{\isadigit{1}}\ {\isacharequal}{\kern0pt}\ {\isacharminus}{\kern0pt}\ w\ {\isacharasterisk}{\kern0pt}\ x{\isadigit{2}}\ {\isasymand}\ d{\isadigit{2}}\ {\isacharequal}{\kern0pt}\ w\ {\isacharasterisk}{\kern0pt}\ x{\isadigit{1}}{\isacharparenright}{\kern0pt}\isactrlsub e\ {\isasymle}\isanewline
\ \ {\isacharbar}{\kern0pt}{\isacharbraceleft}{\kern0pt}x{\isadigit{1}}{\isacharbackquote}{\kern0pt}\ {\isacharequal}{\kern0pt}\ d{\isadigit{1}}{\isacharcomma}{\kern0pt}\ x{\isadigit{2}}{\isacharbackquote}{\kern0pt}\ {\isacharequal}{\kern0pt}\ d{\isadigit{2}}{\isacharcomma}{\kern0pt}\ d{\isadigit{1}}{\isacharbackquote}{\kern0pt}\ {\isacharequal}{\kern0pt}\ {\isacharminus}{\kern0pt}\ w\ {\isacharasterisk}{\kern0pt}\ d{\isadigit{2}}{\isacharcomma}{\kern0pt}\ d{\isadigit{2}}{\isacharbackquote}{\kern0pt}\ {\isacharequal}{\kern0pt}\ w\ {\isacharasterisk}{\kern0pt}\ d{\isadigit{1}}{\isacharbraceright}{\kern0pt}{\isacharbrackright}{\kern0pt}\ \isanewline
\ \ {\isacharparenleft}{\kern0pt}d{\isadigit{1}}\isactrlsup {\isadigit{2}}\ {\isacharplus}{\kern0pt}\ d{\isadigit{2}}\isactrlsup {\isadigit{2}}\ {\isacharequal}{\kern0pt}\ w\isactrlsup {\isadigit{2}}\ {\isacharasterisk}{\kern0pt}\ p\isactrlsup {\isadigit{2}}\ {\isasymand}\ d{\isadigit{1}}\ {\isacharequal}{\kern0pt}\ {\isacharminus}{\kern0pt}\ w\ {\isacharasterisk}{\kern0pt}\ x{\isadigit{2}}\ {\isasymand}\ d{\isadigit{2}}\ {\isacharequal}{\kern0pt}\ w\ {\isacharasterisk}{\kern0pt}\ x{\isadigit{1}}{\isacharparenright}{\kern0pt}{\isachardoublequoteclose}\isanewline
\ \ \isacommand{by}\isamarkupfalse%
\ {\isacharparenleft}{\kern0pt}intro\ fbox{\isacharunderscore}{\kern0pt}invs{\isacharsemicolon}{\kern0pt}\ diff{\isacharunderscore}{\kern0pt}inv{\isacharunderscore}{\kern0pt}on{\isacharunderscore}{\kern0pt}eq{\isacharparenright}
\isanewline
\end{isabellebody}

In the proof above, the application of the lemma 
\isa{fbox{\isacharunderscore}{\kern0pt}invs} as an introduction rule splits
the Hoare-triple in three invariant statements, one for each conjunct. The
semicolon indicates to Isabelle that the subsequent tactic should be applied
to all emerging proof obligations. Therefore, our tactic
\isa{diff{\isacharunderscore}{\kern0pt}inv{\isacharunderscore}{\kern0pt}on{\isacharunderscore}{\kern0pt}eq} implementing differential induction for equalities~(\ref{eq:dinv-eq}) is 
applied to each of the emerging invariant statements, which concludes 
the proof.

Nevertheless, we can also tackle this problem directly via the flow. Despite
the fact that we suspect that the solutions involve trigonometric functions,
obtaining the general solution is time-consuming. Therefore, we call our 
integration between the Wolfram language and Isabelle/HOL
\isa{\textbf{find{\isacharunderscore}local{\isacharunderscore}flow}} to
provide the solution for us. We can use the supplied solution in the proof.

\begin{isabellebody}\isanewline
\isacommand{lemma}\isamarkupfalse%
\ {\isachardoublequoteopen}w\ {\isasymnoteq}\ {\isadigit{0}}\ {\isasymLongrightarrow}\ {\isacharparenleft}{\kern0pt}d{\isadigit{1}}\isactrlsup {\isadigit{2}}\ {\isacharplus}{\kern0pt}\ d{\isadigit{2}}\isactrlsup {\isadigit{2}}\ {\isacharequal}{\kern0pt}\ w\isactrlsup {\isadigit{2}}\ {\isacharasterisk}{\kern0pt}\ p\isactrlsup {\isadigit{2}}\ {\isasymand}\ d{\isadigit{1}}\ {\isacharequal}{\kern0pt}\ {\isacharminus}{\kern0pt}\ w\ {\isacharasterisk}{\kern0pt}\ x{\isadigit{2}}\ {\isasymand}\ d{\isadigit{2}}\ {\isacharequal}{\kern0pt}\ w\ {\isacharasterisk}{\kern0pt}\ x{\isadigit{1}}{\isacharparenright}{\kern0pt}\isactrlsub e\ {\isasymle}\isanewline
\ \ {\isacharbar}{\kern0pt}{\isacharbraceleft}{\kern0pt}x{\isadigit{1}}{\isacharbackquote}{\kern0pt}\ {\isacharequal}{\kern0pt}\ d{\isadigit{1}}{\isacharcomma}{\kern0pt}\ x{\isadigit{2}}{\isacharbackquote}{\kern0pt}\ {\isacharequal}{\kern0pt}\ d{\isadigit{2}}{\isacharcomma}{\kern0pt}\ d{\isadigit{1}}{\isacharbackquote}{\kern0pt}\ {\isacharequal}{\kern0pt}\ {\isacharminus}{\kern0pt}\ w\ {\isacharasterisk}{\kern0pt}\ d{\isadigit{2}}{\isacharcomma}{\kern0pt}\ d{\isadigit{2}}{\isacharbackquote}{\kern0pt}\ {\isacharequal}{\kern0pt}\ w\ {\isacharasterisk}{\kern0pt}\ d{\isadigit{1}}{\isacharbraceright}{\kern0pt}{\isacharbrackright}{\kern0pt}\ \isanewline
\ \ {\isacharparenleft}{\kern0pt}d{\isadigit{1}}\isactrlsup {\isadigit{2}}\ {\isacharplus}{\kern0pt}\ d{\isadigit{2}}\isactrlsup {\isadigit{2}}\ {\isacharequal}{\kern0pt}\ w\isactrlsup {\isadigit{2}}\ {\isacharasterisk}{\kern0pt}\ p\isactrlsup {\isadigit{2}}\ {\isasymand}\ d{\isadigit{1}}\ {\isacharequal}{\kern0pt}\ {\isacharminus}{\kern0pt}\ w\ {\isacharasterisk}{\kern0pt}\ x{\isadigit{2}}\ {\isasymand}\ d{\isadigit{2}}\ {\isacharequal}{\kern0pt}\ w\ {\isacharasterisk}{\kern0pt}\ x{\isadigit{1}}{\isacharparenright}{\kern0pt}{\isachardoublequoteclose}\isanewline
\ \ \textbf{find{\isacharunderscore}local{\isacharunderscore}flow}\isanewline
\ \ \isacommand{apply}\isamarkupfalse%
\ {\isacharparenleft}{\kern0pt}wlp{\isacharunderscore}{\kern0pt}solve\ {\isachardoublequoteopen}{\isasymlambda}t{\isachardot}{\kern0pt}\ {\isacharbrackleft}{\kern0pt}d{\isadigit{1}}\ {\isasymleadsto}\ {\isachardollar}{\kern0pt}d{\isadigit{1}}\ {\isacharasterisk}{\kern0pt}\ cos\ {\isacharparenleft}{\kern0pt}t\ {\isacharasterisk}{\kern0pt}\ w{\isacharparenright}{\kern0pt}\ {\isacharplus}{\kern0pt}\ {\isacharminus}{\kern0pt}\ {\isadigit{1}}\ {\isacharasterisk}{\kern0pt}\ {\isachardollar}{\kern0pt}d{\isadigit{2}}\ {\isacharasterisk}{\kern0pt}\ sin\ {\isacharparenleft}{\kern0pt}t\ {\isacharasterisk}{\kern0pt}\ w{\isacharparenright}{\kern0pt}{\isacharcomma}{\kern0pt}\ \isanewline
\ \ \ \ d{\isadigit{2}}\ {\isasymleadsto}\ {\isachardollar}{\kern0pt}d{\isadigit{2}}\ {\isacharasterisk}{\kern0pt}\ cos\ {\isacharparenleft}{\kern0pt}t\ {\isacharasterisk}{\kern0pt}\ w{\isacharparenright}{\kern0pt}\ {\isacharplus}{\kern0pt}\ {\isachardollar}{\kern0pt}d{\isadigit{1}}\ {\isacharasterisk}{\kern0pt}\ sin\ {\isacharparenleft}{\kern0pt}t\ {\isacharasterisk}{\kern0pt}\ w{\isacharparenright}{\kern0pt}{\isacharcomma}{\kern0pt}\ \isanewline
\ \ \ \ x{\isadigit{1}}\ {\isasymleadsto}\ {\isachardollar}{\kern0pt}x{\isadigit{1}}\ {\isacharplus}{\kern0pt}\ {\isadigit{1}}{\isacharslash}{\kern0pt}w\ {\isacharasterisk}{\kern0pt}\ {\isachardollar}{\kern0pt}d{\isadigit{2}}\ {\isacharasterisk}{\kern0pt}\ {\isacharparenleft}{\kern0pt}{\isacharminus}{\kern0pt}\ {\isadigit{1}}\ {\isacharplus}{\kern0pt}\ cos\ {\isacharparenleft}{\kern0pt}t\ {\isacharasterisk}{\kern0pt}\ w{\isacharparenright}{\kern0pt}{\isacharparenright}{\kern0pt}\ {\isacharplus}{\kern0pt}\ {\isadigit{1}}{\isacharslash}{\kern0pt}w\ {\isacharasterisk}{\kern0pt}\ {\isachardollar}{\kern0pt}d{\isadigit{1}}\ {\isacharasterisk}{\kern0pt}\ sin\ {\isacharparenleft}{\kern0pt}t\ {\isacharasterisk}{\kern0pt}\ w{\isacharparenright}{\kern0pt}{\isacharcomma}{\kern0pt}\ \isanewline
\ \ \ \ x{\isadigit{2}}\ {\isasymleadsto}\ {\isachardollar}{\kern0pt}x{\isadigit{2}}\ {\isacharplus}{\kern0pt}\ {\isadigit{1}}{\isacharslash}{\kern0pt}w\ {\isacharasterisk}{\kern0pt}\ {\isachardollar}{\kern0pt}d{\isadigit{1}}\ {\isacharasterisk}{\kern0pt}\ {\isacharparenleft}{\kern0pt}{\isadigit{1}}\ {\isacharplus}{\kern0pt}\ {\isacharminus}{\kern0pt}\ cos\ {\isacharparenleft}{\kern0pt}t\ {\isacharasterisk}{\kern0pt}\ w{\isacharparenright}{\kern0pt}{\isacharparenright}{\kern0pt}\ {\isacharplus}{\kern0pt}\ {\isadigit{1}}{\isacharslash}{\kern0pt}w\ {\isacharasterisk}{\kern0pt}\ {\isachardollar}{\kern0pt}d{\isadigit{2}}\ {\isacharasterisk}{\kern0pt}\ sin\ {\isacharparenleft}{\kern0pt}t\ {\isacharasterisk}{\kern0pt}\ w{\isacharparenright}{\kern0pt}{\isacharbrackright}{\kern0pt}{\isachardoublequoteclose}{\isacharparenright}{\kern0pt}\isanewline
\ \ \isacommand{apply}\isamarkupfalse%
\ {\isacharparenleft}{\kern0pt}expr{\isacharunderscore}{\kern0pt}auto\ add{\isacharcolon}{\kern0pt}\ le{\isacharunderscore}{\kern0pt}fun{\isacharunderscore}{\kern0pt}def\ field{\isacharunderscore}{\kern0pt}simps{\isacharparenright}{\kern0pt}\isanewline
\ \ \isacommand{subgoal}\isamarkupfalse%
\ \isakeyword{for}\ s\ t\isanewline
\ \ \ \ \isacommand{apply}\isamarkupfalse%
\ mon{\isacharunderscore}{\kern0pt}pow{\isacharunderscore}{\kern0pt}simp\isanewline
\ \ \ \ \isacommand{apply}\isamarkupfalse%
\ {\isacharparenleft}{\kern0pt}mon{\isacharunderscore}{\kern0pt}simp{\isacharunderscore}{\kern0pt}vars\ {\isachardoublequoteopen}get\isactrlbsub x{\isadigit{1}}\isactrlesub \ s{\isachardoublequoteclose}\ {\isachardoublequoteopen}get\isactrlbsub x{\isadigit{2}}\isactrlesub \ s{\isachardoublequoteclose}\ {\isacharparenright}{\kern0pt}\isanewline
\ \ \ \ \isacommand{using}\isamarkupfalse%
\ rotational{\isacharunderscore}{\kern0pt}dynamics{\isadigit{3}}{\isacharunderscore}{\kern0pt}arith\ \isacommand{by}\isamarkupfalse%
\ force\isanewline
\ \ \isacommand{done}\isanewline
\end{isabellebody}

The first line of the proof applies our tactic 
\isa{wlp{\isacharunderscore}{\kern0pt}solve} with the suggested solution
from \isa{\textbf{find{\isacharunderscore}local{\isacharunderscore}flow}}.
The solution's syntax specifies one expression per variable in the 
system of ODEs. As explained in Section~\ref{sec:tactics}, the tactic
\isa{wlp{\isacharunderscore}{\kern0pt}solve} applies the 
rule~(\ref{eq:wlp-flow}), certifies that the corresponding vector field
is Lipschitz-continuous by calling our tactic 
\isa{c{\isadigit{1}}{\isacharunderscore}{\kern0pt}lipschitz},
and certifies that it is indeed the solution to the system of ODEs via our
tactic \isa{vderiv}. The \isa{\isacommand{subgoal}} command allows us
to specify the name of the variables in the proof obligation so that we can
use those names in our subsequent tactics. Our new tactic 
\isa{mon{\isacharunderscore}{\kern0pt}pow{\isacharunderscore}{\kern0pt}simp}
simplifies powers in the monomial expressions of the proof obligation.
In the next line, our tactic 
\isa{mon{\isacharunderscore}{\kern0pt}simp{\isacharunderscore}{\kern0pt}vars}
calls \isa{mon{\isacharunderscore}{\kern0pt}pow{\isacharunderscore}{\kern0pt}simp}
twice but reorders factors in the order of its inputs 
\isa{get\isactrlbsub x{\isadigit{1}}\isactrlesub \ s} and 
\isa{get\isactrlbsub x{\isadigit{2}}\isactrlesub \ s}. The last line supplies
the lemma \isa{rotational{\isacharunderscore}{\kern0pt}dynamics{\isadigit{3}}{\isacharunderscore}{\kern0pt}arith}
(below) whose proof was provided by Sledgehammer~\cite{PaulsonB10}.

\begin{isabellebody}
\isanewline
\isacommand{lemma}\isamarkupfalse%
\ rotational{\isacharunderscore}{\kern0pt}dynamics{\isadigit{3}}{\isacharunderscore}{\kern0pt}arith{\isacharcolon}{\kern0pt}\ \isanewline
\ \ {\isachardoublequoteopen}w\isactrlsup {\isadigit{2}}\ {\isacharasterisk}{\kern0pt}\ {\isacharparenleft}{\kern0pt}get\isactrlbsub x{\isadigit{1}}\isactrlesub \ s{\isacharparenright}{\kern0pt}\isactrlsup {\isadigit{2}}\ {\isacharplus}{\kern0pt}\ w\isactrlsup {\isadigit{2}}\ {\isacharasterisk}{\kern0pt}\ {\isacharparenleft}{\kern0pt}get\isactrlbsub x{\isadigit{2}}\isactrlesub \ s{\isacharparenright}{\kern0pt}\isactrlsup {\isadigit{2}}\ {\isacharequal}{\kern0pt}\ p\isactrlsup {\isadigit{2}}\ {\isacharasterisk}{\kern0pt}\ w\isactrlsup {\isadigit{2}}\ \isanewline
\ \ {\isasymLongrightarrow}\ w\isactrlsup {\isadigit{2}}\ {\isacharasterisk}{\kern0pt}\ {\isacharparenleft}{\kern0pt}{\isacharparenleft}{\kern0pt}cos\ {\isacharparenleft}{\kern0pt}t\ {\isacharasterisk}{\kern0pt}\ w{\isacharparenright}{\kern0pt}{\isacharparenright}{\kern0pt}\isactrlsup {\isadigit{2}}\ {\isacharasterisk}{\kern0pt}\ {\isacharparenleft}{\kern0pt}get\isactrlbsub x{\isadigit{1}}\isactrlesub \ s{\isacharparenright}{\kern0pt}\isactrlsup {\isadigit{2}}{\isacharparenright}{\kern0pt}\ \isanewline
\ \ \ \ {\isacharplus}{\kern0pt}\ {\isacharparenleft}{\kern0pt}w\isactrlsup {\isadigit{2}}\ {\isacharasterisk}{\kern0pt}\ {\isacharparenleft}{\kern0pt}{\isacharparenleft}{\kern0pt}sin\ {\isacharparenleft}{\kern0pt}t\ {\isacharasterisk}{\kern0pt}\ w{\isacharparenright}{\kern0pt}{\isacharparenright}{\kern0pt}\isactrlsup {\isadigit{2}}\ {\isacharasterisk}{\kern0pt}\ {\isacharparenleft}{\kern0pt}get\isactrlbsub x{\isadigit{1}}\isactrlesub \ s{\isacharparenright}{\kern0pt}\isactrlsup {\isadigit{2}}{\isacharparenright}{\kern0pt}\ \isanewline
\ \ \ \ {\isacharplus}{\kern0pt}\ {\isacharparenleft}{\kern0pt}w\isactrlsup {\isadigit{2}}\ {\isacharasterisk}{\kern0pt}\ {\isacharparenleft}{\kern0pt}{\isacharparenleft}{\kern0pt}cos\ {\isacharparenleft}{\kern0pt}t\ {\isacharasterisk}{\kern0pt}\ w{\isacharparenright}{\kern0pt}{\isacharparenright}{\kern0pt}\isactrlsup {\isadigit{2}}\ {\isacharasterisk}{\kern0pt}\ {\isacharparenleft}{\kern0pt}get\isactrlbsub x{\isadigit{2}}\isactrlesub \ s{\isacharparenright}{\kern0pt}\isactrlsup {\isadigit{2}}{\isacharparenright}{\kern0pt}\ \isanewline
\ \ \ \ {\isacharplus}{\kern0pt}\ w\isactrlsup {\isadigit{2}}\ {\isacharasterisk}{\kern0pt}\ {\isacharparenleft}{\kern0pt}{\isacharparenleft}{\kern0pt}sin\ {\isacharparenleft}{\kern0pt}t\ {\isacharasterisk}{\kern0pt}\ w{\isacharparenright}{\kern0pt}{\isacharparenright}{\kern0pt}\isactrlsup {\isadigit{2}}\ {\isacharasterisk}{\kern0pt}\ {\isacharparenleft}{\kern0pt}get\isactrlbsub x{\isadigit{2}}\isactrlesub \ s{\isacharparenright}{\kern0pt}\isactrlsup {\isadigit{2}}{\isacharparenright}{\kern0pt}{\isacharparenright}{\kern0pt}{\isacharparenright}{\kern0pt}\ {\isacharequal}{\kern0pt}\ p\isactrlsup {\isadigit{2}}\ {\isacharasterisk}{\kern0pt}\ w\isactrlsup {\isadigit{2}}{\isachardoublequoteclose}\isanewline
\ \ $\langle proof\rangle$\isanewline
\end{isabellebody}

This example shows the versatility of using general-purpose proof assistants
for hybrid systems verification. Users can provide automation methods for
invariant or flow certification. This allows the integration of unverified 
tools into the verification process. Fast certification enables ITPs to quickly validate CAS' inputs. In terms of our contributions, 
this example showcases our shallow embedding's intuitive syntax, our integration of the Wolfram language for suggesting simple
solutions, and our tactics automating VCG,
C1-lipschitz continuity, derivatives certification, and real-arithmetic reasoning.

%%%%%%%%%%%%%% VCG-INVAR %%%%%%%%%%%%%%%
% Example where solutions are complicated but we use invariants
\subsection{Dynamics: Conserved quantity}\label{sec:longsol} We prove that
the inequality $I\Leftrightarrow x_1^4\cdot x_2^2 + x_1^2\cdot x_2^4 - 3\cdot x_1^2\cdot x_2^2 + 1 \leq c$
is an invariant of the system
\begin{equation*}
f=
\begin{cases}
x_1' = 2\cdot x_1^4\cdot x_2 + 4\cdot x_1^2\cdot x_2^3 - 6\cdot x_1^2\cdot x_2, \\
x_2' = -4\cdot x_1^3\cdot x_2^2 - 2\cdot x_1\cdot x_2^4 + 6\cdot x_1\cdot x_2^2,
\end{cases}
\end{equation*}
where we abuse notation and ``equate'' the vector field with its representation as a
system of ODEs. In contrast with the previous benchmark, the solution to this system
of ODEs is not easy to describe analytically. The following is a subexpression
to the solution for $x_2$ according to Wolfram$\mid$Alpha
\begin{equation*}
\left(\int_1^t\frac{1}{\sqrt{\tau^2(\tau^6-6\tau^4+9\tau^2+c_1)}\sqrt{\frac{-\tau^4+3\tau^2-\sqrt{\tau^2(\tau^6-6\tau^4+9\tau^2+c_1)}}{\tau^2}}}d\tau\right)^{-1}_.
\end{equation*}
The solution for $x_2$ involves another four factors with integrals of fractions 
with denominators having square roots of square roots. Instead of computing 
these solutions and certifying them in Isabelle, we perform differential induction. 
We show that $I$ is an invariant by proving that the framed Fr\'echet 
derivatives on both sides of the inequality are $0$. In Isabelle/HOL, certification of 
this reasoning is automatic due to our \emph{dInduct{\isacharunderscore}{\kern0pt}mega} tactic.

\begin{isabellebody}
\isanewline
\isacommand{lemma}\isamarkupfalse%
\ {\isachardoublequoteopen}{\isacharparenleft}{\kern0pt}x{\isadigit{1}}\isactrlsup {\isadigit{4}}{\isacharasterisk}{\kern0pt}x{\isadigit{2}}\isactrlsup {\isadigit{2}}\ {\isacharplus}{\kern0pt}\ x{\isadigit{1}}\isactrlsup {\isadigit{2}}{\isacharasterisk}{\kern0pt}x{\isadigit{2}}\isactrlsup {\isadigit{4}}\ {\isacharminus}{\kern0pt}\ {\isadigit{3}}{\isacharasterisk}{\kern0pt}x{\isadigit{1}}\isactrlsup {\isadigit{2}}{\isacharasterisk}{\kern0pt}x{\isadigit{2}}\isactrlsup {\isadigit{2}}\ {\isacharplus}{\kern0pt}\ {\isadigit{1}}\ {\isasymle}\ c{\isacharparenright}{\kern0pt}\isactrlsub e\ {\isasymle}\isanewline
\ \ {\isacharbar}{\kern0pt}{\isacharbraceleft}{\kern0pt}x{\isadigit{1}}{\isacharbackquote}{\kern0pt}\ {\isacharequal}{\kern0pt}\ {\isadigit{2}}{\isacharasterisk}{\kern0pt}x{\isadigit{1}}\isactrlsup {\isadigit{4}}{\isacharasterisk}{\kern0pt}x{\isadigit{2}}\ {\isacharplus}{\kern0pt}\ {\isadigit{4}}{\isacharasterisk}{\kern0pt}x{\isadigit{1}}\isactrlsup {\isadigit{2}}{\isacharasterisk}{\kern0pt}x{\isadigit{2}}\isactrlsup {\isadigit{3}}\ {\isacharminus}{\kern0pt}\ {\isadigit{6}}{\isacharasterisk}{\kern0pt}x{\isadigit{1}}\isactrlsup {\isadigit{2}}{\isacharasterisk}{\kern0pt}x{\isadigit{2}}{\isacharcomma}{\kern0pt}
\isanewline\ \ \ \ x{\isadigit{2}}{\isacharbackquote}{\kern0pt}\ {\isacharequal}{\kern0pt}\ {\isacharminus}{\kern0pt}{\isadigit{4}}{\isacharasterisk}{\kern0pt}x{\isadigit{1}}\isactrlsup {\isadigit{3}}{\isacharasterisk}{\kern0pt}x{\isadigit{2}}\isactrlsup {\isadigit{2}}\ {\isacharminus}{\kern0pt}\ {\isadigit{2}}{\isacharasterisk}{\kern0pt}x{\isadigit{1}}{\isacharasterisk}{\kern0pt}x{\isadigit{2}}\isactrlsup {\isadigit{4}}\ {\isacharplus}{\kern0pt}\ {\isadigit{6}}{\isacharasterisk}{\kern0pt}x{\isadigit{1}}{\isacharasterisk}{\kern0pt}x{\isadigit{2}}\isactrlsup {\isadigit{2}}{\isacharbraceright}{\kern0pt}{\isacharbrackright}{\kern0pt}\isanewline
\ \ {\isacharparenleft}{\kern0pt}x{\isadigit{1}}\isactrlsup {\isadigit{4}}{\isacharasterisk}{\kern0pt}x{\isadigit{2}}\isactrlsup {\isadigit{2}}\ {\isacharplus}{\kern0pt}\ x{\isadigit{1}}\isactrlsup {\isadigit{2}}{\isacharasterisk}{\kern0pt}x{\isadigit{2}}\isactrlsup {\isadigit{4}}\ {\isacharminus}{\kern0pt}\ {\isadigit{3}}{\isacharasterisk}{\kern0pt}x{\isadigit{1}}\isactrlsup {\isadigit{2}}{\isacharasterisk}{\kern0pt}x{\isadigit{2}}\isactrlsup {\isadigit{2}}\ {\isacharplus}{\kern0pt}\ {\isadigit{1}}\ {\isasymle}\ c{\isacharparenright}{\kern0pt}{\isachardoublequoteclose}\isanewline
\ \ \isacommand{by}\isamarkupfalse%
\ dInduct{\isacharunderscore}{\kern0pt}mega%
\isanewline
\end{isabellebody}

Although the problem is simpler through differential induction, readers 
should be aware that the simplicity of proving this invariance in Isabelle
benefits greatly from our increased automation. We describe below the automated
steps to conclude the proof.
\begin{enumerate}
\item The inequality is transformed into a proof of invariance 
$\diffinv\, (x_1\lplus x_2)\, f\, \top\, \reals_+\, 0\, I$
\item Backward reasoning with (\ref{eq:dinv-leq}) requires 
showing $\fderiv{f}{e}{x_1\lplus x_2}\leq\fderiv{f}{c}{x_1\lplus x_2}$
\item The right hand side reduces to $0$ while the simplifier performs the
 following rewrites on the left-hand side 
 $e=x_1^4\cdot x_2^2 + x_1^2\cdot x_2^4 - 3\cdot x_1^2\cdot x_2^2 + 1$ (abbreviating $x_i'=\fderiv{f}{x_i}{x_1\lplus x_2}$)
\begin{align*}
\fderiv{f}{e}{x_1\lplus x_2}
&= 4\cdot x_1^3\cdot x_2^2\cdot x_1' + 2\cdot x_1^4\cdot x_2\cdot x_2' + 2\cdot x_1\cdot x_2^4\cdot x_1'\\
&+ 4\cdot x_1^2\cdot x_2^3\cdot x_2' - 6\cdot x_1\cdot x_2^2\cdot x_1' - 6\cdot x_1^2\cdot x_2\cdot x_2'\\
&=8\cdot x_1^7\cdot x_2^3 + 16\cdot x_1^5\cdot x_2^5 - 24\cdot x_1^5\cdot x_2^3 
 - 8\cdot x_1^7\cdot x_2^3 - 4\cdot x_1^5\cdot x_2^5\\
&+ 12\cdot x_1^5\cdot x_2^3 + 4\cdot x_1^5\cdot x_2^5 + 8\cdot x_1^3\cdot x_2^7 - 12\cdot x_1^3\cdot x_2^5 
 - 16\cdot x_1^5\cdot x_2^5\\
&- 8\cdot x_1^3\cdot x_2^7 + 24\cdot x_1^3\cdot x_2^5 - 12\cdot x_1^5\cdot x_2^3 - 24\cdot x_1^3\cdot x_2^5\\
&+ 36\cdot x_1^3\cdot x_2^3 
 + 24\cdot x_1^5\cdot x_2^3 + 12\cdot x_1^3\cdot x_2^5 - 36\cdot x_1^3\cdot x_2^3\\
& = 0
\end{align*}
\item Since $\fderiv{f}{e}{x_1\lplus x_2}=0\leq 0 = \fderiv{f}{c}{x_1\lplus x_2}$, 
the proof ends satisfactorily.
\end{enumerate}
Thus, our tactic \isa{dInduct{\isacharunderscore}{\kern0pt}mega} hides 
various logical, algebraic, differential and numerical computations and 
certifications. This example showcases the scalability of hybrid systems
verifications in interactive theorem provers. Namely, with adequate tactic
implementations, ITPs become more automated tools and easier to use
over time.

%%%%%%%%%% REACHABILITY ROCKET %%%%%%%%%%%%
% Example where we certify a forward diamond
\subsection{Reachability of a rocket launch}\label{sec:rocket}
\begin{figure}
%\vspace*{-1.8em}
\begin{center}
\includegraphics[scale=0.3]{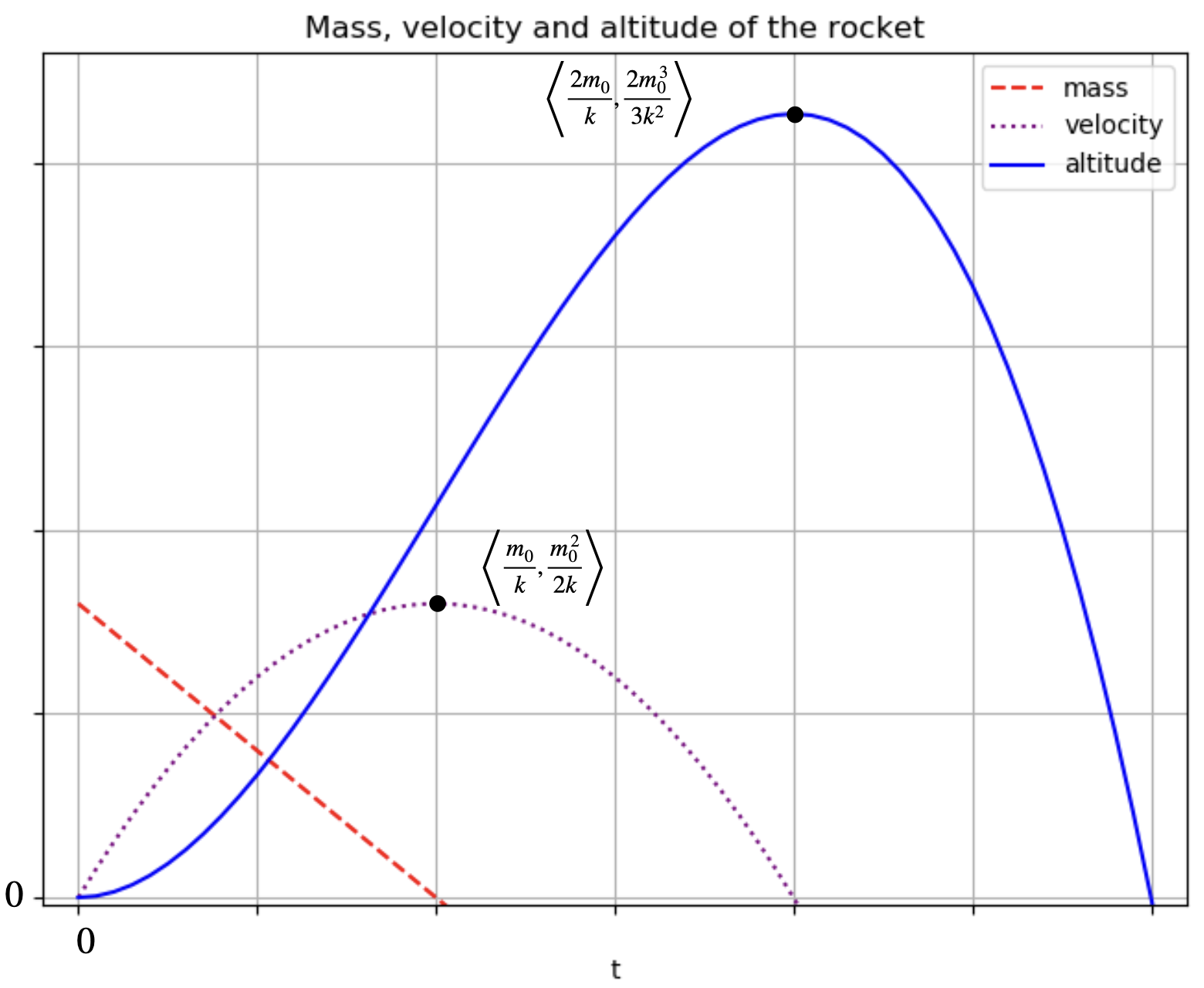} % trim=0 20 0 30 % left bottom right top
\vspace*{-1ex}
\caption{\centering Depiction of the rocket behaviour assuming $m_0>k$}
\label{fig:rocket}
\vspace*{-3ex}
\end{center}
\end{figure}

Consider a rocket's vertical liftoff and assume it loses fuel at a constant 
rate of $k>0$ kilograms per second starting with $m_0>k$ kilograms 
while its acceleration is equal to the amount of fuel left in it. These 
assumptions do not accurately model rockets' liftoff; however, they suffice to 
produce a behaviour approximating the observed phenomena, see 
Figure~\ref{fig:rocket}, and facilitate the presentation of our contributions. 
The corresponding system of ODEs and its solution are\\

\begin{minipage}{0.3\textwidth}
\begin{equation*}
f = \begin{cases}
m' = - k,\\
v' = m,\\
y'= v,\\ 
t' = 1, 
\end{cases}
\end{equation*}
\end{minipage}
\begin{minipage}{0.59\textwidth}\vspace{-2ex}
\begin{equation*}
\varphi\, \tau=\begin{cases}
m\, \tau= - k\cdot \tau + m_0,\\
v\, \tau = - k\cdot \frac{\tau^2}{2} + m_0\cdot \tau = \tau \cdot (- k\cdot\frac{\tau}{2} + m_0),\\
y\, \tau= - k\cdot \frac{\tau^3}{6} + m_0\cdot \frac{\tau^2}{2}= \frac{\tau^2}{2} \cdot (- k\cdot\frac{\tau}{3} + m_0),\\ 
t\, \tau = \tau,
\end{cases}
\end{equation*}
\end{minipage}\\

\noindent where $m$ is the fuel's mass, $v$ is the rocket's
velocity, $y$ is its altitude, and $t$ models time. 

We study the rocket's behaviour before it reaches a maximum altitude 
with its first-stage propulsion because the second stage should begin 
before the rocket starts falling. We therefore prove two things 
about the initial propulsion stage. The first is that no matter which 
height $h$ we consider strictly below the maximum altitude 
$H=2m_0^3/(3k^2)$, there will always be a state of the rocket 
greater than $h$ approaching $H$. The second is that all scenarios with 
a single propulsion from the ground lead to altitudes lower than $H$. 
Using the abbreviations ``\isa{odes}'' for $f$ and ``\isa{flow}'' for $\varphi$, the verification of the 
first specification is now possible thanks to our implementation of forward 
diamonds.
\begin{isabellebody}
\isanewline
\isacommand{lemma}\isamarkupfalse%
\ local{\isacharunderscore}{\kern0pt}flow{\isacharunderscore}{\kern0pt}on{\isacharunderscore}{\kern0pt}rocket{\isacharcolon}{\kern0pt}\isanewline
\ \ {\isachardoublequoteopen}local{\isacharunderscore}{\kern0pt}flow{\isacharunderscore}{\kern0pt}on\ {\isacharbrackleft}{\kern0pt}y{\isasymleadsto}{\isachardollar}{\kern0pt}v{\isacharcomma}{\kern0pt}\ v{\isasymleadsto}{\isachardollar}{\kern0pt}m{\isacharcomma}{\kern0pt}\ t{\isasymleadsto}{\isadigit{1}}{\isacharcomma}{\kern0pt}\ m{\isasymleadsto}{\isacharminus}{\kern0pt}k{\isacharbrackright}{\kern0pt}\ {\isacharparenleft}{\kern0pt}y{\isacharplus}{\kern0pt}\isactrlsub Lv{\isacharplus}{\kern0pt}\isactrlsub Lt{\isacharplus}{\kern0pt}\isactrlsub Lm{\isacharparenright}{\kern0pt}\ UNIV\ UNIV\ flow{\isachardoublequoteclose}\isanewline
\ \ \isacommand{by}\isamarkupfalse%
\ local{\isacharunderscore}{\kern0pt}flow{\isacharunderscore}{\kern0pt}on{\isacharunderscore}{\kern0pt}auto%

\isanewline
\isacommand{lemma}\isamarkupfalse%
\ {\isachardoublequoteopen}{\isacharparenleft}{\kern0pt}{\isadigit{0}}{\isasymle}h\ {\isasymand}\ h{\isacharless}{\kern0pt}H\ {\isasymand}\ m{\isacharequal}{\kern0pt}m\isactrlsub {\isadigit{0}}\ {\isasymand}\ m\isactrlsub {\isadigit{0}}{\isachargreater}{\kern0pt}k\ {\isasymand}\ t{\isacharequal}{\kern0pt}{\isadigit{0}}\ {\isasymand}\ v{\isacharequal}{\kern0pt}{\isadigit{0}}\ {\isasymand}\ y{\isacharequal}{\kern0pt}{\isadigit{0}}{\isacharparenright}{\kern0pt}\isactrlsub e\ {\isasymle}\ {\isacharbar}{\kern0pt}odes{\isasymrangle}\ {\isacharparenleft}{\kern0pt}h{\isasymle}y{\isacharparenright}{\kern0pt}{\isachardoublequoteclose}\isanewline
\ \ \isacommand{using}\isamarkupfalse%
\ k{\isacharunderscore}{\kern0pt}ge{\isacharunderscore}{\kern0pt}{\isadigit{1}}\isanewline
\ \ \isacommand{by}\isamarkupfalse%
\ {\isacharparenleft}{\kern0pt}subst\ fdia{\isacharunderscore}{\kern0pt}g{\isacharunderscore}{\kern0pt}ode{\isacharunderscore}{\kern0pt}frame{\isacharunderscore}{\kern0pt}flow{\isacharbrackleft}{\kern0pt}OF\ local{\isacharunderscore}{\kern0pt}flow{\isacharunderscore}{\kern0pt}on{\isacharunderscore}{\kern0pt}rocket{\isacharbrackright}{\kern0pt}{\isacharsemicolon}{\kern0pt}\ expr{\isacharunderscore}{\kern0pt}simp{\isacharparenright}{\kern0pt}\isanewline
\ \ \ \ {\isacharparenleft}{\kern0pt}auto\ simp{\isacharcolon}{\kern0pt}\ field{\isacharunderscore}{\kern0pt}simps\ power{\isadigit{3}}{\isacharunderscore}{\kern0pt}eq{\isacharunderscore}{\kern0pt}cube\ intro{\isacharbang}{\kern0pt}{\isacharcolon}{\kern0pt}\ exI{\isacharbrackleft}{\kern0pt}of\ {\isacharunderscore}{\kern0pt}\ {\isachardoublequoteopen}{\isadigit{2}}{\isacharasterisk}{\kern0pt}m\isactrlsub {\isadigit{0}}{\isacharslash}{\kern0pt}k{\isachardoublequoteclose}{\isacharbrackright}{\kern0pt}{\isacharparenright}{\kern0pt}%
\isanewline
\end{isabellebody}

The proof starts with the law~(\ref{eq:fdia-flow}) using the lemma 
\isa{local\_flow\_on\_rocket} asserting that $\varphi$ is the flow for 
$f$. The proof of this is now automatic due to our tactic 
\isa{local{\isacharunderscore}{\kern0pt}flow{\isacharunderscore}{\kern0pt}on{\isacharunderscore}{\kern0pt}auto}.
The second line of the proof for our specification culminates with some
arithmetical reasoning, where we provide the time $2m_0/k$ as the 
witness for the existential quantifier in the law (\ref{eq:fdia-flow}). That is,
 the time for the second $0$-intercept of the velocity when the maximum 
 altitude is reached.
 
 The second specification looks equally simple and its three line proof is
 deceiving (see below). Our contributed tactics automatically handle VCG, 
 derivative certifications, and uniqueness. However, as reported in
 Section~\ref{sec:eval}, emerging arithmetic proof obligations often have 
 to be checked separately. We do this with the lemma 
 \isa{rocket{\isacharunderscore}{\kern0pt}arith} proved using our
 derivative tests from Section~\ref{subsec:deriv-tests}.
 
\begin{isabellebody}
\isanewline
\isacommand{lemma}\isamarkupfalse%
\ {\isachardoublequoteopen}{\isacharparenleft}{\kern0pt}m\ {\isacharequal}{\kern0pt}\ m\isactrlsub {\isadigit{0}}\ {\isasymand}\ m\isactrlsub {\isadigit{0}}\ {\isachargreater}{\kern0pt}\ k\ {\isasymand}\ t\ {\isacharequal}{\kern0pt}\ {\isadigit{0}}\ {\isasymand}\ v\ {\isacharequal}{\kern0pt}\ {\isadigit{0}}\ {\isasymand}\ y\ {\isacharequal}{\kern0pt}\ {\isadigit{0}}{\isacharparenright}{\kern0pt}\isactrlsub e
\isanewline\ \  {\isasymle}\ {\isacharbar}{\kern0pt}ode{\isacharbrackright}{\kern0pt}\ {\isacharparenleft}{\kern0pt}y\ {\isasymle}\ {\isadigit{2}}{\isacharasterisk}{\kern0pt}m\isactrlsub {\isadigit{0}}\isactrlsup {\isadigit{3}}{\isacharslash}{\kern0pt}{\isacharparenleft}{\kern0pt}{\isadigit{3}}{\isacharasterisk}{\kern0pt}k\isactrlsup {\isadigit{2}}{\isacharparenright}{\kern0pt}{\isacharparenright}{\kern0pt}{\isachardoublequoteclose}\isanewline
\ \ \isacommand{apply}\isamarkupfalse%
\ {\isacharparenleft}{\kern0pt}wlp{\isacharunderscore}{\kern0pt}solve\ {\isachardoublequoteopen}flow{\isachardoublequoteclose}{\isacharparenright}{\kern0pt}\isanewline
\ \ \isacommand{using}\isamarkupfalse%
\ k{\isacharunderscore}{\kern0pt}ge{\isacharunderscore}{\kern0pt}{\isadigit{1}}\ rocket{\isacharunderscore}{\kern0pt}arith\isanewline
\ \ \isacommand{by}\isamarkupfalse%
\ {\isacharparenleft}{\kern0pt}expr{\isacharunderscore}{\kern0pt}simp\ add{\isacharcolon}{\kern0pt}\ le{\isacharunderscore}{\kern0pt}fun{\isacharunderscore}{\kern0pt}def{\isacharparenright}{\kern0pt}%
\isanewline

\isacommand{lemma}\isamarkupfalse%
\ rocket{\isacharunderscore}{\kern0pt}arith{\isacharcolon}{\kern0pt}\isanewline
\ \ \isakeyword{assumes}\ {\isachardoublequoteopen}{\isacharparenleft}{\kern0pt}k{\isacharcolon}{\kern0pt}{\isacharcolon}{\kern0pt}real{\isacharparenright}{\kern0pt}\ {\isachargreater}{\kern0pt}\ {\isadigit{1}}{\isachardoublequoteclose}\ \isakeyword{and}\ {\isachardoublequoteopen}m\isactrlsub {\isadigit{0}}\ {\isachargreater}{\kern0pt}\ k{\isachardoublequoteclose}\ \isakeyword{and}\ {\isachardoublequoteopen}x\ {\isasymin}\ {\isacharbraceleft}{\kern0pt}{\isadigit{0}}{\isachardot}{\kern0pt}{\isachardot}{\kern0pt}{\isacharbraceright}{\kern0pt}{\isachardoublequoteclose}\isanewline
\ \ \isakeyword{shows}\ {\isachardoublequoteopen}{\isacharminus}{\kern0pt}\ k{\isacharasterisk}{\kern0pt}x\isactrlsup {\isadigit{3}}{\isacharslash}{\kern0pt}{\isadigit{6}}\ {\isacharplus}{\kern0pt}\ m\isactrlsub {\isadigit{0}}{\isacharasterisk}{\kern0pt}x\isactrlsup {\isadigit{2}}{\isacharslash}{\kern0pt}{\isadigit{2}}\ {\isasymle}\ {\isadigit{2}}{\isacharasterisk}{\kern0pt}m\isactrlsub {\isadigit{0}}\isactrlsup {\isadigit{3}}{\isacharslash}{\kern0pt}{\isacharparenleft}{\kern0pt}{\isadigit{3}}{\isacharasterisk}{\kern0pt}k\isactrlsup {\isadigit{2}}{\isacharparenright}{\kern0pt}{\isachardoublequoteclose}\ {\isacharparenleft}{\kern0pt}\isakeyword{is}\ {\isachardoublequoteopen}{\isacharquery}{\kern0pt}lhs\ {\isasymle}\ {\isacharunderscore}{\kern0pt}{\isachardoublequoteclose}{\isacharparenright}{\kern0pt}\isanewline
\isacommand{proof}\isamarkupfalse%
{\isacharminus}{\kern0pt}\isanewline
\ \ \isacommand{let}\isamarkupfalse%
\ {\isacharquery}{\kern0pt}f\ {\isacharequal}{\kern0pt}\ {\isachardoublequoteopen}{\isasymlambda}t{\isachardot}{\kern0pt}\ {\isacharminus}{\kern0pt}k{\isacharasterisk}{\kern0pt}t\isactrlsup {\isadigit{3}}{\isacharslash}{\kern0pt}{\isadigit{6}}\ {\isacharplus}{\kern0pt}\ m\isactrlsub {\isadigit{0}}{\isacharasterisk}{\kern0pt}t\isactrlsup {\isadigit{2}}{\isacharslash}{\kern0pt}{\isadigit{2}}{\isachardoublequoteclose}\ \isanewline
\ \ \ \ \isakeyword{and}\ {\isacharquery}{\kern0pt}f{\isacharprime}{\kern0pt}\ {\isacharequal}{\kern0pt}\ {\isachardoublequoteopen}{\isasymlambda}t{\isachardot}{\kern0pt}\ {\isacharminus}{\kern0pt}k{\isacharasterisk}{\kern0pt}t\isactrlsup {\isadigit{2}}{\isacharslash}{\kern0pt}{\isadigit{2}}\ {\isacharplus}{\kern0pt}\ m\isactrlsub {\isadigit{0}}{\isacharasterisk}{\kern0pt}t{\isachardoublequoteclose}\isanewline
\ \ \ \ \isakeyword{and}\ {\isacharquery}{\kern0pt}f{\isacharprime}{\kern0pt}{\isacharprime}{\kern0pt}\ {\isacharequal}{\kern0pt}\ {\isachardoublequoteopen}{\isasymlambda}t{\isachardot}{\kern0pt}\ {\isacharminus}{\kern0pt}k{\isacharasterisk}{\kern0pt}t\ {\isacharplus}{\kern0pt}\ m\isactrlsub {\isadigit{0}}{\isachardoublequoteclose}\isanewline
\ \ \isacommand{have}\isamarkupfalse%
\ {\isachardoublequoteopen}{\isadigit{2}}{\isacharasterisk}{\kern0pt}m\isactrlsub {\isadigit{0}}\isactrlsup {\isadigit{3}}{\isacharslash}{\kern0pt}{\isacharparenleft}{\kern0pt}{\isadigit{3}}{\isacharasterisk}{\kern0pt}k\isactrlsup {\isadigit{2}}{\isacharparenright}{\kern0pt}\ {\isacharequal}{\kern0pt}\ {\isacharminus}{\kern0pt}k{\isacharasterisk}{\kern0pt}{\isacharparenleft}{\kern0pt}{\isadigit{2}}{\isacharasterisk}{\kern0pt}m\isactrlsub {\isadigit{0}}{\isacharslash}{\kern0pt}k{\isacharparenright}{\kern0pt}\isactrlsup {\isadigit{3}}{\isacharslash}{\kern0pt}{\isadigit{6}}\ {\isacharplus}{\kern0pt}\ m\isactrlsub {\isadigit{0}}{\isacharasterisk}{\kern0pt}{\isacharparenleft}{\kern0pt}{\isadigit{2}}{\isacharasterisk}{\kern0pt}m\isactrlsub {\isadigit{0}}{\isacharslash}{\kern0pt}k{\isacharparenright}{\kern0pt}\isactrlsup {\isadigit{2}}{\isacharslash}{\kern0pt}{\isadigit{2}}{\isachardoublequoteclose}\ {\isacharparenleft}{\kern0pt}\isakeyword{is}\ {\isachardoublequoteopen}{\isacharunderscore}{\kern0pt}\ {\isacharequal}{\kern0pt}\ {\isacharquery}{\kern0pt}rhs{\isachardoublequoteclose}{\isacharparenright}{\kern0pt}\isanewline
\ \ \ \ \isacommand{by}\isamarkupfalse%
\ {\isacharparenleft}{\kern0pt}auto\ simp{\isacharcolon}{\kern0pt}\ field{\isacharunderscore}{\kern0pt}simps\ power{\isadigit{3}}{\isacharunderscore}{\kern0pt}eq{\isacharunderscore}{\kern0pt}cube{\isacharparenright}{\kern0pt}\isanewline
\ \ \isacommand{moreover}\isamarkupfalse%
\ \isacommand{have}\isamarkupfalse%
\ {\isachardoublequoteopen}{\isacharquery}{\kern0pt}lhs\ {\isasymle}\ {\isacharquery}{\kern0pt}rhs{\isachardoublequoteclose}\isanewline
\ \ \isacommand{proof}\isamarkupfalse%
{\isacharparenleft}{\kern0pt}cases\ {\isachardoublequoteopen}x\ {\isasymle}\ {\isadigit{2}}\ {\isacharasterisk}{\kern0pt}\ m\isactrlsub {\isadigit{0}}\ {\isacharslash}{\kern0pt}\ k{\isachardoublequoteclose}{\isacharparenright}{\kern0pt}\isanewline
\ \ \ \ \isacommand{case}\isamarkupfalse%
\ True\isanewline
\ \ \ \ \isacommand{have}\isamarkupfalse%
\ ge{\isacharunderscore}{\kern0pt}{\isadigit{0}}{\isacharunderscore}{\kern0pt}left{\isacharcolon}{\kern0pt}\ {\isachardoublequoteopen}{\isadigit{0}}\ {\isasymle}\ y\ {\isasymLongrightarrow}\ y\ {\isasymle}\ m\isactrlsub {\isadigit{0}}{\isacharslash}{\kern0pt}k\ {\isasymLongrightarrow}\ {\isacharquery}{\kern0pt}f{\isacharprime}{\kern0pt}\ {\isadigit{0}}\ {\isasymle}\ {\isacharquery}{\kern0pt}f{\isacharprime}{\kern0pt}\ y{\isachardoublequoteclose}\ \isakeyword{for}\ y\isanewline
\ \ \ \ \ \ \isacommand{apply}\isamarkupfalse%
\ {\isacharparenleft}{\kern0pt}rule\ has{\isacharunderscore}{\kern0pt}vderiv{\isacharunderscore}{\kern0pt}mono{\isacharunderscore}{\kern0pt}test{\isacharparenleft}{\kern0pt}{\isadigit{1}}{\isacharparenright}{\kern0pt}{\isacharbrackleft}{\kern0pt}of\ {\isachardoublequoteopen}{\isacharbraceleft}{\kern0pt}{\isadigit{0}}{\isachardot}{\kern0pt}{\isachardot}{\kern0pt}m\isactrlsub {\isadigit{0}}{\isacharslash}{\kern0pt}k{\isacharbraceright}{\kern0pt}{\isachardoublequoteclose}\ {\isacharquery}{\kern0pt}f{\isacharprime}{\kern0pt}\ {\isacharquery}{\kern0pt}f{\isacharprime}{\kern0pt}{\isacharprime}{\kern0pt}\ {\isadigit{0}}{\isacharbrackright}{\kern0pt}{\isacharparenright}{\kern0pt}\isanewline
\ \ \ \ \ \ \isacommand{using}\isamarkupfalse%
\ {\isacartoucheopen}k\ {\isachargreater}{\kern0pt}\ {\isadigit{1}}{\isacartoucheclose}\ {\isacartoucheopen}m\isactrlsub {\isadigit{0}}\ {\isachargreater}{\kern0pt}\ k{\isacartoucheclose}\isanewline
\ \ \ \ \ \ \isacommand{by}\isamarkupfalse%
\ {\isacharparenleft}{\kern0pt}auto\ intro{\isacharbang}{\kern0pt}{\isacharcolon}{\kern0pt}\ vderiv{\isacharunderscore}{\kern0pt}intros\ simp{\isacharcolon}{\kern0pt}\ field{\isacharunderscore}{\kern0pt}simps{\isacharparenright}{\kern0pt}\isanewline
\ \ \ \ \isacommand{moreover}\isamarkupfalse%
\ \isacommand{have}\isamarkupfalse%
\ ge{\isacharunderscore}{\kern0pt}{\isadigit{0}}{\isacharunderscore}{\kern0pt}right{\isacharcolon}{\kern0pt}\ {\isachardoublequoteopen}m\isactrlsub {\isadigit{0}}{\isacharslash}{\kern0pt}k{\isasymle}y\ {\isasymLongrightarrow}\ y{\isasymle}{\isadigit{2}}{\isacharasterisk}{\kern0pt}m\isactrlsub {\isadigit{0}}{\isacharslash}{\kern0pt}k
\isanewline \ \ \ \ \ \ 
{\isasymLongrightarrow}\ {\isacharquery}{\kern0pt}f{\isacharprime}{\kern0pt}\ {\isacharparenleft}{\kern0pt}{\isadigit{2}}{\isacharasterisk}{\kern0pt}m\isactrlsub {\isadigit{0}}{\isacharslash}{\kern0pt}k{\isacharparenright}{\kern0pt}\ {\isasymle}\ {\isacharquery}{\kern0pt}f{\isacharprime}{\kern0pt}\ y{\isachardoublequoteclose}\ \isakeyword{for}\ y\isanewline
\ \ \ \ \ \ \isacommand{apply}\isamarkupfalse%
{\isacharparenleft}{\kern0pt}rule\ has{\isacharunderscore}{\kern0pt}vderiv{\isacharunderscore}{\kern0pt}mono{\isacharunderscore}{\kern0pt}test{\isacharparenleft}{\kern0pt}{\isadigit{2}}{\isacharparenright}\isanewline
\ \ \ \ \ \ \ \ {\kern0pt}{\isacharbrackleft}{\kern0pt}of\ {\isachardoublequoteopen}{\isacharbraceleft}{\kern0pt}m\isactrlsub {\isadigit{0}}{\isacharslash}{\kern0pt}k{\isachardot}{\kern0pt}{\isachardot}{\kern0pt}{\isadigit{2}}{\isacharasterisk}{\kern0pt}m\isactrlsub {\isadigit{0}}{\isacharslash}{\kern0pt}k{\isacharbraceright}{\kern0pt}{\isachardoublequoteclose}\ {\isacharquery}{\kern0pt}f{\isacharprime}{\kern0pt}\ {\isacharquery}{\kern0pt}f{\isacharprime}{\kern0pt}{\isacharprime}{\kern0pt}\ {\isacharunderscore}{\kern0pt}\ {\isachardoublequoteopen}{\isadigit{2}}{\isacharasterisk}{\kern0pt}m\isactrlsub {\isadigit{0}}{\isacharslash}{\kern0pt}k{\isachardoublequoteclose}{\isacharbrackright}{\kern0pt}{\isacharparenright}{\kern0pt}\isanewline
\ \ \ \ \ \ \isacommand{using}\isamarkupfalse%
\ {\isacartoucheopen}k\ {\isachargreater}{\kern0pt}\ {\isadigit{1}}{\isacartoucheclose}\ {\isacartoucheopen}m\isactrlsub {\isadigit{0}}\ {\isachargreater}{\kern0pt}\ k{\isacartoucheclose}\isanewline
\ \ \ \ \ \ \isacommand{by}\isamarkupfalse%
\ {\isacharparenleft}{\kern0pt}auto\ intro{\isacharbang}{\kern0pt}{\isacharcolon}{\kern0pt}\ vderiv{\isacharunderscore}{\kern0pt}intros\ simp{\isacharcolon}{\kern0pt}\ field{\isacharunderscore}{\kern0pt}simps{\isacharparenright}{\kern0pt}\isanewline
\ \ \ \ \isacommand{ultimately}\isamarkupfalse%
\ \isacommand{have}\isamarkupfalse%
\ ge{\isacharunderscore}{\kern0pt}{\isadigit{0}}{\isacharcolon}{\kern0pt}\ {\isachardoublequoteopen}{\isasymforall}y{\isasymin}{\isacharbraceleft}{\kern0pt}{\isadigit{0}}{\isachardot}{\kern0pt}{\isachardot}{\kern0pt}{\isadigit{2}}{\isacharasterisk}{\kern0pt}m\isactrlsub {\isadigit{0}}{\isacharslash}{\kern0pt}k{\isacharbraceright}{\kern0pt}{\isachardot}{\kern0pt}\ {\isadigit{0}}\ {\isasymle}\ {\isacharquery}{\kern0pt}f{\isacharprime}{\kern0pt}\ y{\isachardoublequoteclose}\isanewline
\ \ \ \ \ \ \isacommand{using}\isamarkupfalse%
\ {\isacartoucheopen}k\ {\isachargreater}{\kern0pt}\ {\isadigit{1}}{\isacartoucheclose}\ {\isacartoucheopen}m\isactrlsub {\isadigit{0}}\ {\isachargreater}{\kern0pt}\ k{\isacartoucheclose}\isanewline
\ \ \ \ \ \ \isacommand{by}\isamarkupfalse%
\ {\isacharparenleft}{\kern0pt}fastforce\ simp{\isacharcolon}{\kern0pt}\ field{\isacharunderscore}{\kern0pt}simps{\isacharparenright}{\kern0pt}\isanewline
\ \ \ \ \isacommand{show}\isamarkupfalse%
\ {\isacharquery}{\kern0pt}thesis\isanewline
\ \ \ \ \ \ \isacommand{apply}\isamarkupfalse%
\ {\isacharparenleft}{\kern0pt}rule\ has{\isacharunderscore}{\kern0pt}vderiv{\isacharunderscore}{\kern0pt}mono{\isacharunderscore}{\kern0pt}test{\isacharparenleft}{\kern0pt}{\isadigit{1}}{\isacharparenright}{\kern0pt}{\isacharbrackleft}{\kern0pt}of\ {\isachardoublequoteopen}{\isacharbraceleft}{\kern0pt}{\isadigit{0}}{\isachardot}{\kern0pt}{\isachardot}{\kern0pt}{\isadigit{2}}{\isacharasterisk}{\kern0pt}m\isactrlsub {\isadigit{0}}{\isacharslash}{\kern0pt}k{\isacharbraceright}{\kern0pt}{\isachardoublequoteclose}\ {\isacharquery}{\kern0pt}f\ {\isacharquery}{\kern0pt}f{\isacharprime}{\kern0pt}\ {\isacharunderscore}{\kern0pt}\ {\isachardoublequoteopen}{\isadigit{2}}{\isacharasterisk}{\kern0pt}m\isactrlsub {\isadigit{0}}{\isacharslash}{\kern0pt}k{\isachardoublequoteclose}{\isacharbrackright}{\kern0pt}{\isacharparenright}{\kern0pt}\isanewline
\ \ \ \ \ \ \isacommand{using}\isamarkupfalse%
\ ge{\isacharunderscore}{\kern0pt}{\isadigit{0}}\ True\ {\isacartoucheopen}x\ {\isasymin}\ {\isacharbraceleft}{\kern0pt}{\isadigit{0}}{\isachardot}{\kern0pt}{\isachardot}{\kern0pt}{\isacharbraceright}{\kern0pt}{\isacartoucheclose}\ {\isacartoucheopen}k\ {\isachargreater}{\kern0pt}\ {\isadigit{1}}{\isacartoucheclose}\ {\isacartoucheopen}m\isactrlsub {\isadigit{0}}\ {\isachargreater}{\kern0pt}\ k{\isacartoucheclose}\isanewline
\ \ \ \ \ \ \isacommand{by}\isamarkupfalse%
\ {\isacharparenleft}{\kern0pt}auto\ intro{\isacharbang}{\kern0pt}{\isacharcolon}{\kern0pt}\ vderiv{\isacharunderscore}{\kern0pt}intros\ simp{\isacharcolon}{\kern0pt}\ field{\isacharunderscore}{\kern0pt}simps{\isacharparenright}{\kern0pt}\isanewline
\ \ \isacommand{next}\isamarkupfalse%
\isanewline
\ \ \ \ \isacommand{case}\isamarkupfalse%
\ False\isanewline
\ \ \ \ \isacommand{have}\isamarkupfalse%
\ {\isachardoublequoteopen}{\isadigit{2}}{\isacharasterisk}{\kern0pt}m\isactrlsub {\isadigit{0}}{\isacharslash}{\kern0pt}k\ {\isasymle}\ y\ {\isasymLongrightarrow}\ {\isacharquery}{\kern0pt}f{\isacharprime}{\kern0pt}\ y\ {\isasymle}\ {\isacharquery}{\kern0pt}f{\isacharprime}{\kern0pt}\ {\isacharparenleft}{\kern0pt}{\isadigit{2}}{\isacharasterisk}{\kern0pt}m\isactrlsub {\isadigit{0}}{\isacharslash}{\kern0pt}k{\isacharparenright}{\kern0pt}{\isachardoublequoteclose}\ \isakeyword{for}\ y\isanewline
\ \ \ \ \ \ \isacommand{apply}\isamarkupfalse%
\ {\isacharparenleft}{\kern0pt}rule\ has{\isacharunderscore}{\kern0pt}vderiv{\isacharunderscore}{\kern0pt}mono{\isacharunderscore}{\kern0pt}test{\isacharparenleft}{\kern0pt}{\isadigit{2}}{\isacharparenright}{\kern0pt}{\isacharbrackleft}{\kern0pt}of\ {\isachardoublequoteopen}{\isacharbraceleft}{\kern0pt}m\isactrlsub {\isadigit{0}}{\isacharslash}{\kern0pt}k{\isachardot}{\kern0pt}{\isachardot}{\kern0pt}{\isacharbraceright}{\kern0pt}{\isachardoublequoteclose}\ {\isacharquery}{\kern0pt}f{\isacharprime}{\kern0pt}\ {\isacharquery}{\kern0pt}f{\isacharprime}{\kern0pt}{\isacharprime}{\kern0pt}{\isacharbrackright}{\kern0pt}{\isacharparenright}{\kern0pt}\isanewline
\ \ \ \ \ \ \isacommand{using}\isamarkupfalse%
\ {\isacartoucheopen}k\ {\isachargreater}{\kern0pt}\ {\isadigit{1}}{\isacartoucheclose}\ {\isacartoucheopen}m\isactrlsub {\isadigit{0}}\ {\isachargreater}{\kern0pt}\ k{\isacartoucheclose}\ \isacommand{by}\isamarkupfalse%
\ {\isacharparenleft}{\kern0pt}auto\ intro{\isacharbang}{\kern0pt}{\isacharcolon}{\kern0pt}\ vderiv{\isacharunderscore}{\kern0pt}intros\ simp{\isacharcolon}{\kern0pt}\ field{\isacharunderscore}{\kern0pt}simps{\isacharparenright}{\kern0pt}\isanewline
\ \ \ \ \isacommand{hence}\isamarkupfalse%
\ obs{\isacharcolon}{\kern0pt}\ {\isachardoublequoteopen}{\isasymforall}y{\isasymin}{\isacharbraceleft}{\kern0pt}{\isadigit{2}}{\isacharasterisk}{\kern0pt}m\isactrlsub {\isadigit{0}}{\isacharslash}{\kern0pt}k{\isachardot}{\kern0pt}{\isachardot}{\kern0pt}{\isacharbraceright}{\kern0pt}{\isachardot}{\kern0pt}\ {\isacharquery}{\kern0pt}f{\isacharprime}{\kern0pt}\ y\ {\isasymle}\ {\isadigit{0}}{\isachardoublequoteclose}\isanewline
\ \ \ \ \ \ \isacommand{using}\isamarkupfalse%
\ {\isacartoucheopen}k\ {\isachargreater}{\kern0pt}\ {\isadigit{1}}{\isacartoucheclose}\ {\isacartoucheopen}m\isactrlsub {\isadigit{0}}\ {\isachargreater}{\kern0pt}\ k{\isacartoucheclose}\isanewline
\ \ \ \ \ \ \isacommand{by}\isamarkupfalse%
\ {\isacharparenleft}{\kern0pt}clarsimp\ simp{\isacharcolon}{\kern0pt}\ field{\isacharunderscore}{\kern0pt}simps{\isacharparenright}{\kern0pt}\isanewline
\ \ \ \ \isacommand{show}\isamarkupfalse%
\ {\isacharquery}{\kern0pt}thesis\isanewline
\ \ \ \ \ \ \isacommand{apply}\isamarkupfalse%
\ {\isacharparenleft}{\kern0pt}rule\ has{\isacharunderscore}{\kern0pt}vderiv{\isacharunderscore}{\kern0pt}mono{\isacharunderscore}{\kern0pt}test{\isacharparenleft}{\kern0pt}{\isadigit{2}}{\isacharparenright}{\kern0pt}{\isacharbrackleft}{\kern0pt}of\ {\isachardoublequoteopen}{\isacharbraceleft}{\kern0pt}{\isadigit{2}}{\isacharasterisk}{\kern0pt}m\isactrlsub {\isadigit{0}}{\isacharslash}{\kern0pt}k{\isachardot}{\kern0pt}{\isachardot}{\kern0pt}{\isacharbraceright}{\kern0pt}{\isachardoublequoteclose}\ {\isacharquery}{\kern0pt}f\ {\isacharquery}{\kern0pt}f{\isacharprime}{\kern0pt}{\isacharbrackright}{\kern0pt}{\isacharparenright}{\kern0pt}\isanewline
\ \ \ \ \ \ \isacommand{using}\isamarkupfalse%
\ False\ {\isacartoucheopen}k\ {\isachargreater}{\kern0pt}\ {\isadigit{1}}{\isacartoucheclose}\ obs\isanewline
\ \ \ \ \ \ \isacommand{by}\isamarkupfalse%
\ {\isacharparenleft}{\kern0pt}auto\ intro{\isacharbang}{\kern0pt}{\isacharcolon}{\kern0pt}\ vderiv{\isacharunderscore}{\kern0pt}intros\ simp{\isacharcolon}{\kern0pt}\ field{\isacharunderscore}{\kern0pt}simps{\isacharparenright}{\kern0pt}\isanewline
\ \ \isacommand{qed}\isamarkupfalse%
\isanewline
\ \ \isacommand{ultimately}\isamarkupfalse%
\ \isacommand{show}\isamarkupfalse%
\ {\isacharquery}{\kern0pt}thesis\isanewline
\ \ \ \ \isacommand{by}\isamarkupfalse%
\ simp\isanewline
\isacommand{qed}
\isanewline
\end{isabellebody}

Our proof of \isa{rocket{\isacharunderscore}{\kern0pt}arith}
analyses the altitude behaviour to the left ($x\leq 2m_0/k$) and
right ($x> 2m_0/k$) of the maximum altitude. Using the first derivative 
test, we show that, to the left, the altitude is increasing, and to the right,
it is decreasing. Thus, the rocket achieves a maximum altitude at 
$\langle 2m_0/k,2m_0^3/(3k^2)\rangle$. Accordingly, we repeat our use 
of the derivative test to show that the velocity remains above $0$ in the 
interval $[0,2m_0/k]$. 

This example illustrates the relevance of our additions of forward 
diamonds and derivative tests. The diamonds allow us to do reachability
proofs and show progress for hybrid systems. The tests enable us to 
prove emerging real-arithmetic proof obligations after VCG and
provide the basis for increasing automation in this subtask
of the verification process.

\looseness=-1
We have showcased various examples illustrating the automation
added to our hybrid system verification framework. We refer readers 
interested in more examples using, for instance, our 
integration of vectors and matrices to previous publications and our
participation in ARCH competition 
reports~\cite{FosterMGS21,MitschMJZWZ20,ARCH22,Munive20}.

\section{Related Work}\label{sec:related-work}

We know of two complementary approaches to the verification of hybrid systems: reachability analysis and deductive verification. Reachability analysis~\cite{Adimoolam022,Althoff15, BogomolovFFPS19,BuLWCL10,ChenAS13,
  FrehseGDCRLRGDM11, FrehseKRM06, LiZBSM23} approximates the set of all reachable states of a hybrid system via the
iteration of its transition relation until reaching a fixed point or a specification-violating state. This approach iteratively
explores a hybrid system's state space and finds states that violate specified properties.

Our focus is on the deductive verification of hybrid systems with interactive theorem provers (ITPs)~\cite{Abraham-MummSH01,
  AnandK15,FosterMGS21,KeYmaera,Platzer10,RickettsMAGL15,ShengBZ23,WangZZ15}.
%\cite{Abraham-MummSH01, AnandK15,FosterMGS21,KeYmaera,MiaoCZZCWJR23,Platzer10,RickettsMAGL15, Schirmer2006,ShengBZ23,WangZZ15}. 
It uses mathematical proofs to establish
adherence to safety specifications for all system states. Examples of this approach
in relevant general-purpose ITPs include a formalisation of hybrid automata and 
invariant reasoning for them in the PVS prover~\cite{Abraham-MummSH01}, a
shallowly embedded implementation of a Logic of Events for hybrid systems reasoning
in the Coq prover~\cite{AnandK15}, or the use of the Coquelicot library for formalising
a temporal logic of actions with the same purpose~\cite{RickettsMAGL15}. All these
approaches have different semantics from our predicate transformer one and arguably employ less automated ITPs than our choice. In Isabelle/HOL, % work towards verification of control theory has started~\cite{JasimS17}. More broadly, 
Hoare-style verification
and refinement frameworks have appeared~\cite{ArmstrongGS16, Lammich19, Schirmer2006} and fewer have been specialised to hybrid systems~\cite{FosterMS20}. These frameworks could be combined with our own
in the spirit of working towards a full hybrid systems development environment within
Isabelle/HOL.

The formalisms to describe and prove hybrid systems' correctness specifications are as
diverse as the tools to implement such deductive verification systems. The \textsf{HHL}
and \textsf{HHLPy} provers~\cite{ShengBZ23,WangZZ15} employ a Hybrid Hoare 
Logic (HHL)~\cite{ZhanZWGJ23} for reasoning about Hybrid Communicating Sequential 
Processes (HCSPs) with the duration calculus in Isabelle/HOL and Python respectively.
Analogously, the KeYmaera and KeYmaeara X provers implement versions of differential
dynamic logic \dL, a logic to reason about hybrid systems~\cite{KeYmaera,Platzer10}.
Our work has been compared with these provers in hybrid system verification 
competitions~\cite{MitschMJZWZ20,ARCH22}. Yet, the tools are very different in nature and 
implementation.
While both families of provers implement specific logics, our development is flexible and
includes a differential Hoare logic, a refinement calculus~\cite{FosterMS20}, rules from 
differential dynamic logic~\cite{FosterMGS21}, and linear systems (matrix) integrations~\cite{Munive20}. 
Moreover, both the HHL and KeYmaera families 
have integrated unverified tools into their certification process. The \textsf{HHLPy} prover
is mainly written in Python, while KeYmaera X uses the Wolfram Engine and/or Z3 
as black-box solvers for quantifying elimination procedures. Albeit, the first HHL prover 
is verified with Isabelle/HOL, and there is work towards verifying real arithmetic algorithms 
for integration into KeYmaera X~\cite{ScharagerCMP21}. In contrast, our work has taken
a stricter approach where every input from external tools must be certified by Isabelle/HOL. 
This illustrates our long-term goal of enabling general-purpose ITPs with fully automated 
% hybrid systems 
verification capabilities. % This is in contrast with a different trend in proof assistants' use where they serve for the formalisation and verification of tools that can help in the safety assurance processes~\cite{BohrerRVVP17,afp:dgl}.

%In comparison to our concrete contributions on this paper, 
%\sfin{Missing related work on lenses, e.g. how does this work compare with RoboChart}
%There are various formalisations of separation logic in general purpose proof 
%assistants~\cite{JungSSSTBD15,LammichM12,MartiA09,Tuerk11,VarmingB08}.
%\sfin{Missing related work on implementation of separation logic in proof assistants}

We built our IsaVODEs framework on top of the Archive of Formal Proofs (AFP) entry for Ordinary Differential 
Equations~\cite{ImmlerT19} and our own extensions to it~\cite{Munive20, MuniveS22}. Together
with Isabelle's HOL-Analysis library, they provide a thorough basis for stating real analysis theorems
in Isabelle/HOL. In particular, the library already has a different formalisation of the fact that
$C^1$-differentiation implies Lipschitz continuity. Nevertheless, that version depends on a type
of bounded linear continuous functions while our implementation avoids creating a new type and the 
corresponding abstraction functions. As a result, our version is more manageable within IsaVODEs.

Formalisations of \dL and related logics have recently appeared in the 
AFP~\cite{BohrerRVVP17,afp:dgl} but they are not intended as verification tools and, therefore, they are incomparable with our framework. % They have served to increase trust in previous \dL-related work rather than being part of a CPS verification framework inside a general-purpose ITP. A direct comparison of these works with our formalisation of the \dL rules is difficult because those implementations are deeply embedded while ours have added lens-conditions due to our shallow embedding.
Despite the implementation differences, we compare KeYmaera X and our support for \dL reasoning. Our personal experience indicates that using (one-step) differential cuts, weakenings, and inductions is similar to their use in KeYmaera X. However, supplying solutions works differently because our framework requires certifying or assuming their uniqueness. In contrast, the uniqueness of solutions is guaranteed by \dL's syntax. Our work in this paper automates this certification process. Finally, our verification of the soundness of the differential ghost rule presented here could be generalised further to match all the cases prescribed by \dL's syntactic implementation.
\section{Conclusions and Future Work}\label{sec:conclusions}

% We invite readers to contribute to the extensibility of ITPs for CPS verification.

In this paper, we have described IsaVODEs, our framework for verifying cyber-physical systems in Isabelle/HOL. This substantial development includes both strong theoretical foundations and practical verification support, provided by automated theorem provers and an integration with computer algebra systems. Our language and verification technique extends \dL's hybrid programs in several ways, notably with matrices to support engineering mathematics, and frames to support modular reasoning. We have validated our tool with a substantial library of benchmarks and examples. 

Here we have improved our framework by formalising and proving VCG laws about forward diamonds which enable reasoning about the reachability or progress of hybrid systems. We have generalised our frame laws and \dL-style differential ghosts rule, allowing us to derive related Darboux rules. We have formalised foundational theorems like the fact that differentiable functions are Lipschitz-continuous, and the first and second derivative test laws. These support the practical goal of increasing automation via our proof methods for performing differential induction or VCG via supplying flows. Our integration of CASs into this process makes the verification experience with flows seamless. Finally, we have evaluated the benefits of all these additions with various verification examples.
%\sfin{Summarise contributions, and benefits. Highlight future work.}

Our Isabelle-based approach is inherently extensible. We can add syntax and semantics for bespoke program operators and associated Hoare-logic rules to support tailoring for particular models. Overall, verifying CPSs using IsaVODEs benefits from the wealth of technology provided by Isabelle, notably the frontend, asynchronous document processing, the theory library, proof automation, and support for code generation. We need not be limited to a single notation but can provide semantics for established 
engineering notations. IsaVODEs benefits from the fact that Isabelle is a gateway for a variety of other verification tools through ``hammers'', 
such as SMT solvers, model generators, and computer algebra systems. Our additions in this paper increase
IsaVODEs' usability for complex verifications. We believe these advantages can allow the integration of our technology into software engineering workflows. 

A limitation of our current approach occurs when the arithmetic obligations at the 
end of the verification are too complex for SMT solvers~\cite{PaulsonB10}. Currently,
there are two options: users can manually prove these obligations themselves, or they
can assert them at the cost of increasing uncertainty in their verification. In these 
cases, the ideal approach would connect tools deciding these expressions, e.g. 
CAS systems or domain-specific automated provers~\cite{AkbarpourP10} in a way that IsaVODEs certifies the underlying reasoning. We leave this
development for future work.

Another avenue of improvement, following our introduction of the forward diamond
in our framework, is the addition of the remaining modal operators and their VCG rules~\cite{afp:transem}. 
Currently, we have only formalised a backward diamond but its VCG rules remain to be 
proved. Their implementation could lead to a framework for incorrectness analysis~\cite{OHearn20} 
of hybrid systems complementing current testing and simulation techniques. We will also consider supporting further extensions to hybrid programs, such as quantified \dL and differential-algebraic logic (DAL), which can both provide additional modelling capacity.

In terms of alternative uses of our framework, we expect \dL-style security analysis 
about hybrid systems~\cite{XiangFC21} to be easily done in IsaVODEs too. 
Similarly, IsaVODEs foundations have been used as semantics for other model-based
robot development technologies~\cite{BaxterCCJ23,CavalcantiABMR23}. Therefore, 
IsaVODEs proofs could be integrated into these tools for increased confidence in the 
performed analysis.

\paragraph{Acknowledgements} This preprint has not undergone peer review (when applicable) or any post-submission improvements or corrections. 

\paragraph{Funding statements}
A Novo Nordisk Fonden Start Package Grant (NNF20OC0063462) and a Horizon MSCA 2022 Postdoctoral Fellowship (project acronym DeepIsaHOL and number 101102608) partially supported the first author during the development of this article. Views and opinions expressed are however those of the author(s) only and do not necessarily reflect those of the European Union or the European Research Executive Agency. Neither the European Union nor the European Research Executive Agency can be held responsible for them. The work was also funded by UKRI-EPSRC project CyPhyAssure (grant reference EP/S001190/1), the Assuring Autonomy International Programme (AAIP; grant CSI:Cobot), a partnership between Lloyd’s Register Foundation and the University of York, and Labex DigiCosme through an invited professorship of the fourth author at the Laboratoire d'informatique de l'\'Ecole polytechnique.

\bibliographystyle{abbrv}
\bibliography{main.bib}

\begin{thebibliography}{10}

\bibitem{Abraham-MummSH01}
E.~{\'{A}}brah{\'{a}}m{-}Mumm, M.~Steffen, and U.~Hannemann.
\newblock Verification of hybrid systems: Formalization and proof rules in
  {PVS}.
\newblock In {\em {ICECCS} 2001}, pages 48--57, New Jersey, 2001. {IEEE}
  Computer Society.

\bibitem{AckermanGRM65}
E.~Ackerman, L.~C. Gatewood, J.~W. Rosevear, and G.~D. Molnar.
\newblock Model studies of blood-glucose regulation.
\newblock {\em The bulletin of mathematical biophysics}, 27(5):21--37, 1965.
\newblock https://doi.org/10.1007/BF02477259.

\bibitem{Adimoolam022}
A.~S. Adimoolam and T.~Dang.
\newblock Safety verification of networked control systems by complex
  zonotopes.
\newblock {\em Leibniz Trans. Embed. Syst.}, 8(2):01:1--01:22, 2022.

\bibitem{AkbarpourP10}
B.~Akbarpour and L.~C. Paulson.
\newblock {MetiTarski}: An automatic theorem prover for real-valued special
  functions.
\newblock {\em JAR}, 44(3):175--205, 2010.

\bibitem{AlmeidaBBBGLOPS17}
J.~B. Almeida, M.~Barbosa, G.~Barthe, A.~Blot, B.~Gr{\'{e}}goire, V.~Laporte,
  T.~Oliveira, H.~Pacheco, B.~Schmidt, and P.~Strub.
\newblock Jasmin: High-assurance and high-speed cryptography.
\newblock In {\em {CCS} 2017}, pages 1807--1823, New York, 2017. {ACM}.

\bibitem{Althoff15}
M.~Althoff.
\newblock An introduction to {CORA} 2015.
\newblock In {\em {ARCH} 2015}, volume~34, pages 120--151, EasyChair, 2015.
  EasyChair.

\bibitem{AnandK15}
A.~Anand and R.~A. Knepper.
\newblock Roscoq: Robots powered by constructive reals.
\newblock In {\em {ITP}}, volume 9236 of {\em LNCS}, pages 34--50, Heidelberg,
  2015. Springer.

\bibitem{AransayD17}
J.~Aransay and J.~Divas{\'{o}}n.
\newblock A formalisation in {HOL} of the fundamental theorem of linear algebra
  and its application to the solution of the least squares problem.
\newblock {\em JAR}, 58(4):509--535, 2017.

\bibitem{ArmstrongGS16}
A.~Armstrong, V.~B.~F. Gomes, and G.~Struth.
\newblock Building program construction and verification tools from algebraic
  principles.
\newblock {\em Formal Aspects of Computing}, 28(2):265--293, 2016.

\bibitem{BackW98}
R.~Back and J.~von Wright.
\newblock {\em Refinement Calculus---{A} Systematic Introduction}.
\newblock Springer, Heidelberg, 1998.

\bibitem{BasinDHHMKKMRST22}
D.~A. Basin, T.~Dardinier, N.~Hauser, L.~Heimes, J.~J.~H. y~Munive,
  N.~Kaletsch, S.~Krstic, E.~Marsicano, M.~Raszyk, J.~Schneider, D.~L. Tirore,
  D.~Traytel, and S.~Zingg.
\newblock {VeriMon}: {A} formally verified monitoring tool.
\newblock In H.~Seidl, Z.~Liu, and C.~S. Pasareanu, editors, {\em {ICTAC}
  2022}, volume 13572 of {\em LNCS}, pages 1--6, Heidelberg, 2022. Springer.

\bibitem{BaxterCCJ23}
J.~Baxter, G.~Carvalho, A.~Cavalcanti, and F.~R. J\'{u}nior.
\newblock Roboworld: Verification of robotic systems with environment in the
  loop.
\newblock {\em FAC}, 2023.
\newblock Just Accepted.

\bibitem{Blanchette2016Hammers}
J.~C. Blanchette, C.~Kaliszyk, L.~C. Paulson, and J.~Urban.
\newblock Hammering towards {QED}.
\newblock {\em Journal of Formalized Reasoning}, 9(1), 2016.

\bibitem{BodinCFGMNSS14}
M.~Bodin, A.~Chargu{\'{e}}raud, D.~Filaretti, P.~Gardner, S.~Maffeis,
  D.~Naudziuniene, A.~Schmitt, and G.~Smith.
\newblock A trusted mechanised {J}ava{S}cript specification.
\newblock In {\em {POPL} 2014}, pages 87--100, New York, 2014. {ACM}.

\bibitem{BogomolovFFPS19}
S.~Bogomolov, M.~Forets, G.~Frehse, K.~Potomkin, and C.~Schilling.
\newblock {JuliaReach}: a toolbox for set-based reachability.
\newblock In N.~Ozay and P.~Prabhakar, editors, {\em {HSCC} 2019}, pages
  39--44, New York, 2019. {ACM}.

\bibitem{BohrerRVVP17}
R.~Bohrer, V.~Rahli, I.~Vukotic, M.~V{\"{o}}lp, and A.~Platzer.
\newblock Formally verified differential dynamic logic.
\newblock In {\em CPP}, pages 208--221, New York, 2017. {ACM}.

\bibitem{BuLWCL10}
L.~Bu, Y.~Li, L.~Wang, X.~Chen, and X.~Li.
\newblock {BACH} 2 : Bounded reachability checker for compositional linear
  hybrid systems.
\newblock In {\em {DATE} 2010}, pages 1512--1517, New Jersey, 2010. {IEEE}
  Computer Society.

\bibitem{CavalcantiABMR23}
A.~Cavalcanti, Z.~Attala, J.~Baxter, A.~Miyazawa, and P.~Ribeiro.
\newblock {\em Model-Based Engineering for Robotics with {RoboChart}
  and {RoboTool}}, pages 106--151.
\newblock Springer, Heidelberg, 2023.

\bibitem{ChenAS13}
X.~Chen, E.~{\'{A}}brah{\'{a}}m, and S.~Sankaranarayanan.
\newblock Flow*: An analyzer for non-linear hybrid systems.
\newblock In {\em {CAV} 2013}, volume 8044 of {\em LNCS}, pages 258--263,
  Heidelberg, 2013. Springer.

\bibitem{Cordwell2021BKR}
K.~Cordwell, K.~T. Yong, and P.~A.
\newblock A verified decision procedure for univariate real arithmetic with the
  {BKR} algorithm.
\newblock In L.~Cohen and C.~Kaliszyk, editors, {\em ITP}, volume 193 of {\em
  Leibniz International Proceedings in Informatics (LIPIcs)}, pages
  14:1--14:20, Germany, 2021. Schloss Dagstuhl -- Leibniz-Zentrum f{\"u}r
  Informatik.

\bibitem{Foster09}
J.~Foster.
\newblock {\em Bidirectional programming languages}.
\newblock PhD thesis, University of Pennsylvania, 2009.

\bibitem{Foster2020-LocalVars}
S.~Foster and J.~Baxter.
\newblock Automated algebraic reasoning for collections and local variables
  with lenses.
\newblock In {\em {RAMiCS}}, volume 12062 of {\em LNCS}, Heidelberg, 2020.
  Springer.

\bibitem{Foster2020-IsabelleUTP}
S.~Foster, J.~Baxter, A.~Cavalcanti, J.~Woodcock, and F.~Zeyda.
\newblock Unifying semantic foundations for automated verification tools in
  {Isabelle/UTP}.
\newblock {\em Science of Computer Programming}, 197, October 2020.

\bibitem{FosterGC20}
S.~Foster, M.~Gleirscher, and R.~Calinescu.
\newblock Towards deductive verification of control algorithms for autonomous
  marine vehicles.
\newblock In {\em {ICECCS}}, New Jersey, October 2020. IEEE.

\bibitem{FosterMS20}
S.~Foster, J.~J. {Huerta y Munive}, and G.~Struth.
\newblock Differential {H}oare logics and refinement calculi for hybrid systems
  with {I}sabelle/{HOL}.
\newblock In {\em RAMiCS[postponed]}, volume 12062 of {\em LNCS}, pages
  169--186, 2020.

\bibitem{Foster2021-IsaSACM}
S.~Foster, Y.~Nemouchi, M.~Gleirscher, R.~Wei, and T.~Kelly.
\newblock Integration of formal proof into unified assurance cases with
  {Isabelle/SACM}.
\newblock {\em Formal Aspects of Computing}, 2021.

\bibitem{FosterMGS21}
S.~Foster, J.~J.~H. y~Munive, M.~Gleirscher, and G.~Struth.
\newblock Hybrid systems verification with {I}sabelle/{HOL}: Simpler syntax,
  better models, faster proofs.
\newblock In {\em {FM} 2021}, volume 13047 of {\em LNCS}, pages 367--386,
  Heidelberg, 2021. Springer.

\bibitem{Optics-AFP}
S.~Foster and F.~Zeyda.
\newblock Optics.
\newblock {\em Archive of Formal Proofs}, May 2017.

\bibitem{FrehseGDCRLRGDM11}
G.~Frehse, C.~L. Guernic, A.~Donz{\'{e}}, S.~Cotton, R.~Ray, O.~Lebeltel,
  R.~Ripado, A.~Girard, T.~Dang, and O.~Maler.
\newblock {SpaceEx}: Scalable verification of hybrid systems.
\newblock In {\em {CAV} 2011}, volume 6806 of {\em LNCS}, pages 379--395,
  Heidelberg, 2011. Springer.

\bibitem{FrehseKRM06}
G.~Frehse, B.~H. Krogh, R.~A. Rutenbar, and O.~Maler.
\newblock Time domain verification of oscillator circuit properties.
\newblock In O.~Maler, editor, {\em {FAC} 2005}, volume 153 of {\em ENTCS},
  pages 9--22, Amsterdam, 2005. Elsevier.

\bibitem{KeYmaera}
N.~Fulton, S.~Mitsch, J.-D. Quesel, M.~V{\"o}lp, and A.~Platzer.
\newblock {KeYmaera X}: An axiomatic tactical theorem prover for hybrid
  systems.
\newblock In A.~P. Felty and A.~Middeldorp, editors, {\em CADE}, volume 9195 of
  {\em LNCS}, pages 527--538, Heidelberg, 2015. Springer.

\bibitem{GallicchioTMP22}
J.~Gallicchio, Y.~K. Tan, S.~Mitsch, and A.~Platzer.
\newblock Implicit definitions with differential equations for keymaera {X} -
  (system description).
\newblock In {\em {IJCAR} 2022}, pages 723--733, 2022.

\bibitem{GomesS16}
V.~B.~F. Gomes and G.~Struth.
\newblock Modal {K}leene algebra applied to program correctness.
\newblock In {\em {FM}}, volume 9995 of {\em LNCS}, pages 310--325, 2016.

\bibitem{Haftman2010-CodeGen}
F.~Haftmann and T.~Nipkow.
\newblock Code generation via higher-order rewrite systems.
\newblock In {\em 10th Intl. Symp. on Functional and Logic Programming
  (FLOPS)}, volume 6009 of {\em LNCS}, pages 103--117, Heidelberg, 2010.
  Springer.

\bibitem{HarelKT00}
D.~Harel, D.~Kozen, and J.~Tiuryn.
\newblock {\em Dynamic Logic}.
\newblock MIT Press, Massachusetts, 2000.

\bibitem{Holzl2009-Approximate}
J.~H\"{o}lzl.
\newblock Proving inequalities over reals with computation in {I}sabelle/{HOL}.
\newblock In {\em PLMMS}, pages 38--45, New York, 2009. ACM.

\bibitem{HolzlIH13}
J.~H{\"{o}}lzl, F.~Immler, and B.~Huffman.
\newblock Type classes and filters for mathematical analysis in
  {I}sabelle/{HOL}.
\newblock In {\em ITP}, volume 7998 of {\em LNCS}, pages 279--294, Heidelberg,
  2013. Springer.

\bibitem{Munive20}
J.~J. {Huerta y Munive}.
\newblock Affine systems of {ODE}s in {I}sabelle/{HOL} for hybrid-program
  verification.
\newblock In {\em {SEFM}}, volume 12310 of {\em LNCS}, pages 77--92,
  Heidelberg, 2020. Springer.

\bibitem{Munive21}
J.~J. Huerta~y Munive.
\newblock {\em {Algebraic verification of hybrid systems in Isabelle/HOL}}.
\newblock PhD thesis, The University of Sheffield, 2021.

\bibitem{MuniveS22}
J.~J. Huerta~y Munive and G.~Struth.
\newblock Predicate transformer semantics for hybrid systems.
\newblock {\em JAR}, 66(1):93--139, 2022.

\bibitem{afp:odes}
F.~Immler and J.~H{\"{o}}lzl.
\newblock Ordinary differential equations.
\newblock {\em Archive of Formal Proofs}, 2012.

\bibitem{ImmlerT16}
F.~Immler and C.~Traut.
\newblock The flow of {ODE}s: Formalization of variational equation and
  poincaré.
\newblock In {\em ITP 2016}, volume 9807 of {\em LNCS}, pages 184--199,
  Heidelberg, 2016. Springer.

\bibitem{ImmlerT19}
F.~Immler and C.~Traut.
\newblock The flow of {ODE}s: Formalization of variational equation and
  {P}oincar{\'{e}} map.
\newblock {\em JAR}, 62(2):215--236, 2019.

\bibitem{JeanninGKSGMP17}
J.~Jeannin, K.~Ghorbal, Y.~Kouskoulas, A.~Schmidt, R.~Gardner, S.~Mitsch, and
  A.~Platzer.
\newblock A formally verified hybrid system for safe advisories in the
  next-generation airborne collision avoidance system.
\newblock {\em {STTT}}, 19(6):717--741, 2017.

\bibitem{Klein2009}
G.~Klein et~al.
\newblock {seL4}: Formal verification of an {OS} kernel.
\newblock In {\em Proc. 22nd Symp. on Operating Systems Principles (SOSP)},
  pages 207--220, New York, 2009. ACM.

\bibitem{Kuncar2019-IsabelleFoundation}
O.~Kuncar and A.~Popescu.
\newblock A consistent foundation for {Isabelle/HOL}.
\newblock {\em JAR}, 62:531--555, 2019.

\bibitem{Lammich19LLVM}
P.~Lammich.
\newblock Generating verified {LLVM} from {Isabelle/HOL}.
\newblock In {\em {ITP} 2019}, volume 141 of {\em LIPIcs}, pages 22:1--22:19,
  Germany, 2019. Schloss Dagstuhl - Leibniz-Zentrum f{\"{u}}r Informatik.

\bibitem{Lammich19}
P.~Lammich.
\newblock Refinement to imperative {HOL}.
\newblock {\em JAR}, 62(4):481--503, 2019.

\bibitem{LeroyBKSPF2016}
X.~Leroy, S.~Blazy, D.~K\"astner, B.~Schommer, M.~Pister, and C.~Ferdinand.
\newblock Compcert -- a formally verified optimizing compiler.
\newblock In {\em ERTS 2016: Embedded Real Time Software and Systems}, France,
  2016. SEE.

\bibitem{Li2017-Poly}
W.~Li, G.~Passmore, and L.~Paulson.
\newblock Deciding univariate polynomial problems using untrusted certificates
  in {Isabelle/HOL}.
\newblock {\em J. Autom. Reasoning}, 62:29--91, 2019.

\bibitem{LiZBSM23}
Y.~Li, H.~Zhu, K.~Braught, K.~Shen, and S.~Mitra.
\newblock Verse: {A} python library for reasoning about multi-agent hybrid
  system scenarios.
\newblock In {\em {CAV} 2023}, volume 13964 of {\em LNCS}, pages 351--364,
  Heidelberg, 2023. Springer.

\bibitem{LoosPN11}
S.~M. Loos, A.~Platzer, and L.~Nistor.
\newblock Adaptive cruise control: Hybrid, distributed, and now formally
  verified.
\newblock In {\em {FM} 2011}, volume 6664 of {\em LNCS}, pages 42--56,
  Heidelberg, 2011. Springer.

\bibitem{MatichukMW16}
D.~Matichuk, T.~C. Murray, and M.~Wenzel.
\newblock Eisbach: {A} proof method language for {I}sabelle.
\newblock {\em J. Automated Reasoning}, 56(3):261--282, 2016.

\bibitem{MitschMJZWZ20}
S.~Mitsch, J.~J. {Huerta y Munive}, X.~Jin, B.~Zhan, S.~Wang, and N.~Zhan.
\newblock {ARCH-COMP20} category report: Hybrid systems theorem proving.
\newblock In {\em {ARCH20.}}, volume~74, pages 141--161, EasyChair, 2020.
  EasyChair.

\bibitem{Mitsch2018}
S.~Mitsch and A.~Platzer.
\newblock Verified runtime validation for partially observable hybrid systems.
\newblock {\em CoRR}, abs/1811.06502, 2018.

\bibitem{ARCH22}
S.~Mitsch, B.~Zhan, H.~Sheng, A.~Bentkamp, X.~Jin, S.~Wang, S.~Foster, C.~P.
  Laursen, and J.~J.~H. y~Munive.
\newblock {ARCH-COMP22} category report: Hybrid systems theorem proving.
\newblock In {\em {ARCH22}}, volume~90 of {\em EPiC Series in Computing}, pages
  185--203, Munich, 2022. EasyChair.

\bibitem{Miyazawa2019-RoboChart}
A.~Miyazawa, P.~Ribeiro, W.~Li, A.~Cavalcanti, J.~Timmis, and J.~Woodcock.
\newblock {RoboChart}: modelling and verification of the functional behaviour
  of robotic applications.
\newblock {\em Software and Systems Modelling}, 18:3097--3149, January 2019.

\bibitem{OHearn20}
P.~W. O'Hearn.
\newblock Incorrectness logic.
\newblock {\em Proc. {ACM} Program. Lang.}, 4({POPL}):10:1--10:32, 2020.

\bibitem{Oles82}
F.~Oles.
\newblock {\em A Category-Theoretic Approach to the Semantics of Programming
  Languages}.
\newblock PhD thesis, Syracuse University, 1982.

\bibitem{Paulson96}
L.~C. Paulson.
\newblock {\em {ML} for the working programmer {(2.} ed.)}.
\newblock Cambridge University Press, Cambridge, 1996.
\newblock \url{https://www.cl.cam.ac.uk/~lp15/MLbook/}.

\bibitem{PaulsonB10}
L.~C. Paulson and J.~C. Blanchette.
\newblock Three years of experience with {Sledgehammer}, a practical link
  between automatic and interactive theorem provers.
\newblock In G.~Sutcliffe, S.~Schulz, and E.~Ternovska, editors, {\em {IWIL}
  2010}, volume~2 of {\em EPiC Series in Computing}, pages 1--11, EasyChair,
  2010. EasyChair.

\bibitem{Platzer12}
A.~Platzer.
\newblock The structure of differential invariants and differential cut
  elimination.
\newblock {\em Logical Methods in Computer Science}, 8(4), 2008.

\bibitem{Platzer10}
A.~Platzer.
\newblock {\em Logical Analysis of Hybrid Systems}.
\newblock Springer, Heidelberg, 2010.

\bibitem{Platzer12a}
A.~Platzer.
\newblock Logics of dynamical systems.
\newblock In {\em LICS}, pages 13--24, 2012.
\newblock https://doi.org/10.1109/LICS.2012.13.

\bibitem{Platzer18}
A.~Platzer.
\newblock {\em Logical Foundations of Cyber-Physical Systems}.
\newblock Springer, Heidelberg, 2018.

\bibitem{afp:dgl}
A.~Platzer.
\newblock Differential game logic.
\newblock {\em Archive of Formal Proofs}, 2019, 2019.

\bibitem{PlatzerQ09}
A.~Platzer and J.~Quesel.
\newblock European train control system: {A} case study in formal verification.
\newblock In {\em {ICFEM}}, volume 5885 of {\em LNCS}, pages 246--265,
  Heidelberg, 2009. Springer.

\bibitem{PlatzerT18}
A.~Platzer and Y.~K. Tan.
\newblock Differential equation axiomatization: The impressive power of
  differential ghosts.
\newblock In {\em {LICS}}, pages 819--828, New Jersey, 2018. {ACM}.

\bibitem{QueselMLAP16}
J.~Quesel, S.~Mitsch, S.~M. Loos, N.~Arechiga, and A.~Platzer.
\newblock How to model and prove hybrid systems with {K}e{Y}maera: a tutorial
  on safety.
\newblock {\em {STTT}}, 18(1):67--91, 2016.

\bibitem{Reynolds2002}
J.~Reynolds.
\newblock Separation logic: a logic for shared mutable data structures.
\newblock In {\em LICS}, New Jersey, 2002. IEEE.

\bibitem{RickettsMAGL15}
D.~Ricketts, G.~Malecha, M.~M. Alvarez, V.~Gowda, and S.~Lerner.
\newblock Towards verification of hybrid systems in a foundational proof
  assistant.
\newblock In {\em {MEMOCODE}}, pages 248--257, New Jersey, 2015. {IEEE}.

\bibitem{ScharagerCMP21}
M.~Scharager, K.~Cordwell, S.~Mitsch, and A.~Platzer.
\newblock Verified quadratic virtual substitution for real arithmetic.
\newblock In M.~Huisman, C.~S. Pasareanu, and N.~Zhan, editors, {\em {FM}
  2021}, volume 13047 of {\em LNCS}, pages 200--217, Heidelberg, 2021.
  Springer.

\bibitem{Schirmer2006}
N.~Schirmer.
\newblock {\em Verification of sequential imperative programs in Isabelle-HOL}.
\newblock PhD thesis, Technical University Munich, Germany, 2006.

\bibitem{Schirmer2009}
N.~Schirmer and M.~Wenzel.
\newblock State spaces -- the locale way.
\newblock In {\em SSV 2009}, volume 254 of {\em ENTCS}, pages 161--179, 2009.

\bibitem{ShengBZ23}
H.~Sheng, A.~Bentkamp, and B.~Zhan.
\newblock {HHLPy}: Practical verification of hybrid systems using hoare logic.
\newblock In M.~Chechik, J.~Katoen, and M.~Leucker, editors, {\em {FM} 2023},
  volume 14000 of {\em LNCS}, pages 160--178, Heidelberg, 2023. Springer.

\bibitem{Stanescu09}
D.~Stanescu and B.~M. Chen-Charpentier.
\newblock Random coefficient differential equation models for bacterial growth.
\newblock {\em Mathematical and Computer Modelling}, 50(5):885--895, 2009.
\newblock https://doi.org/10.1016/j.mcm.2009.05.017.

\bibitem{afp:transem}
G.~Struth.
\newblock Transformer semantics.
\newblock {\em Archive of Formal Proofs}, 2018.

\bibitem{Teschl12}
G.~Teschl.
\newblock {\em Ordinary Differential Equations and Dynamical Systems}.
\newblock AMS, Rhode Island, 2012.

\bibitem{sagemath}
{The Sage Developers}.
\newblock {\em {SageMath, the Sage Mathematics Software System (Version 9.0)}},
  2020.

\bibitem{WangZZ15}
S.~Wang, N.~Zhan, and L.~Zou.
\newblock An improved {HHL} prover: An interactive theorem prover for hybrid
  systems.
\newblock In {\em {ICFEM}}, volume 9407 of {\em LNCS}, pages 382--399, 2015.

\bibitem{Weinert09}
F.~Weinert.
\newblock {\em Radioactive Decay Law (Rutherford--Soddy)}.
\newblock Springer, Berlin, Heidelberg, 2009.
\newblock https://doi.org/10.1007/978-3-540-70626-7\_183.

\bibitem{MakariusIsarManual}
M.~Wenzel, C.~Ballarin, S.~Berghofer, J.~Blanchette, T.~Bourke, L.~Bulwahn,
  A.~Chaieb, L.~Dixon, F.~Haftmann, B.~Huffman, L.~Hupel, G.~Klein, A.~Krauss,
  O.~Kuncar, A.~Lochbihler, T.~Nipkow, L.~Noschinski, D.~von Oheimb,
  L.~Paulson, S.~Skalberg, C.~Sternagel, and D.~Traytel.
\newblock The {Isabelle/Isar} reference manual, 2023.
\newblock \url{https://isabelle.in.tum.de/doc/isar-ref.pdf}.

\bibitem{wolfram_expression}
{Wolfram Research}.
\newblock Expressions - wolfram language.

\bibitem{wolfram_engine}
{Wolfram Research}.
\newblock Wolfram engine.

\bibitem{XiangFC21}
J.~Xiang, N.~Fulton, and S.~Chong.
\newblock Relational analysis of sensor attacks on cyber-physical systems.
\newblock In {\em {CSF} 2021}, pages 1--16, New Jersey, 2021. {IEEE}.

\bibitem{ZhanZWGJ23}
N.~Zhan, B.~Zhan, S.~Wang, D.~P. Guelev, and X.~Jin.
\newblock A generalized hybrid {Hoare} logic.
\newblock {\em CoRR}, abs/2303.15020, 2023.

\end{thebibliography}

\end{document}